%% file: main.tex
\documentclass[preprint,twocolumn]{aastex63}
\usepackage{float, bm, graphicx, amsmath, morefloats}
\usepackage[caption=false]{subfig}
\usepackage{booktabs}
\PassOptionsToPackage{table}{xcolor}   

\usepackage[table]{xcolor}
\usepackage{placeins}

\shorttitle{Low Luminosity Type IIP Supernovae}
\shortauthors{Das et al.}

\usepackage{float,graphicx,amsmath,tabularx,booktabs,natbib,threeparttable}
\usepackage{float, bm, graphicx, amsmath, morefloats}
\usepackage[caption=false]{subfig}
\usepackage{float,graphicx,amsmath,tabularx,booktabs,natbib,threeparttable}
\usepackage{subfig}
\usepackage{longtable}
\usepackage{booktabs}

\definecolor{ecsnAgree}{HTML}{B8E186}    

\definecolor{ecsnDisagree}{HTML}{FDBF6F}  
\definecolor{ecsnMaybe}{HTML}{D9D9D9}     

\usepackage{rotating}
\usepackage[table]{xcolor}
\usepackage{booktabs}
\usepackage{colortbl}
\usepackage{array}

\usepackage{booktabs}

\usepackage{comment}
\definecolor{dark-red}{rgb}{0.4,0.15,0.15}
\definecolor{dark-blue}{rgb}{0.15,0.15,0.4}
\definecolor{medium-blue}{rgb}{0,0,0.5}
\hypersetup{
    colorlinks, linkcolor={dark-red},
    citecolor={dark-blue}, urlcolor={medium-blue}
}

\newcommand{\beqa}{\begin{eqnarray}} 
\newcommand{\eeqa}{\end{eqnarray}}

\newcommand{\bsub}{\begin{subequations}}
\newcommand{\esub}{\end{subequations}}
\newcommand{\beal}{\begin{align}}
\newcommand{\ealn}{\end{align}}

\newcommand{\Msun}{{\ensuremath{\mathrm{M}_{\odot}}}}

\graphicspath{{./}{figures/}}

\begin{document}

\title{Low-Luminosity Type IIP Supernovae from the Zwicky Transient Facility Census of the Local Universe. III: Hunting for electron-capture supernovae using nebular spectroscopy}

\author[0000-0001-8372-997X]{Kaustav K.~Das}\thanks{E-mail: kdas@astro.caltech.edu}\affiliation{Cahill Center for Astrophysics, California Institute of Technology, MC 249-17, 
1200 E California Boulevard, Pasadena, CA, 91125, USA}

\author{Anders Jerkstrand}
\affiliation{The Oskar Klein Centre, Department of Astronomy, Stockholm University, AlbaNova, SE-10691 Stockholm, Sweden}

\author[0000-0002-5619-4938]{Mansi~M.~Kasliwal}
\affiliation{Cahill Center for Astrophysics, 
California Institute of Technology, MC 249-17, 
1200 E California Boulevard, Pasadena, CA, 91125, USA}

\author[0000-0003-1546-6615]{Jesper Sollerman}
\affiliation{The Oskar Klein Centre, Department of Astronomy, Stockholm University, AlbaNova, SE-10691 Stockholm, Sweden}

\author[0000-0002-4223-103X]{Christoffer Fremling}
\affil{Caltech Optical Observatories, California Institute of Technology, Pasadena, CA 91125, USA}

\author[0000-0001-6797-1889]{Steve Schulze}
\affiliation{Center for Interdisciplinary Exploration and Research in Astrophysics (CIERA), 1800 Sherman Ave., Evanston, IL 60201, USA}

\author{Avishay Gal-Yam}
\affiliation{Department of Particle Physics and Astrophysics, Weizmann Institute of Science, 234 Herzl St, 76100 Rehovot, Israel}

\author{Tomas Ahumuda}
\affiliation{Cahill Center for Astrophysics, 
California Institute of Technology, MC 249-17, 
1200 E California Boulevard, Pasadena, CA, 91125, USA}
\affiliation{Cerro Tololo Inter-American Observatory/NSF NOIRLab,
Casilla 603, La Serena, Chile}

\author[0000-0003-3768-7515]{Shreya Anand}
\affiliation{Kavli Institute for Particle Astrophysics and Cosmology, Stanford University, Stanford, CA 94305-4085, USA}

\author[0009-0001-3767-942X]{Bart van Baal}
\affiliation{The Oskar Klein Centre, Department of Astronomy, Stockholm University, AlbaNova, SE-10691 Stockholm, Sweden}

\author[0000-0002-8262-2924]{Michael W. Coughlin}
\affiliation{School of Physics and Astronomy, University of Minnesota, Minneapolis, MN 55455, USA}    

\author{Sofia Covarrubias}
\affiliation{Cahill Center for Astrophysics, 
California Institute of Technology, MC 249-17, 
1200 E California Boulevard, Pasadena, CA, 91125, USA}

\author[0000-0002-5884-7867]{Richard Dekany}
\affiliation{Caltech Optical Observatories, California Institute of Technology, Pasadena, CA  91125}

\author{Nicholas Earley}
\affiliation{Cahill Center for Astrophysics, 
California Institute of Technology, MC 249-17, 
1200 E California Boulevard, Pasadena, CA, 91125, USA}

\author[0000-0002-3934-2644]{W. V. Jacobson-Galán}
\altaffiliation{NASA Hubble Fellow}
\affiliation{Cahill Center for Astrophysics, 
California Institute of Technology, MC 249-17, 
1200 E California Boulevard, Pasadena, CA, 91125, USA}

\author[0000-0002-0987-3372]{Joahan Castaneda Jaimes}
\affiliation{Cahill Center for Astrophysics, 
California Institute of Technology, MC 249-17, 
1200 E California Boulevard, Pasadena, CA, 91125, USA}

\author[0000-0002-8532-9395]{Frank J. Masci}
\affiliation{IPAC, California Institute of Technology, 1200 E. California
             Blvd, Pasadena, CA 91125, USA}

\author[0000-0003-3658-6026]{Yu-Jing Qin}
\affiliation{Cahill Center for Astrophysics, 
California Institute of Technology, MC 249-17, 
1200 E California Boulevard, Pasadena, CA, 91125, USA}

\author{Reed Riddle}
\affiliation{Cahill Center for Astrophysics, 
California Institute of Technology, MC 249-17, 
1200 E California Boulevard, Pasadena, CA, 91125, USA}    

\author[0000-0003-4725-4481]{Sam Rose}
\affiliation{Cahill Center for Astrophysics, 
California Institute of Technology, MC 249-17, 
1200 E California Boulevard, Pasadena, CA, 91125, USA}

\author{Yashvi Sharma}
\affiliation{Caltech Optical Observatories, California Institute of Technology, Pasadena, CA  91125}

\begin{abstract}
Electron-capture supernovae (ECSNe) may arise from ONeMg-core collapse in super-asymptotic giant branch (sAGB) stars near the low-mass core-collapse limit ($\approx\!8$--$10$\,\Msun). At early times, models predict that ECSNe resemble low-mass red supergiant iron-core-collapse SNe (FeCCSNe), making the two channels difficult to distinguish. Nebular spectroscopy, however, can reveal differences in ejecta composition. We present a systematic sample of 19 nebular spectra of low-luminosity Type IIP (LLIIP) SNe from the ZTF CLU survey, obtained 115$-$450\,d after explosion. Their low velocities expose narrow lines blended in brighter SNe, which we identify and model to constrain progenitor properties. We find a strong correlation between the FWHM of H\,\textsc{i}\,$\lambda$6563 and peak luminosity, showing that LLIIP SNe occupy the low-energy end of the core-collapse population, but no correlation with plateau duration, suggesting that envelope and core properties are not tightly linked. Only one SN reaches the extremely low H\,\textsc{i}\,$\lambda$6563 widths predicted for the weakest $\sim$9\,M$_\odot$ explosion models, implying that such low-energy events are intrinsically rare. Combining our sample with 118 literature nebular spectra of Type II SNe, we infer an IMF slope of $2.1\pm1.2$. We also introduce an `ECSN score'' based on the absence of He- and O-shell emission lines, and identify two plausible ECSN candidates, SN~2023bvj and SN~2024btj. However, neither shows the extremely narrow nebular lines predicted by current ECSN models. If ECSNe arise predominantly through the LLIIP channel, we infer an upper limit on the ECSN rate of $\lesssim (5$--$8)\times10^{2}\,\mathrm{Gpc^{-3}\,yr^{-1}}$, corresponding to a narrow sAGB progenitor mass window of $\Delta M_{\rm sAGB} \lesssim 0.02$--$0.06\,\mathrm{M_\odot}$.

\end{abstract}

\keywords{}

\section{Introduction} \label{sec:intro}

The final fate of a star is predominantly determined by its initial mass: stars with lower masses become white dwarfs, while more massive ones undergo core collapse, producing neutron stars or black holes. However, the fate of stars in the intermediate-mass regime, near the threshold between white dwarf and neutron star formation, is not well understood. Stellar-evolution models suggest that stars within the $8$--$12$\,M$_\odot$ interval possess distinct core characteristics compared to their more massive counterparts ($>$ 12\,M$_\odot$), particularly showing much lower core compactness \citep{Sukhbold2016}. Modeling their evolution is more challenging than for stars above 12\,M$_\odot$ due to complex degeneracy effects \citep{Miyaji1980, Gutierrez2005, Woosley2015}. Owing to the steep slope of the initial mass function (IMF), a substantial portion, over 40\%, of all massive star explosions should occur in the $8$--$12$\,M$_\odot$ mass range, with 25\% originating from stars between $8$--$10$\,M$_\odot$. If many stars above 20\,M$_\odot$ collapse into black holes without producing visible supernovae, the $8$--$12$\,M$_\odot$ contribution could rise to 50--60\% \citep{Sukhbold2016}. 

Massive stars with initial masses just below the threshold for forming Fe cores are expected to follow a different evolutionary path: instead of producing Fe cores, they may form electron-degenerate ONeMg cores that collapse through electron-capture reactions. Such events are thought to give rise to electron-capture supernovae (ECSNe) \citep{Miyaji1980, Nomoto1984, Kitaura2006, Janka2008, Takahashi2013, Jones2013, 
Hiramatsu2021, Podsiadlowski2004, Wang2026a}. 
In the single-star channel, ECSN progenitors are thought to be super-asymptotic giant branch (sAGB) stars by the time collapse occurs. The mass range capable of producing ECSNe is predicted to be narrow ($<$ 1\,M$_\odot$) \citep{Siess2007, Poelarends2008, Doherty2015}, and could be even smaller depending on choices for mass-loss rates and third dredge-up efficiency. However, even a $1$\,M$_\odot$-wide interval would account for $\sim$15\% of CCSNe under a Salpeter IMF, falling to $\sim$8\% for a $0.5$\,M$_\odot$ range. Whether ECSNe actually occur and what fraction of core-collapse SNe they represent remain key open questions.

While the exact mass range or even the existence of ECSN progenitors remains uncertain,  ECSNe are predicted to have low explosion energies ($\sim$$10^{50}$\,erg) and small $^{56}$Ni yields ($\sim$$10^{-3}$\,M$_\odot$) from the neutrino-driven mechanism \citep{Kitaura2006, Janka2008, Wanajo2009, Jones2013, Wang2026a}. However, this holds true also for the lowest-mass Fe CCSNe, and are therefore not sufficient diagnostics. \citet{Jerkstrand2018} showed that nebular spectra can break the degeneracy, concluding that none of the candidates SN 1997D, SN 2005cs or SN 2008bk were ECSNe but rather low-mass Fe CCSNe. With zero of three events confirmed, the ECSN progenitor mass window was preliminarily constrained to $\Delta M_{ZAMS} \lesssim\ 1\ M_\odot$.

The first observational candidate with spectroscopic properties in better agreement with ECSNe than Fe CCSNe was SN 2016bkv \citep{Hosseinzadeh2018}. This supernova also had the very low expansion velocities expected. The main dissonance point lies in the $^{56}$Ni mass, estimated at $\sim$0.02 $M_\odot$, much higher than ECSN predictions.

An second candidate, SN\,2018zd, was proposed by \citet{Hiramatsu2021}. This event had an estimated $^{56}$Ni mass ($8 \times 10^{-3}\ M_\odot)$ in better agreement with LLIIP SNe, and lacked [C I] lines as predicted for ECSNe. It did however show He lines (weak in ECSNe) and lacked the strong Bowen fluorecence O I 8448 (strong in ECSNe).
Later studies questioned this classification, arguing that the extreme properties of SN 2018zd were overestimated due to an incorrect distance, and that the event is consistent with a faint Fe core-collapse SN instead \citep{Zhang2020, Callis2021}. 

Other proposed ECSN-like explosions include the class of Intermediate Luminosity Red Transients, such as, AT\,2019abn, NGC\,300 2008OT-1, and SN\,2008S \citep{Botticella2009, Adams2016, Jencson2019b, Rose2024, Valerin2025}. They exhibit luminous, dusty progenitors and low-velocity outflows suggestive of sAGB-like systems. Although their physical origins remain debated, ILRTs may represent the non-terminal or weakly-terminal end of the same low-mass core-collapse pathway \citep{Pumo2009}.

The most promising observational analogs of low-mass core-collapse events are low-luminosity Type IIP supernovae (LLIIP SNe), defined here as those with $M_{\mathrm{r,peak}} \ge -16$ mag, as motivated by prior literature \citep[e.g.,][]{Spiro2014}. Recent studies based on pre-explosion imaging of SNe 2005cs, 2008bk, 2018aoq, and 2022acko \citep{Maund2005, Li2006, Mattila2008, ONeill2019, VanDyk2023} show that SNe with $r$-band peaks fainter than $-16$ mag have progenitor mass estimates below $\sim$11\,M$_\odot$, consistent with low-mass red supergiants (RSGs) and sAGB progenitors 
(see Appendix A in \citealt{Das2025a}; see also the recent review by \citealt{VanDyk2025}). \textcolor{black}{Previous progenitor studies have also explored possible observational distinctions between these channels from, including differences in the expected spectral energy distributions of sAGB and RSG progenitors \citep{Eldridge2017} and the implications of unusually low progenitor luminosities inferred for some LLIIP SNe \citep{Fraser2011}.}
LLIIP SNe exhibit faint plateau luminosities, low expansion velocities, and low explosion energies and nickel masses \citep[e.g.,][]{Turatto1998, Pastorello2004, Spiro2014, Jager2020, Muller2020, Sheng2021, Valerin2022, Kozyreva2022, Bostroem2023, Teja2024, Dastidar2025, Das2025b}, linking their light curves to low-mass progenitors ($\sim$8--12\,M$_\odot$) through both direct imaging and hydrodynamical modeling. Although these SNe trace the low-mass end of core collapse, photospheric observables cannot distinguish between models with low mass RSG and those with sAGB progenitors. Both retain extended hydrogen envelopes and have similar pre-SN masses, producing photospheric light curves that are nearly indistinguishable despite their different core structures.

The nebular phase offers unique diagnostics to distinguish between these scenarios \citep{Jerkstrand2018}. Once the ejecta become optically thin ($\sim$100 d), the luminosity is powered by $^{56}$Co decay and the cooling is dominated by forbidden transitions. As the hydrogen-rich material recombines and becomes transparent, the metal-rich inner layers are exposed, revealing the products of hydrostatic and explosive nucleosynthesis. Nebular spectra therefore provide direct constraints on ejecta composition, nucleosynthetic yields, explosion geometry and the Zero-Age Main Sequence (ZAMS) mass \citep{Fransson1989, Jerkstrand2012, Jerkstrand2014, Jerkstrand2015, Dessart2020, Dessart2021, Dessart2021b, Dessart2025}. For ECSNe, progenitor structures at the point of collapse have been calculated by \citet{Nomoto1984}, \citet{Takahashi2013}, and \citet{Jones2013}, while earlier evolutionary models predicted to reach the ECSN stage were developed by \citet{Ritossa1996}, \citet{Ritossa1999}, \citet{Siess2006}, and \citet{Woosley2015}. These studies show that second dredge-up or dredge-out episodes remove nearly the entire helium shell \citep{Nomoto1987, Ritossa1996}, leaving an ONe core enveloped by a thin O/C shell (a few $\times 10^{-2}$\,M$_\odot$) and a dilute H/He envelope. At collapse, the ONe core and part of the O/C shell undergo explosive burning to iron-group elements \citep{Kitaura2006, Janka2008}. Longer-term simulations of the neutrino-driven wind show no major modifications to these yields \citep{Wanajo2009, Wanajo2011, Pllumbi2015}. An important consequence is that ECSNe from single-star progenitors lack both an O-rich shell and a He shell in their ejecta. Their nebular spectra should therefore be dominated by lines from the extended H/He envelope, with the possible appearance of neutron-rich iron-group species (e.g., stable Ni and Zn) if the innermost Fe-rich material is visible through the $\sim$8\,M$_\odot$ envelope \citep{Wanajo2009}. In contrast, low-mass Fe-core CCSNe retain thin Si, O, and He shells, which generate unique signatures. Thus, nebular spectroscopy may distinguish ECSNe primarily through the absence of these layers \citep{Jerkstrand2018}.

There has been no uniform sample of LLIIP SN nebular spectra from a systematic survey. Earlier nebular studies focused on individual objects \citep{Turatto1998, Pastorello2006, VanDyk2012, Muller2020, Jager2020, Bostroem2023, VanDyk2023}, and although \citet{Fang2025} collected 50 Type II nebular spectra from the literature, only three were LLIIP SNe. In this work (Paper III), we present a uniform sample of nebular spectra for 19 LLIIP SNe spanning phases from $\sim$115 to 450 days after explosion, the largest such dataset obtained to date. This work is the third in a series on LLIIP SNe. In Paper I \citep{Das2025a}, we used the Zwicky Transient Facility Census of the Local Universe \citep[ZTF CLU;][]{Bellm2019, Graham2019, Dekany20, Masci2019, De2020} to measure the luminosity function and volumetric rate of Type IIP SNe, using the largest sample of 330 Type IIP SNe to date. In Paper II \citep{Das2025b}, we analyzed 129 Type IIP SNe with semi-analytic and radiation-hydrodynamical modeling to infer explosion and progenitor properties, finding that LLIIP SNe predominantly arise from the low-mass RSG/sAGB population.

Along with our LLIIP sample, we also assemble a comparison set of 120 nebular-phase spectra of canonical Type II SNe from the literature, providing a broader context for interpreting the late-time properties of faint events. The intrinsically narrow nebular emission lines of LLIIP SNe reduce blending and enable secure identification of weak features, allowing us to measure key nebular diagnostics, compare them to theoretical models, assess potential ECSN candidates, and infer ZAMS mass values to constrain the low-mass end of the core-collapse IMF.

The paper is structured as follows. Section~\ref{sec:sample} describes the sample selection, and Section~\ref{sec:spectroscopy} summarizes the spectroscopic observations and  reduction. Section~\ref{sec:nebular_methods} details the spectral processing and analysis methods, and correlations with lightcurve observables. In Section~\ref{sec:zams_estimation}, we estimate progenitor ZAMS masses and examine the resulting mass distribution and IMF constraints. Section~\ref{sec:ecsn} introduces nebular diagnostics for identifying ECSN candidates, presents ECSN scores for the LLIIP sample, and estimates the ECSN rate. We summarize our conclusions in Section~\ref{sec:conclusion}.

\begin{table}[h!]
\centering
\caption{ZTF CLU LLIIP Sample Summary}
\begin{tabular}{lc}
\toprule
Criteria & Number \\
\midrule
\parbox[t]{5.5cm}{\raggedright LLIIP SNe within 100\,Mpc (2022 Nov--2024 Nov) meeting Paper~I quality criteria} & 28 \\
\midrule
\parbox[t]{5.5cm}{\raggedright LLIIP SNe with usable nebular spectra ($\geq$\,110\,d)} & 19 (47 spectra) \\
\bottomrule
\end{tabular}
\end{table}

\section{Sample}
\label{sec:sample}

The ZTF CLU experiment aims to build a comprehensive spectroscopic sample of transients in the local Universe within 150 Mpc. The CLU experiment is designed as a volume-limited SN survey where sources at less than 150 Mpc ($z\leq0.033$) and spatially coincident with (within 30 kpc) or visibly associated with a galaxy in the CLU catalog were assigned for spectroscopic follow-up for classification. In Paper I, we selected Type~IIP SNe from the ZTF CLU experiment with well-sampled $r$-band light curves and a clearly defined plateau phase. The selection required (i) a Gaussian Process regression–derived peak apparent magnitude of $m_{r,\rm peak}<20$~mag, (ii) sufficient temporal coverage around peak and during the plateau (more than ten total detections and at least one detection on both sides of peak, or an equivalent set of conditions ensuring a constrained rise), and (iii) a plateau lasting at least 40~days with $<1$~mag decline.  We refer to Paper I \citep{Das2025a} for the complete sample selection criteria. 

For this work, we focus on ZTF CLU LLIIP SNe that exploded within a volume of 100\,Mpc between 2022 November 1 and 2024 November 1, yielding a sample of 28 events that satisfy the Paper~I quality criteria in this period. Among these, we obtained usable late-time spectroscopy for 19 SNe. The remaining events could not be followed spectroscopically at nebular phases either because they were behind the Sun, outside observable windows, or had spectra with too low signal-to-noise ratio to extract robust diagnostics. The final LLIIP spectroscopic sample consists of 19 SNe for which we obtained a total of 47 nebular spectra spanning 115--450 days after explosion. Throughout this paper, all quoted phases are measured relative to the estimated explosion epoch of each SN. This sample provides the first systematic sample of nebular spectra of LLIIP SNe from a volume-limited survey. SN\,2022aaad has the most extensive coverage, with eight spectra between 150\,d and 450\,d. Multi-epoch nebular observations were also secured for SN\,2022zmb, SN\,2023wcr, SN\,2023kmk, SN\,2023vci, SN\,2024jxm, SN\,2024btj, SN\,2024cro, and SN\,2024abfl, typically with coverage between 150\,d and 350\,d. Additional LLIIP events observed at nebular phases include SN\,2022jzc, SN\,2022aang, SN\,2022acko, SN\,2023mpz, SN\,2023ywa, and SN\,2024wp, which were generally observed once between 150\,d and 350\,d. The full spectral log of the LLIIP SN sample is listed in Table~\ref{table:LLIIP_spectral_log}.

For comparison and context, we also include nebular spectra of a small set of more luminous ZTF Type~II SNe drawn from the same CLU volume. This bright-ZTF comparison sample consists of 6 SNe with 10 spectra, obtained either as part of routine CLU follow-up or serendipitously during other spectroscopic programs (see Table \ref{table:bright_spectral_lo} for the log.) In addition, we assemble a literature comparison sample of Type~II SNe using the dataset summarized in Table~\ref{table:literature_log}, which contains 54 SNe with a total of 120 nebular spectra. Following the selection criteria of \citet{Fang2025}, we require (1) spectral coverage spanning 5000--8500\,\AA; (2) observations obtained between 110 and 450\,d post-explosion to ensure a well-defined nebular phase; and (3) public availability of the spectra through the Open Supernova Catalog \citep{Guillochon2017} or WISeREP \citep{Yaron2012}. 

\input{ZTF_LLIIP_table}

\section{Spectroscopy Data}
\label{sec:spectroscopy}

We obtained late-time spectra primarily using the Double Beam Spectrograph (DBSP) on the 5.1\,m Hale Telescope at Palomar Observatory and the Low Resolution Imaging Spectrometer (LRIS) on the 10\,m Keck I Telescope. Additional spectra were acquired with the Alhambra Faint Object Spectrograph and Camera (ALFOSC) on the 2.56\,m Nordic Optical Telescope (NOT). DBSP observations used the 600/4000 grating (blue) and the 316/7500 grating (red), providing $R\sim1000$ over 3200–10000\,Å and were reduced following \citet{Bellm2016, Roberson2022}. LRIS spectra were taken with the 600/4000 grism (blue) and 400/8500 grating (red), achieving resolutions of $\sim$5–9\,Å and were reduced with the \texttt{lpipe} pipeline \citep{Perley2019}. ALFOSC spectra were obtained with a $1\farcs0$ wide slit and grism 4, covering 3500–9000\,Å at $R\approx360$ and were reduced using PyNOT and PypeIt \citep{Prochaska2020a}. We also include one spectrum from the new NGPS spectrograph \citep{Kasliwal24_first_science} on the Hale Telescope, using 2x3 binning and $1\farcs5$ wide slit ($R\!\sim\!2000$ over 5600–10000\,Å), reduced using a tool based on PypeIt. The spectra were gathered on the \texttt{Fritz} Marshal \citep{van2019, Coughlin2023}, a web portal designed for vetting and coordinating transient follow-up observations. All spectroscopy data will be made public through Zenodo and WISeREP \citep{Yaron2012}.

\begin{figure*}[htbp]
    \centering
    \includegraphics[width=1.0\textwidth]{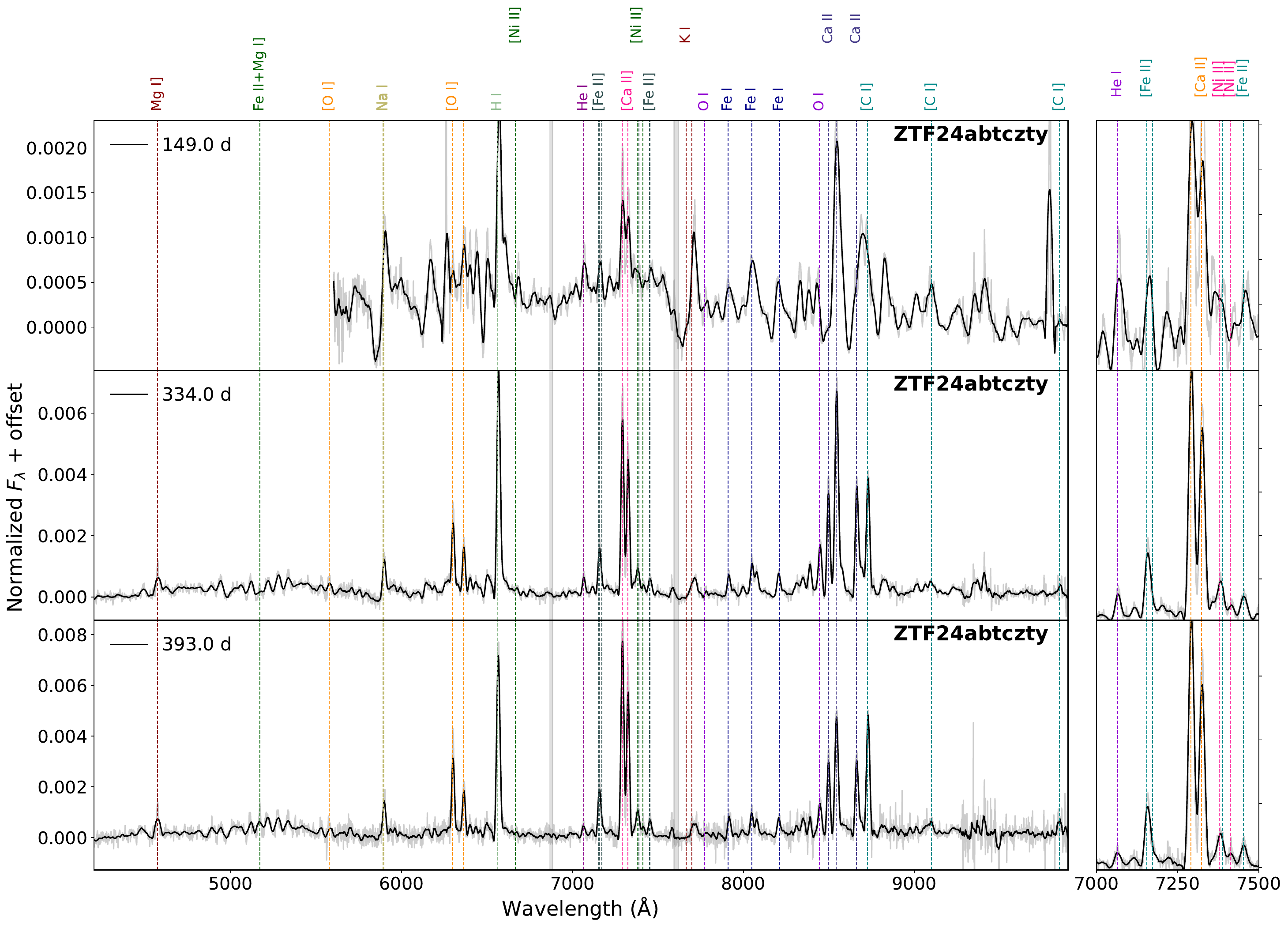}
\caption{Keck/LRIS nebular spectra of ZTF24abtczty/SN\,2024abfl at 149\,d, 334\,d and 393\,d after explosion.  The black curve shows the spectrum smoothed with a Savitzky--Golay filter, while the faint gray curve indicates the raw data. Colored vertical dashed lines mark the rest wavelengths
of prominent nebular features, as labeled at the top (e.g.\ Mg\,\textsc{i}]\,$\lambda4571$, 
[O\,\textsc{i}]\,$\lambda\lambda6300,6364$, H\,\textsc{i}\,$\lambda6563$, 
[Ca\,\textsc{ii}]\,$\lambda\lambda7291,7323$, and [C\,\textsc{i}]\,$\lambda8727$). 
The right-hand panels zoom in on the 7000–7500\,\AA\ region around the 
[Ca\,\textsc{ii}]\,$\lambda\lambda7291,7323$ complex, highlighting the narrow 
[Ca\,\textsc{ii}] emission and the neighboring Fe- and Ni-peak features.}

    \label{fig:ZTF24aaejecr_obs}
\end{figure*}

\section{Spectra processing and analysis}

\label{sec:nebular_methods}
\begin{figure*}
    \centering
    \includegraphics[width=1.0\textwidth]{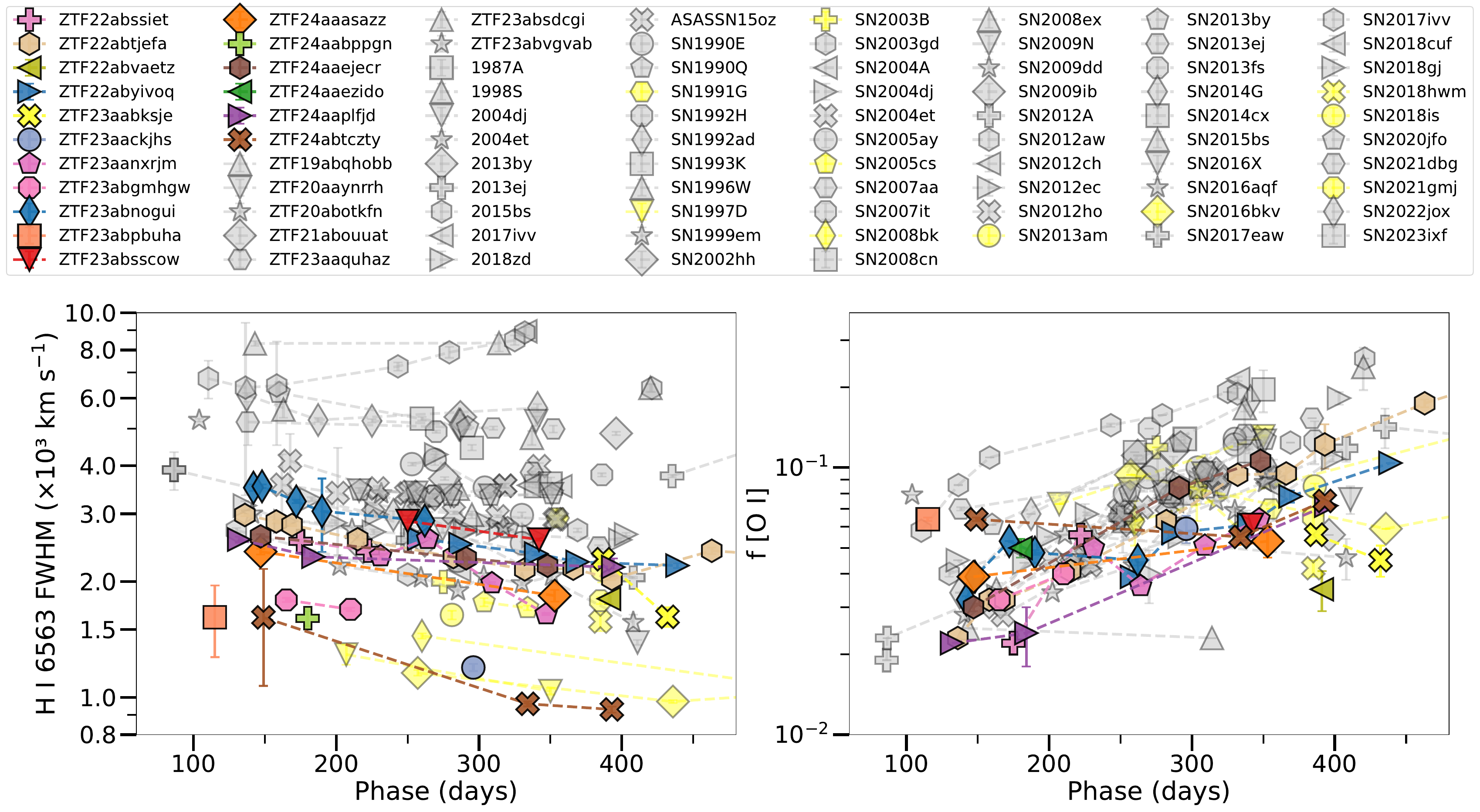}
\caption{Left: H\,\textsc{i}\,$\lambda6563$ FWHM for all objects in our sample as a 
function of phase. The colored points represent the ZTF CLU LLIIP sample, the 
literature comparison sample is shown in gray, and low-luminosity literature 
events are highlighted in yellow. Right: Same objects but showing the fractional 
[O\,\textsc{i}] luminosity $f_{\rm [O\,I]}$, as defined in Section~\ref{sec:fractional_flux}. 
Dashed lines connect measurements of the same SN across different epochs. Throughout this paper, all quoted phases are measured relative to the estimated explosion epoch of each SN.}
    \label{fig:ha_fwhm_panel}
\end{figure*}

We measure the nebular emission-line properties for all Type~II SNe in our sample using rest-frame, extinction-corrected spectra. Milky Way and host-galaxy extinction values are adopted from Paper~I and Paper~II. Because our analysis focuses on identifying and quantifying the presence or absence of specific nebular features, comparing fractional line fluxes, and measuring velocity widths, modest uncertainties in extinction do not affect our conclusions. The code and data used in this analysis will be made publicly available on GitHub and Zenodo after publication.

\subsection{Host-Galaxy Continuum Subtraction}\label{sec:host_subtraction}

We first subtract the underlying host-galaxy continuum from each spectrum. The reduced blue and red arms of the spectra often have different instrumental responses and can require an artificial splice near the overlap region. To construct a single smooth host continuum, we fit the blue and red sides separately. On the red side we identify five relatively line-poor windows: 5530--5830~\AA, 6010--6210~\AA, 6750--6950~\AA, 7900--8150~\AA, and 9000--9450~\AA. We compute the mean flux in each available window and fit a piecewise linear curve through at least three such points. On the blue side we select the 
4300--5200~\AA\ region, where nebular contamination is minimal, and fit a straight line. The blue and red continuum models are then joined smoothly across the overlap region to form the final continuum, which is subtracted from the observed spectrum. All line measurements below are performed on these host-subtracted data.

\subsection{Measuring \textup{[O}\,\textup{\textsc{i}}\textup{]}\,$ \lambda\lambda6300,6363$ flux}

The [O\,\textsc{i}]\, $\lambda\lambda6300,6363$ doublet lies within a broad emission complex that also includes H\,\textsc{i} $\lambda6563$. To isolate the oxygen feature, we first apply slight Savitzky–Golay smoothing to the host-subtracted spectrum and use the smoothed curve only to identify the local minima that bracket the [O\,\textsc{i}]/H$\alpha$ blend. A straight line connecting these minima defines the local pseudo-continuum. Subtracting this baseline isolates the emission-line structure. Although the nebular background is not a true continuum but a blend of many faint lines, the same procedure is applied uniformly to every SN, ensuring negligible systematic bias for population-level statistics.

We integrate the continuum-subtracted flux across the [O\,\textsc{i}] feature to obtain the total line flux, and estimate the velocity width by fitting a Gaussian profile (or two components when needed for visibly separated peaks).

\subsection{Measuring \textup{[Ca}\,\textup{\textsc{ii}}\textup{]}\,$ \lambda\lambda7291,7323$ flux and the Ni/Fe flux ratio diagnostic}

The region around 7000--7500~\AA\ contains several blended nebular lines. We model the spectrum in this range using eight Gaussian components at the rest wavelengths of He\,\textsc{i}~$\lambda7065$, [Fe\,\textsc{ii}]~$\lambda7155$, [Fe\,\textsc{ii}]~$\lambda7172$, [Ca\,\textsc{ii}]~$\lambda7291$, [Ca\,\textsc{ii}]~$\lambda7323$, [Ni\,\textsc{ii}]~$\lambda7378$, [Fe\,\textsc{ii}]~$\lambda7388$, and [Ni\,\textsc{ii}]~$\lambda7412$. All components are fit simultaneously. To extract a clean [Ca\,\textsc{ii}] profile, we subtract the best-fitting contributions of all lines except the two [Ca\,\textsc{ii}] components. The residual yields the [Ca\,\textsc{ii}] flux and FWHM.

The same fit gives fluxes for [Ni\,\textsc{ii}]~$\lambda7378$ and [Fe\,\textsc{ii}]~$\lambda7155$. Their ratio serves as a diagnostic of the Ni/Fe mass ratio in the inner ejecta. In ordinary Fe-core CCSNe, [Fe\,\textsc{ii}] dominates, whereas electron-capture SNe are expected to produce a Ni-rich, neutron-rich inner zone, yielding predicted [Ni\,\textsc{ii}]/[Fe\,\textsc{ii}] ratios of $\sim1.3$--3.0 \citep{Jerkstrand2015}.

\subsection{Fractional [O\,\textsc{i}] Flux}
\label{sec:fractional_flux}
To enable direct comparison between spectra of different objects and epochs, we also compute the fractional flux of the [O\,\textsc{i}] line following the  prescription in Equation~A2 of \citet{Barmentloo2024} and  \citet{Fang2025}. The total observable flux is defined as
\begin{equation}
    \int_{5500\,\text{\AA}}^{8000\,\text{\AA}} \left( F_\lambda - F_{\mathrm{pseudo}} \right)\, d\lambda,
\end{equation}
where $F_\lambda$ is the observed spectral flux density and $F_{\mathrm{pseudo}}$ is the constant pseudo-continuum level derived from the line-poor windows identified in our host-continuum subtraction procedure (Section~\ref{sec:host_subtraction}). The fractional [O\,\textsc{i}] flux is then computed as the ratio of the [O\,\textsc{i}] line flux to this integrated denominator.  Importantly, the fractional [O\,\textsc{i}] flux has been shown to correlate with ZAMS progenitor mass in nebular-phase models and observations \citep{Fang2025}, making it a powerful diagnostic of progenitor properties. We also find a strong correlation in our sample (see Figures~4 and~5), further supporting the use of fractional [O\,\textsc{i}] emission as an indicator of progenitor mass of Type~IIP SNe. This offers the main advantage that, because both the numerator and denominator are measured from the same extinction-corrected spectrum over a relatively narrow wavelength range, the resulting fractional fluxes are largely insensitive to uncertainties in the extinction correction or absolute flux calibration. In practice, the fractional flux is therefore effectively independent of the photometric calibration, extinction corrections and distance uncertainty, which typically contribute the largest sources of systematic uncertainty in nebular spectral analyses.

\subsection{Uncertainty Estimation}

To estimate uncertainties in fluxes and FWHM, we perform Monte Carlo simulations. After smoothing the host-subtracted spectrum with a Savitzky–Golay filter and subtracting it, we compute the standard deviation of the residuals in the 6000--8000~\AA\ region and adopt this as the noise level. We generate $N=100$ realizations by adding Gaussian noise of this amplitude.

For each realization, we allow the endpoints of all local continua and the central wavelengths of all fitted components (He\,\textsc{i}, [Fe\,\textsc{ii}], [Ca\,\textsc{ii}], [Ni\,\textsc{ii}]) to vary within $\pm600$~km~s$^{-1}$ to account for wavelength-calibration uncertainties and finite spectral resolution. We then repeat the full fitting procedure. The standard deviation of the resulting distribution of fluxes, FWHM values, and ratios is adopted as the measurement uncertainty.

\begin{figure}
    \centering
    \includegraphics[width=1.0\columnwidth]{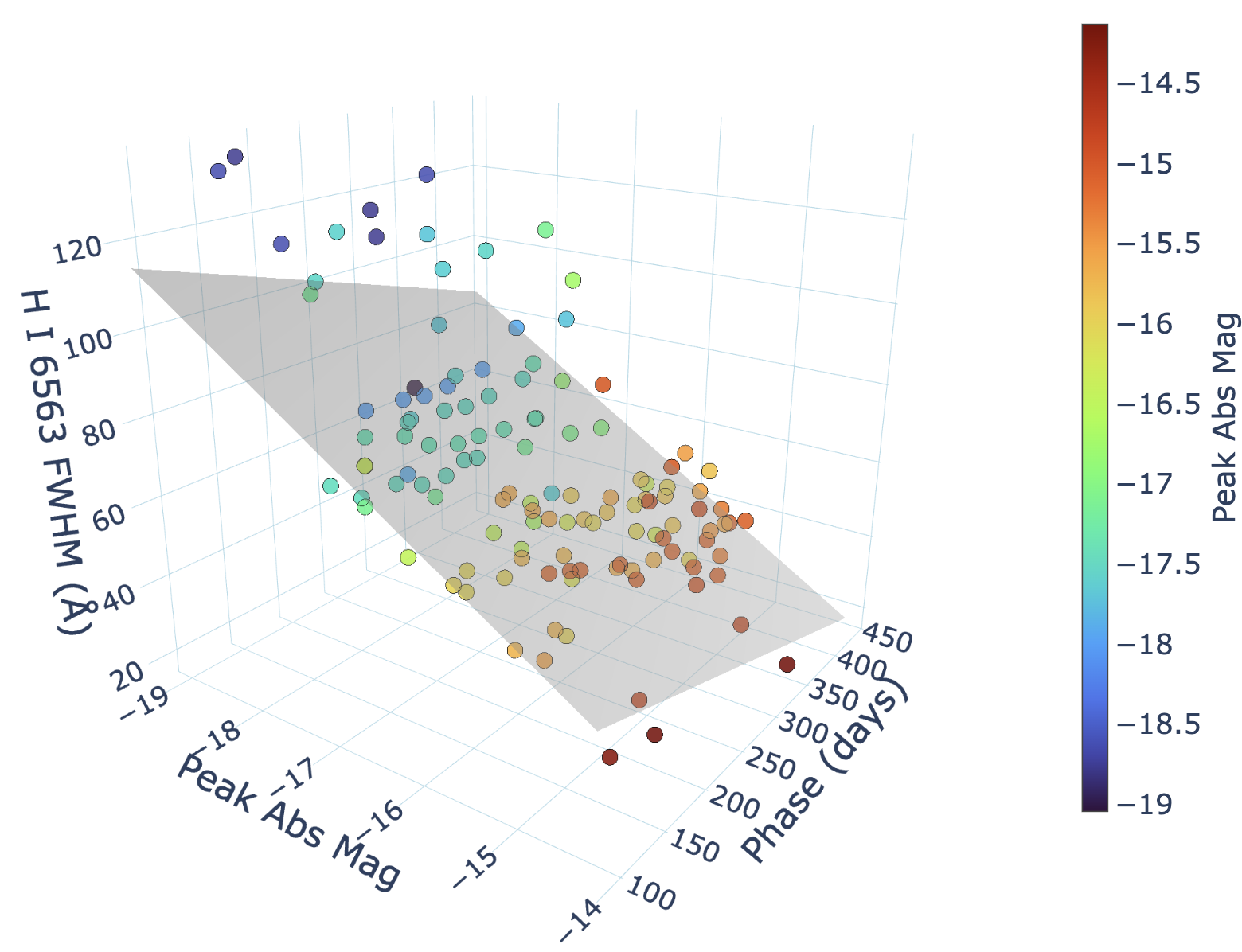}
    \caption{Nebular H\,\textsc{i}\,$\lambda6563$ FWHM as a function of phase and peak absolute
magnitude for our sample. Points are color-coded by $M_{\rm peak}$, and the
best-fit weighted plane (Section~\ref{eq:fitplane}) is shown in gray. The trend indicates that fainter
SNe exhibit systematically narrower nebular H\,\textsc{i} lines. A two-dimensional projection of this relation is shown in the top-left panel of Figure~\ref{fig:lc_correlation}.}
    \label{fig:plane}
\end{figure}

\subsection{Correlation of nebular spectra with lightcurve observables}

\label{sec:correlation}


We measure the key $r$-band light-curve parameters: peak absolute magnitude, plateau onset, plateau duration, and plateau-end slope, using the same methodology developed in Paper~I and Paper~II \citep{Das2025b}. Briefly, we determine $t_{\rm peak}$ and $M_{r,\rm peak}$ from a Gaussian Process regression fit to the observed light curve, which provides a smooth, noise-robust representation of the data. The plateau onset ($t_{\rm plateau,start}$) is defined as the earlier of (i) the epoch at which the instantaneous decline rate becomes shallower than 0.01\,mag\,day$^{-1}$, or (ii) the epoch within 60\,d of first detection at which the change in slope ($\Delta_{\rm slope}$) reaches a local maximum. The plateau end ($t_{\rm plateau,end}$) is identified after $\sim$40\,d, when the light curve begins its characteristic steep decline, and is measured as the epoch at which $\Delta_{\rm slope}$ first drops below $-0.003$\,mag\,day$^{-2}$. The plateau duration is then defined as $t_{\rm plateau,end} - t_{\rm plateau,start}$. 

For non-ZTF SNe drawn from the literature, we apply exactly the same GP-based procedure to the published $r$-band (or $R$-band) photometry, ensuring uniformity across the combined sample. All light curves are corrected for both Milky Way extinction and host-galaxy extinction using the extinction values reported in the original references. The resulting light-curve parameters, together with the adopted extinction corrections and literature sources, are listed for each of the 54 SNe in Table~\ref{tab:nebular_master}.


\begin{figure*}
    \centering
    \includegraphics[width=1.0\textwidth]{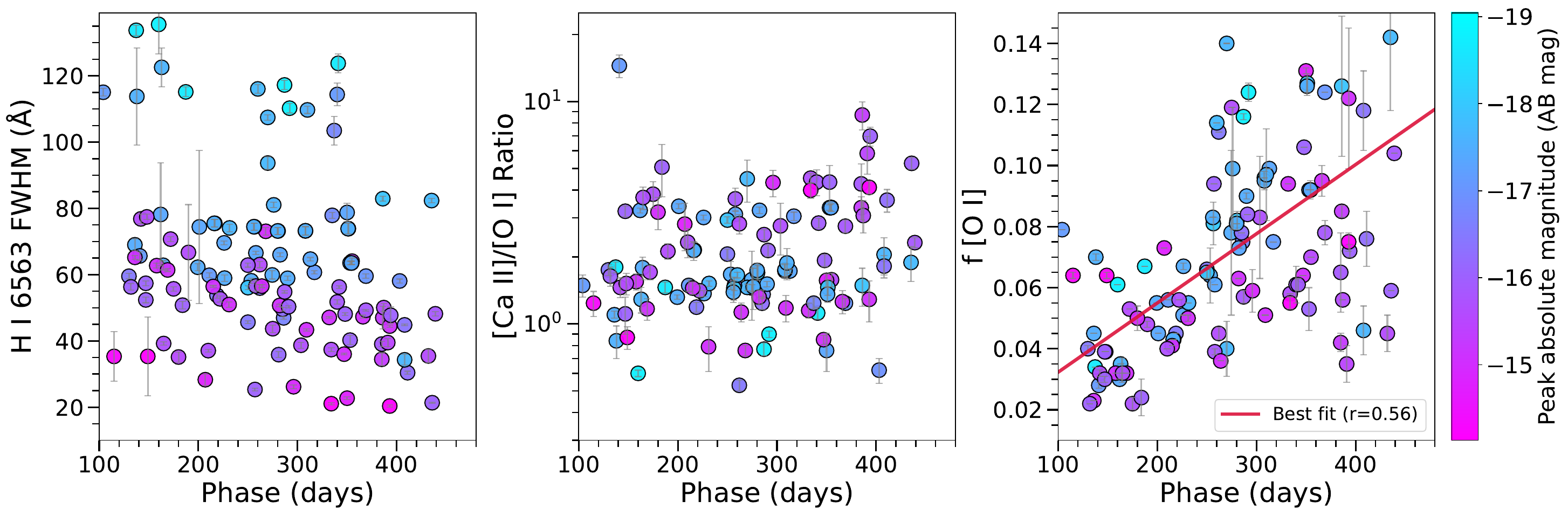}
    \includegraphics[width=1.0\textwidth]{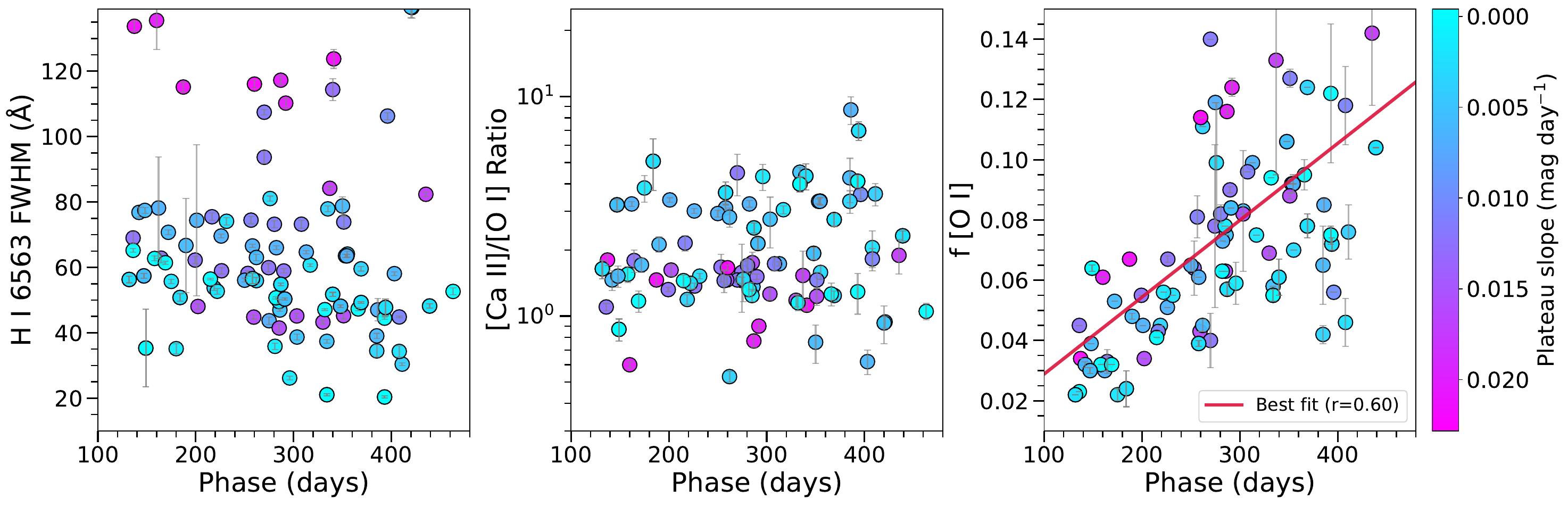}
    \includegraphics[width=1.0\textwidth]{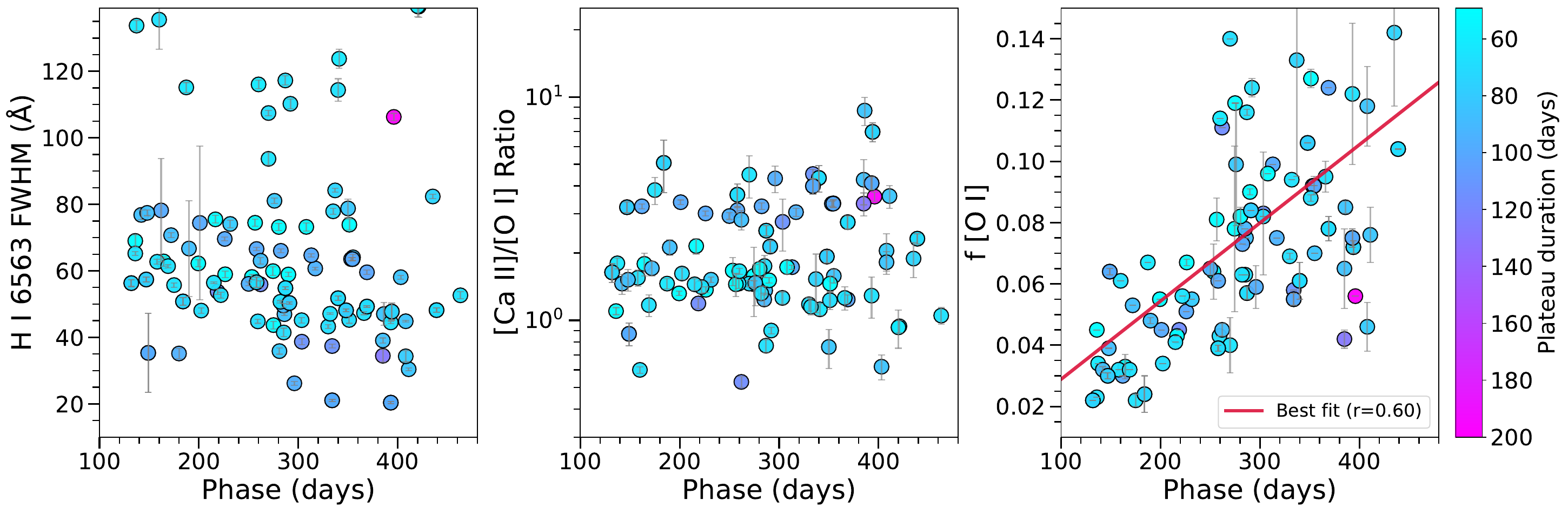}

\caption{Correlation of nebular observables with $r$-band light-curve properties. 
Each column shows a different nebular diagnostic as a function of phase: 
H\,\textsc{i}\,$\lambda6563$ FWHM (left), the [Ca\,\textsc{ii}]/[O\,\textsc{i}] ratio (middle), 
and the fractional [O\,\textsc{i}] luminosity $f_{\rm [O\,I]}$ (right). 
The three rows use different photometric quantities for the color coding: 
peak absolute magnitude (top), plateau slope (middle), and plateau duration (bottom). 
SNe with steeper plateau slopes, which are associated with higher luminosity 
\citep{Das2025b}, tend to have broader nebular H\,\textsc{i} emission. 
In contrast, we do not see any clear dependence of the nebular line widths or flux ratios 
on the plateau duration. A positive correlation between $f_{\rm [O\,I]}$ and phase is also 
visible, as discussed in Section~\ref{sec:correlation}.}

    \label{fig:lc_correlation}
\end{figure*}

We now investigate correlations between the nebular spectra properties and the lightcurve observables. As shown in Figure~\ref{fig:ha_fwhm_panel}, the LLIIP SNe in our sample exhibit systematically narrower H\,\textsc{i} $\lambda$6563 emission lines than the brighter comparison SNe from the literature. To quantify this trend, we examine how the nebular H\,\textsc{i} $\lambda$6563 FWHM correlates with the optical peak magnitude and the epoch of the spectrum. We model the observed velocities with a planar relation and the best-fit weighted plane is
\begin{equation*}
\label{eq:fitplane}
\begin{aligned}
\mathrm{FWHM\ (\AA)} = &\, (-72.76 \pm 1.15) \\
&+ (-0.05 \pm 0.01)\,\mathrm{Phase\ (d)} \\
&+ (-9.05 \pm 0.07)\,M_{\mathrm{peak}},
\end{aligned}
\end{equation*}
with a Pearson correlation coefficient $r = 0.65$ and $p < 10^{-3}$ (Figure~\ref{fig:plane}). This relation shows that fainter SNe exhibit lower nebular H\,\textsc{i} velocities at fixed phase, likely because LLIIP events arise from intrinsically low-energy explosions.

When we compare these measurements to nebular spectral models for low-mass progenitors, we find that only a small subset of events reach the very low H\,\textsc{i} $\lambda$6563 widths predicted for extreme low-energy explosions (Figure~\ref{fig:analysis_models}). In particular, the \citet{Jerkstrand2018} models for a $9$\,M$_\odot$ progenitor with an explosion energy of $E_{\mathrm{exp}} \simeq 1.1\times10^{50}$\,erg produce extremely narrow nebular lines, with H\,\textsc{i} FWHM values of order $1000$\,km\,s$^{-1}$. In our data, only SN~2024abfl in the LLIIP sample, together with SN~1997D and SN~2016bkv from the literature, approach similarly small H\,\textsc{i} FWHM values. Most of our LLIIP SNe have significantly broader nebular lines at all observed epochs, typically $\gtrsim 2000$\,km\,s$^{-1}$, even though they occupy the low-energy, low-$^{56}$Ni end of the explosion parameter space discussed in Paper~II. For a Salpeter IMF, stars in the $\sim 8$–$10$\,M$_\odot$ range constitute roughly one quarter of all CCSN progenitors, so the apparent scarcity of events with such narrow nebular lines is somewhat surprising if every star in this mass interval produced a $\sim10^{50}$\,erg explosion.

One possible interpretation is that the effective lower mass limit for successful core collapse is higher than in those calculations, so that only a subset of the nominal 8–9.5\,M$_\odot$ population actually reaches core collapse and produces SNe. 
Alternatively, real explosions at the low-mass end may not be as uniformly weak as seen in some neutrino-driven explosion models \citep[e.g.,][]{Burrows2019,Janka2008,Burrows2024}. Classic ONeMg and related low-mass neutrino-driven models often yield explosion energies of $\sim10^{50}$\,erg \citep{Janka2008,Burrows2019}. However, more recent three-dimensional simulations indicate that somewhat higher energies are possible in this regime: for example, Wang \& Burrows (2024) obtain $E_{\mathrm{exp}} = 1.71\times10^{50}$ and $2.07\times10^{50}$~erg for the u8.1 and z9.6 progenitors in 3D, compared with $1.7\times10^{49}$ and $1.29\times10^{50}$~erg for the corresponding one-dimensional models. We note that alternative explosion scenarios, such as the jittering-jet mechanism, also predict somewhat higher explosion energies in this mass regime \citep[e.g.,][]{Papish2011}.

In addition to the comparison with explosion models, we also examined how the nebular observables correlate with photometric properties of the $r$-band lightcurve (Figure~\ref{fig:lc_correlation}). We find that SNe with steeper plateau slopes tend to have larger H\,\textsc{i}\,$\lambda 6563$ widths, consistent with the idea that more rapidly declining plateaus correspond to higher luminosities and larger envelope velocities (Paper~II). By contrast, we do not observe any significant correlation between nebular-phase line widths or flux ratios and the plateau duration, indicating that the length of the recombination phase is largely decoupled from the core properties such as explosion energy.

We also detect a modest correlation between the fractional [O\,\textsc{i}] flux and the phase of the nebular spectrum.  Across the full sample, the Pearson correlation coefficient between $f_{\rm [OI]}$ and phase is $r \simeq 0.6$ (with $p < 10^{-5}$), and the best-fitting linear relation is
\[
f_{\rm [OI]} \;=\; (2.0 \pm 0.1)\!\times\!10^{-4}\,t \;+\; (0.0112 \pm 0.0003),
\]
where $t$ is the phase in days. Although the scatter remains substantial, the values of $f_{\rm [OI]}$ increase slowly over time, as expected once the ejecta become more transparent and the oxygen-rich core contributes a larger fraction of the emergent flux.

We also find a moderate anti-correlation between $f_{\rm [OI]}$ and $M_{r,\mathrm{peak}}$, with a Pearson coefficient $r = -0.50$ (and $p \simeq 8.4\times10^{-4}$), following
\[
f_{\rm [OI]} \;\approx\; (-0.016 \pm 0.006)\,M_{r,\mathrm{peak}} - (0.18 \pm 0.05),
\]
such that fainter SNe exhibit systematically smaller fractional [O\,\textsc{i}] contributions at fixed phase. This behaviour is consistent with the picture that most LLIIP SNe synthesize smaller oxygen masses and/or have reduced excitation in their oxygen-rich cores.

\section{Estimating ZAMS mass}
\label{sec:zams_estimation}

We estimated the zero-age main sequence (ZAMS) masses for our sample of LLIIP SNe using the nebular [O~\textsc{i}]~$\lambda\lambda6300,6364$ emission as a proxy for the oxygen core mass and hence the progenitor ZAMS mass. For each SN with nebular spectra, we measured the [O~\textsc{i}] doublet flux in spectra obtained at phases $\geq 150$~days after explosion, when the ejecta are optically thin and nebular lines directly trace the nucleosynthetic yields. The [O~\textsc{i}] flux was normalized by the integrated pseudo-continuum-subtracted flux between 5500 and 8000\,\AA\ to form the dimensionless fractional [O~\textsc{i}] flux, $f_{\rm [OI]}$, as defined in Section~\ref{sec:fractional_flux}. This fractional quantity is closely related to the fractional [O~\textsc{i}] flux used in previous work and has been shown to correlate with ZAMS mass in both radiative-transfer models and observed nebular spectra \citep[e.g.,][]{Jerkstrand2014,Jerkstrand2015,Fang2025}.

\begin{figure*}
    \centering
    \includegraphics[width=1.0\textwidth]{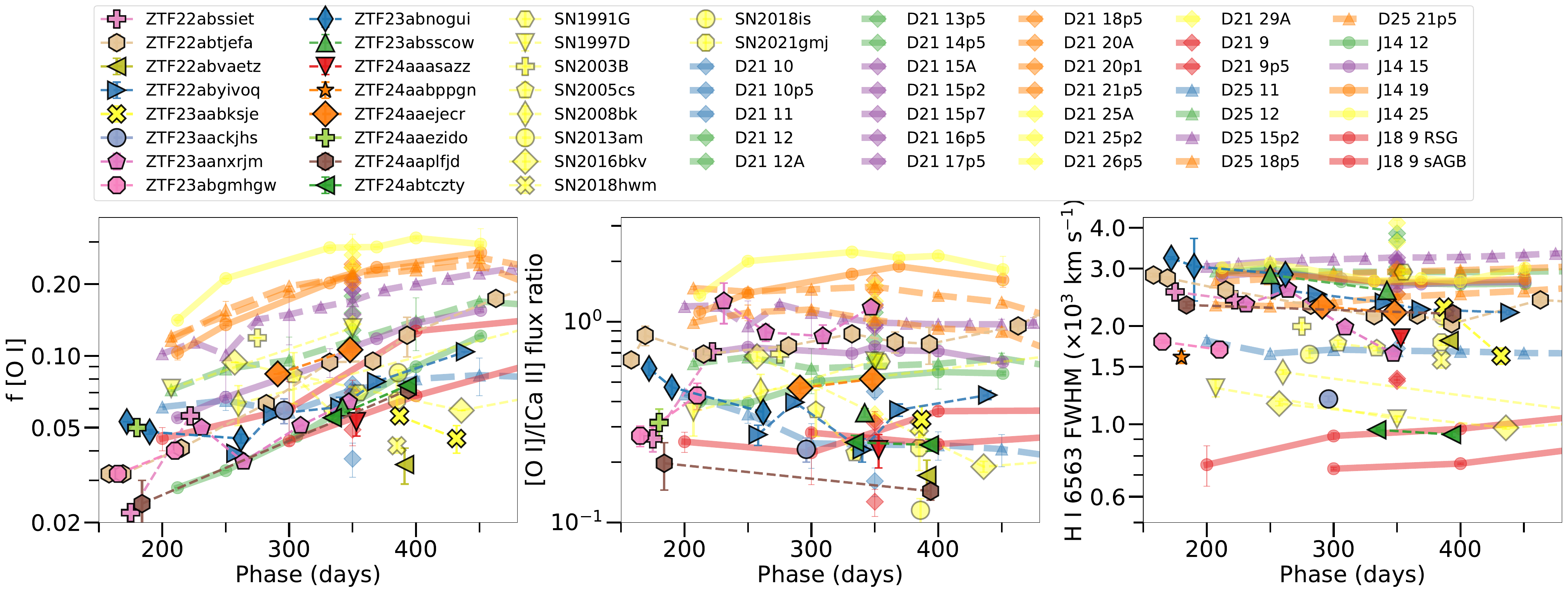}
\caption{Comparison of the ZTF LLIIP sample with literature LLIIP SNe and published 
nebular models. The colored points show the ZTF+CLU LLIIP SNe, and the yellow points 
show low-luminosity LLIIP events from the literature. 
Model tracks from \citeauthor{Dessart2021b} \citeyear{Dessart2021b} (D21),
\citeauthor{Dessart2025} \citeyear{Dessart2025} (D25),
\citeauthor{Jerkstrand2014} \citeyear{Jerkstrand2014} (J14), and \citeauthor{Jerkstrand2018} \citeyear{Jerkstrand2018} (J18) are shown for reference.
The three panels present the evolution of the fractional [O\,\textsc{i}] 
luminosity (left), the [O\,\textsc{i}]/[Ca\,\textsc{ii}] flux ratio (middle), and the 
H\,\textsc{i}\,$\lambda6563$ FWHM (right) as functions of phase.}    \label{fig:analysis_models}
\end{figure*}

To map the fractional [O~\textsc{i}] flux and phase to ZAMS mass, we compared our measurements to the grid of nebular red-supergiant explosion models presented by \citet{Jerkstrand2014}. These models 
use the \texttt{SUMO} spectral synthesis code \citep{Jerkstrand2011} and span a range of progenitor ZAMS masses for hydrogen-rich core-collapse SNe. We do not combine the \citet{Jerkstrand2018} 9~$M_\odot$ models with the \citet{Jerkstrand2014} grid because the underlying progenitor and explosion implementations differ substantially. This is also evident in Figure~\ref{fig:analysis_models}, where the 9~$M_\odot$ models exhibit non-monotonic behavior in $f_{\rm [OI]}$ relative to the 12~$M_\odot$ \citet{Jerkstrand2014} model.

\begin{figure}[ht]
    \centering
    \includegraphics[width=0.99\columnwidth]{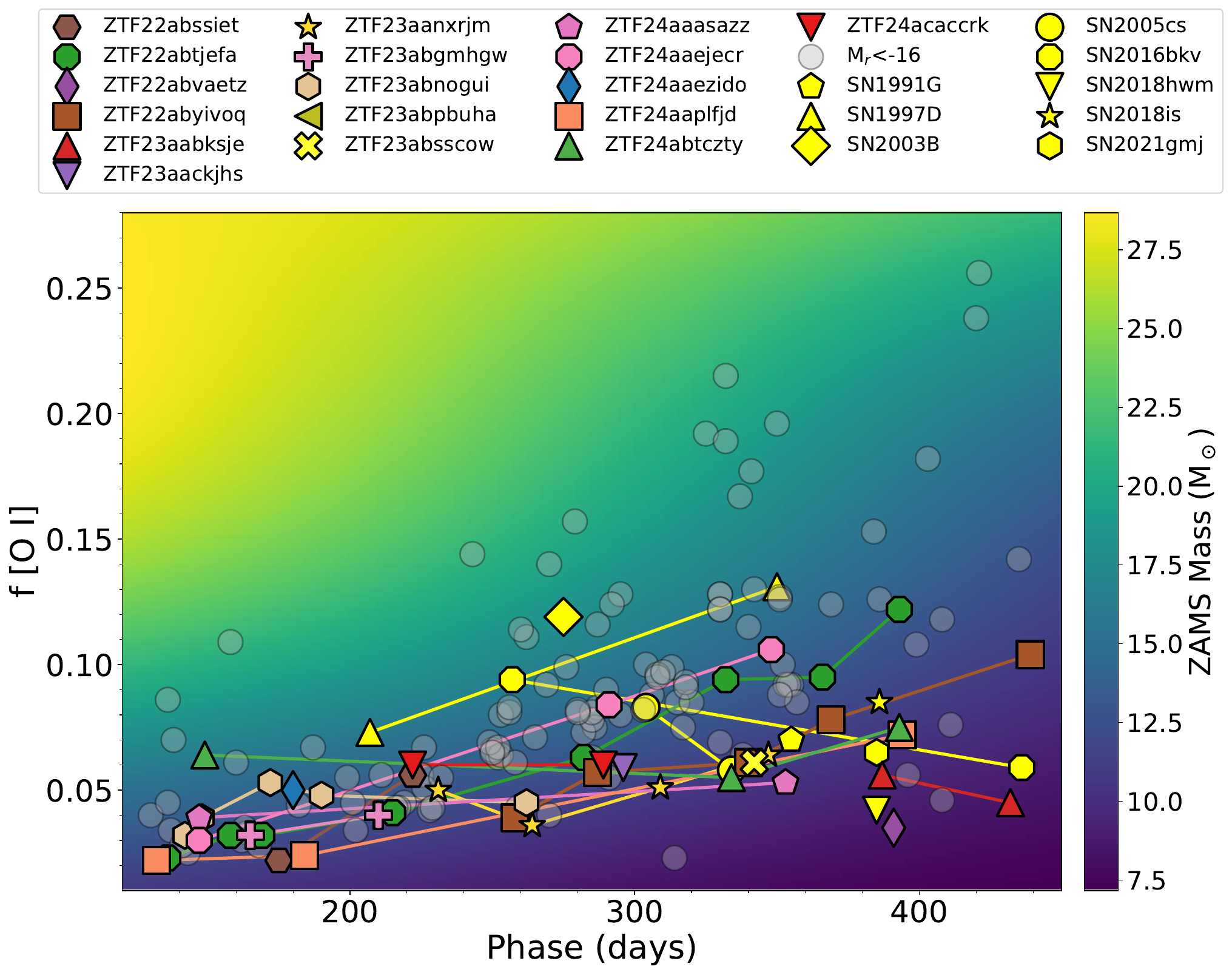}
\caption{Fractional [O\,\textsc{i}] flux as a function of phase for the ZTF LLIIP 
sample (colored symbols) and for literature comparison SNe (gray points). 
The background color shows the ZAMS mass predicted by our Gaussian Process regression 
model, which maps phase and fractional [O\,\textsc{i}] flux to $M_{\rm ZAMS}$ 
as described in Section~\ref{sec:zams_estimation}.}
    \label{fig:mzams_gpr}
\end{figure}

For each model spectrum, we computed $f_{\rm [OI]}$ in exactly the same way as for the data, and we restricted the training set to model epochs with phases $\geq 150$~days in order to minimise systematic biases from residual photospheric contamination. We then trained a Gaussian Process regression (GPR) model on the two-dimensional space of phase and fractional [O~\textsc{i}] flux, $(t, f_{\rm [OI]}) \rightarrow M_{\rm ZAMS}$. The input training vectors consisted of the model phases and $f_{\rm [OI]}$ values, and the targets were the corresponding numerical ZAMS masses parsed from the model labels. The GPR used a squared-exponential kernel of the form
\begin{equation}
    k(\mathbf{x}, \mathbf{x}') = C \, \exp\left[-\frac{(t - t')^2}{2\ell_t^2} - \frac{(f_{\rm [OI]} - f_{\rm [OI]}' )^2}{2\ell_f^2}\right],
\end{equation}
with characteristic length scales of $\ell_t \approx 50$\,d in phase and $\ell_f \approx 0.05$ in fractional [O~\textsc{i}] flux. This provides a smooth, non-parametric interpolation between the discrete model points in the $(t, f_{\rm [OI]})$ plane and yields a continuous mapping to ZAMS mass that is consistent with the radiative-transfer model grid (Figure~\ref{fig:mzams_gpr}). 

For each observed spectrum with phase $t \geq 150$~days, we then propagated the measurement uncertainties in $f_{\rm [OI]}$ into the inferred ZAMS mass using a Monte Carlo procedure. Given the measured fractional [O~\textsc{i}] flux $\hat{f}_{\rm [OI]}$ and its uncertainty $\sigma_{f}$ (constructed as described in Section~\ref{sec:nebular_methods}), we drew $N = 500$ realizations of $f_{\rm [OI]}$ from a Gaussian distribution $\mathcal{N}(\hat{f}_{\rm [OI]}, \sigma_f)$ at the observed phase $t$. For each draw we evaluated the trained GPR at $(t, f_{\rm [OI]})$ to obtain a sample of $M_{\rm ZAMS}$ values. The 16th, 50th, and 84th percentiles of this Monte Carlo distribution were adopted as the inferred median ZAMS mass and its asymmetric uncertainties for that particular spectrum.  We note that we find a moderate correlation between the inferred progenitor masses and the optical peak magnitude. Across the whole sample, the Pearson coefficient between $M_{\rm ZAMS}$ and $M_{r,\mathrm{peak}}$ is $r = -0.58$ (with $p \simeq 6.8\times10^{-5}$), and the best-fitting linear relation,
\[
M_{\rm ZAMS} \;\approx\; (-1.5 \pm 0.4)\,M_{r,\mathrm{peak}} - (11.7 \pm 1.2)\ \mathrm{M_\odot},
\]
shows that fainter peaks correspond to systematically lower progenitor masses. The resulting ZAMS mass estimates for all SNe are listed in Table~\ref{tab:nebular_master}. 

\textcolor{black}{We caution that nebular [O~\textsc{i}] emission more directly probes the helium/oxygen core structure at the time of explosion rather than the progenitor ZAMS mass itself. The inferred $M_{\rm ZAMS}$ values therefore depend on the mapping between pre-SN core structure and initial mass in stellar-evolution models, including assumptions about metallicity, mass loss, convection, dredge-up, and binary interaction \citep[e.g.,][]{Bryne2025}.} We also note that circumstellar interaction or dust formation at late times can suppress or dilute the nebular [O~\textsc{i}] emission and alter the observed peak luminosity \citep[e.g.,][]{Jacobson2025, Dessart2025}, potentially affecting the inferred $M_{\rm ZAMS}$--$M_{r,\mathrm{peak}}$ correlation. We also emphasize that the \citet{Jerkstrand2014} grid extends down only to 12~$M_\odot$. As a result, for SNe with measured $f_{\rm [OI]}$ values near or below the 12~$M_\odot$ track, the inferred masses rely on extrapolation beyond the lower edge of the model grid and should therefore be interpreted with caution. For SNe in the low-$f_{\rm [OI]}$ regime associated with progenitors at or below the lowest-mass \citet{Jerkstrand2014} model, we can robustly conclude that their ZAMS masses are $\lesssim 12~M_\odot$, even though individual sub-12~$M_\odot$ values are not precisely determined. A more self-consistent determination of ZAMS masses using models spanning the full low- to high-mass range will be presented in future work (A. Bostroem et al., in prep.).

\subsection{ZAMS mass distribution of LLIIP SNe}
\label{sec:lliip_mzams_dist}

To characterise the progenitor population, for every SN we choose the spectrum closest to $\sim$350\,d after explosion. This ensures that the fractional [O\,\textsc{i}] flux is measured during the phase range in which the models show the clearest separation with progenitor mass. We adopt the median ZAMS mass $M_{\rm ZAMS,50}$ from the GPR-based mapping between phase, fractional [O\,\textsc{i}] flux, and progenitor mass. The distribution of $M_{\rm ZAMS,50}$ is strongly weighted toward low progenitor masses: the sample median is $M_{\rm ZAMS} \simeq 11.2$\,M$_\odot$, and the 16th and 84th percentiles of the {distribution of $M_{\rm ZAMS,50}$ lie at $\approx$\,9.9 and 12.7\,M$_\odot$ (Figure~\ref{fig:mzams_hist}). SN\,2024btj has a  median ZAMS masses above 14\,M$_\odot$, however, the 16th--84th percentile range spans $\sim$10-17\,M$_\odot$. 

\subsection{IMF constraints from the Type~IIP ZAMS mass distribution}
\label{sec:imf}

We model the underlying progenitor mass distribution of the entire Type~IIP sample as a power law
\begin{equation*}
    \frac{{\rm d}N}{{\rm d}M} \propto M^{-\alpha},
\end{equation*}
between a lower cutoff $M_{\rm min}$ and an upper cutoff fixed at $M_{\rm max} = 22$\,M$_\odot$. For each SN, we treat the GP–inferred progenitor mass as a Gaussian with mean $M_{\rm ZAMS,50}$ and dispersion $M_{\rm ZAMS,std}$; the latter already includes a 10\% systematic term added in quadrature to account for modeling uncertainties. To propagate these individual mass uncertainties into the inferred population slope, we perform a Monte Carlo experiment. In each realization, we draw one mass for every SN from its Gaussian distribution, discard values outside the adopted mass range, and then resample the drawn masses with probabilities proportional to the approximate maximum observable volume for each event,
\begin{equation*}
    w_i \propto V_{{\rm max},i} \propto 10^{0.6 \left( 20 - M_{r,{\rm peak},i} \right)},
\end{equation*}
where $M_{r,{\rm peak}}$ is the peak absolute $r$-band magnitude. This weighting corrects, in a simple way, for the fact that more luminous SNe are detectable over larger volumes and would otherwise be overrepresented in the raw observed distribution. 

To account for the uncertain lower cutoff of the progenitor mass range, we treat $M_{\rm min}$ as a nuisance parameter and allow it to vary between 8 and 12\,M$_\odot$ in the Monte Carlo. For each realization, we draw a random $M_{\rm min}$ within this interval, select the resampled masses above this threshold, and fit a truncated power law using a maximum-likelihood estimator to obtain the best-fit slope $\alpha_k$. Repeating this procedure for 100 realizations yields a distribution of $\alpha$ values that captures both the individual mass uncertainties and the uncertainty in $M_{\rm min}$. From the volume-weighted ensemble we obtain a mean slope of
\begin{equation*}
    \alpha = 2.1 \pm 1.2,
\end{equation*}
where the quoted uncertainty is the standard deviation of the Monte Carlo ensemble. For comparison, a standard Salpeter-like IMF has $\alpha_{\rm Salpeter} \simeq 2.35$ over the massive-star regime. Within the current uncertainties and the restricted mass range we probe ($\sim$8--22\,M$_\odot$), our inferred slope is statistically consistent with a Salpeter-like progenitor distribution. However, the slope is only weakly constrained owing to the small sample size and the fact that, while our LLIIP subsample is drawn from a systematic, volume-limited SN survey, the full sample (including more luminous Type~II SNe) is heterogeneous and compiled from the literature rather than from a uniform selection. A robust IMF determination for Type~IIP progenitors will ultimately require a larger, systematically selected sample of nebular spectra drawn from a  volume-limited survey with no intrinsic brightness cut.

\begin{figure*}[ht]
    \centering
    \includegraphics[width=0.35\textwidth]{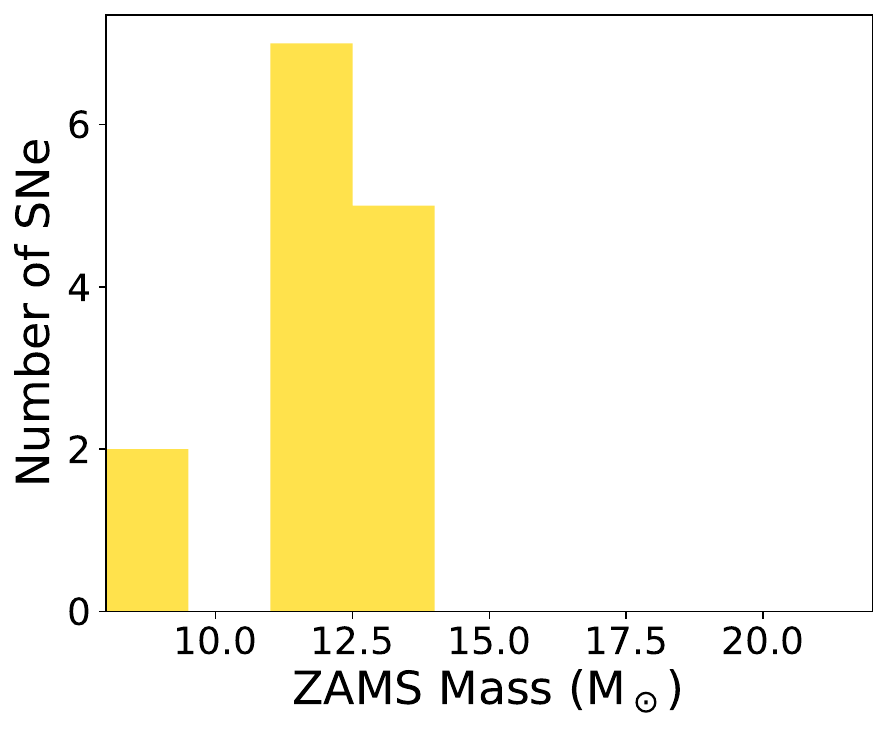}\includegraphics[width=0.4\textwidth]{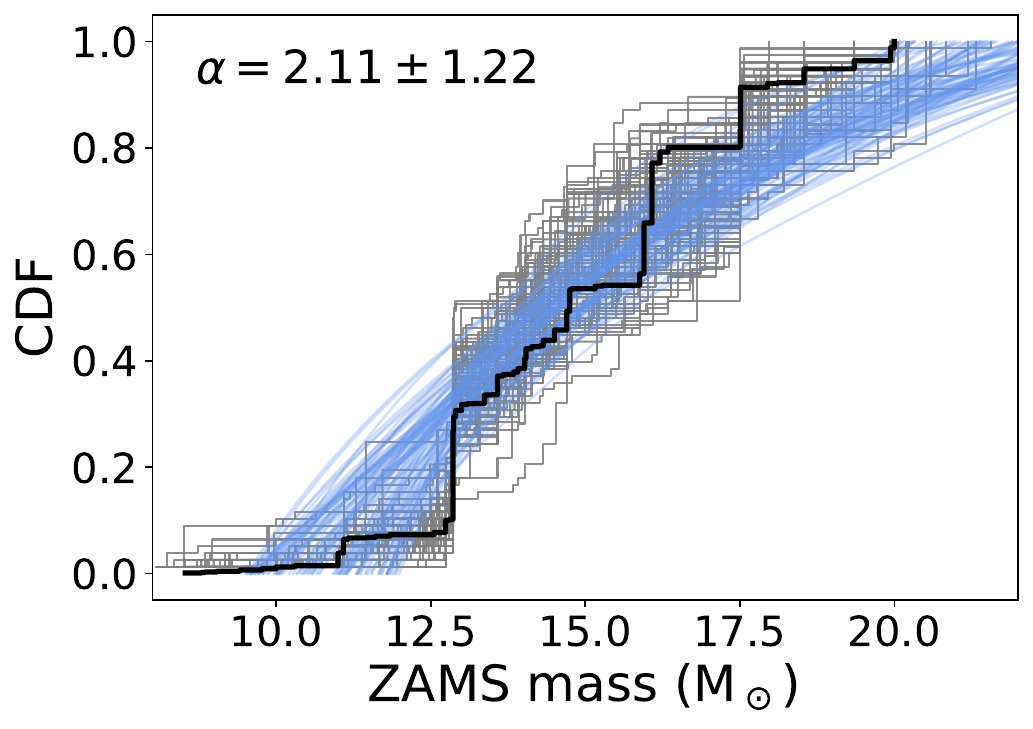}
    \includegraphics[width=0.4\textwidth]{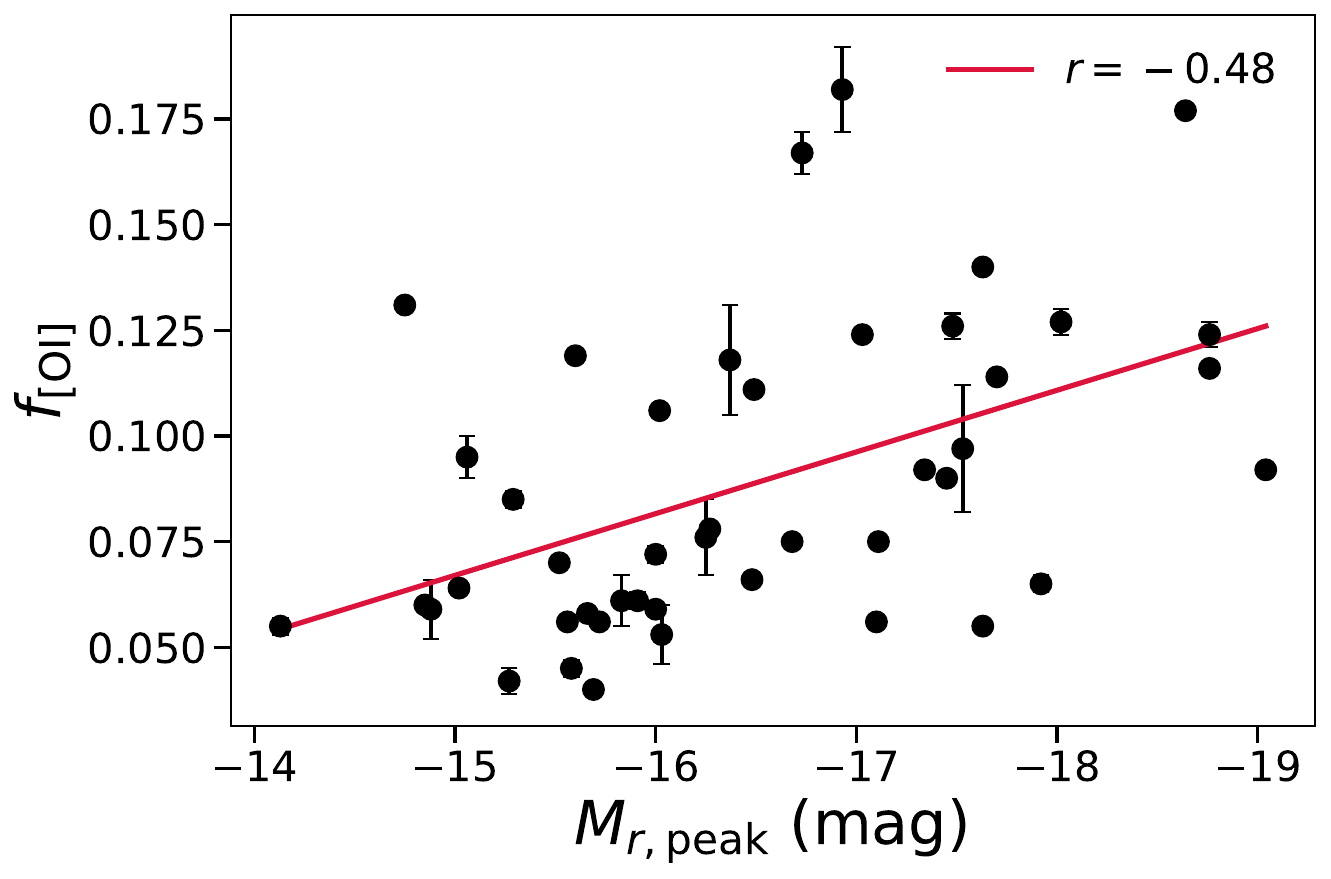}\includegraphics[width=0.4\textwidth]{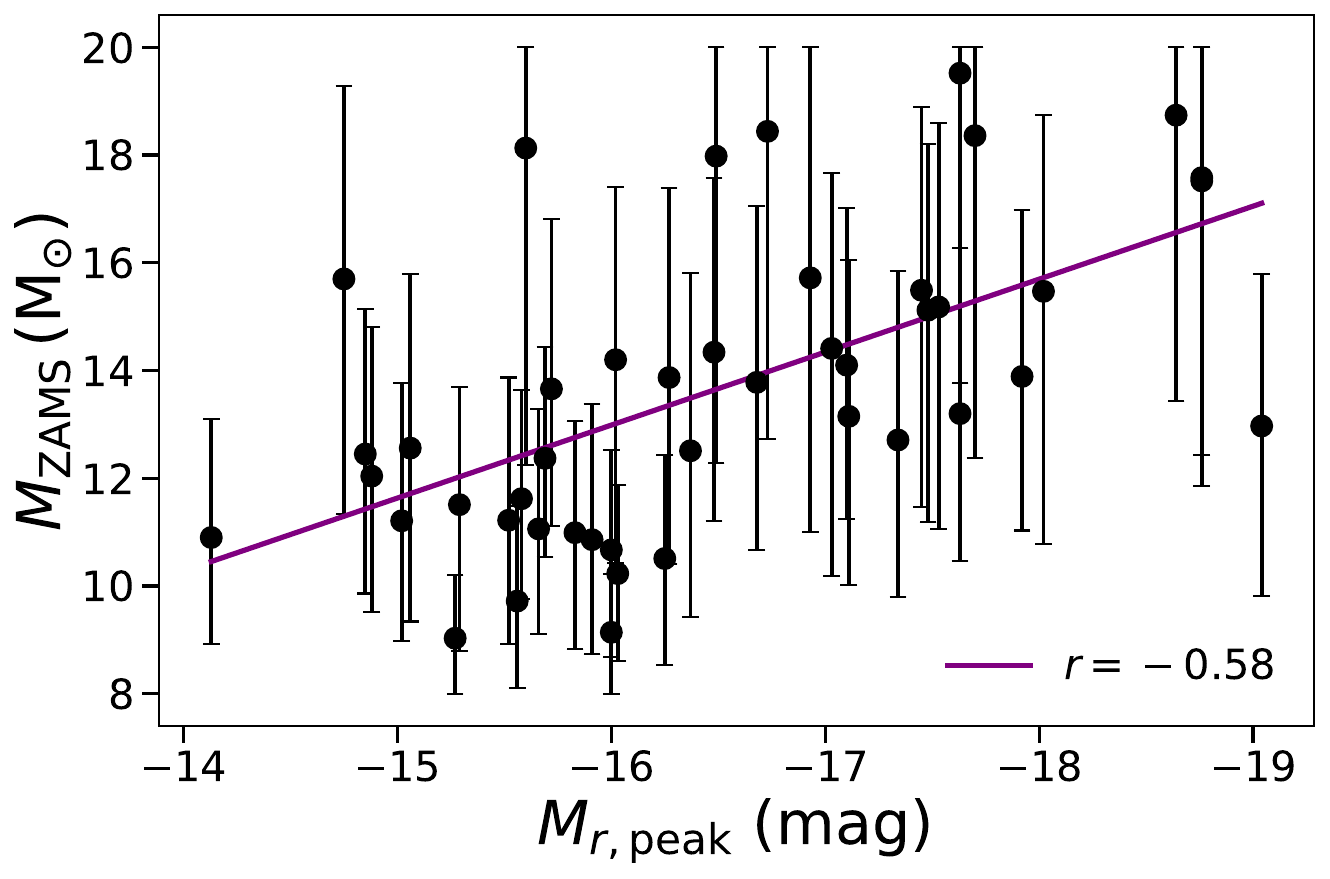}
\caption{ 
Top left: distribution of ZAMS masses inferred for the ZTF LLIIP sample. 
Top right: cumulative distribution of the inferred masses, with Monte Carlo realizations 
shown in blue and the best-fit power-law slope $\alpha$ indicated. 
Bottom left: correlation between the fractional [O\,\textsc{i}] luminosity and the 
$r$-band peak magnitude. 
Bottom right: weak correlation between the inferred ZAMS mass and the $r$-band peak magnitude. 
Pearson coefficients for the two relations are shown in each panel.}
    \label{fig:mzams_hist}
\end{figure*}

\section{Electron-capture SN diagnostics}
\label{sec:ecsn}

\subsection{ECSN score}

An important theoretical motivation for using nebular-line diagnostics comes from the expected ejecta structure of electron-capture supernovae (ECSNe). ECSNe arising from sAGB progenitors lack both an oxygen-rich shell and a helium shell, since their degenerate ONeMg cores collapse before substantial burning can synthesize or expose these layers \citep[e.g.,][]{Jerkstrand2018}. As a result, several nebular features that are prominent in Fe-core-collapse SNe are predicted to be absent or strongly suppressed in ECSNe. These include He I $\lambda7065$, Mg I] $\lambda4571$, and [C I] $\lambda8727$, all of which require the presence of He- or O/C-rich zones. Although [C\,I]\,$\lambda9850$ is also expected to be a useful Fe-core-collapse/ECSN diagnostic, we do not include it in the formal score because the spectra in this wavelength region generally have low SNR across our sample. The absence of an O-rich shell also leads to systematically weak [O\,I]\,$\lambda\lambda$6300,6364. In addition, the innermost ejecta of ECSNe are expected to synthesize large amounts of stable $^{58}$Ni, giving rise to comparatively strong [Ni\,II] emission. Although [Fe\,I] and [Fe\,II] lines should be intrinsically weak due to the lack of an Fe/Si shell, low-level Fe emission may still arise from explosive burning of the residual envelope, and therefore are not used to distinguish ECSNe from Fe-core-collapse SNe. We note that these qualitative nebular distinctions between ECSNe and Fe-core-collapse SNe predicted by one-dimensional models  have also been recently shown to persist in three-dimensional simulations \citep{Bart2025}.

To ensure uniform classification across the sample, we assess each diagnostic quantitatively. For the line-presence diagnostics He\,I\,$\lambda7065$, O\,I\,$\lambda7774$, [C\,I]\,$\lambda8727$, Mg\,I]\,$\lambda4571$, and O\,I\,$\lambda8447$, we extract a 20\,\AA\ window centered on the expected wavelength, subtract a local linear continuum, and compute the standard deviation of the residual flux to estimate the noise and line SNR. We also compute a peak ratio, defined as the line peak relative to the peak of [O\,\textsc{i}]\,$\lambda\lambda6300,6364$ in the same spectrum, providing a luminosity-independent scale for comparing weak features across objects. \textcolor{black}{The `yes' and `no' thresholds are calibrated using the \citet{Jerkstrand2018} $9\,\Msun$ nebular models and are listed in Table~\ref{tab:ecsn_weights_thresholds}; measurements falling between these thresholds are labeled `maybe'. For the ratio diagnostics, we use the measured [Ni\,II]/[Fe\,II] and [Ca\,II]/[O\,I] flux ratios and apply the thresholds in Table~\ref{tab:ecsn_weights_thresholds}. The observational label is then converted to a numerical score according to whether it supports or contradicts the ECSN expectation. For diagnostics expected to be absent or weak in ECSNe, a `no' classification is assigned $s_i=+1$, a `yes' classification is assigned $s_i=-1$, and `maybe' cases are assigned $s_i=0$. For [Ni\,II]/[Fe\,II], the thresholds in Table~\ref{tab:ecsn_weights_thresholds} map directly to $s_i=+1$ and $-1$. For [Ca\,II]/[O\,I], we use the ratio only as a one-sided veto: values below the listed `no' threshold are assigned $s_i=-1$, while all larger values are treated as inconclusive ($s_i=0$). We also perform a visual inspection to ensure the reliability of each classification, particularly for blended or noisy regions.}

Each diagnostic is assigned a weight $w_i$ that reflects its physical robustness
and discriminatory power for distinguishing ECSNe from Fe-core-collapse SNe. He\,I\,$\lambda7065$, O\,I\,$\lambda7774$, [C\,I]\,$\lambda8727$, and [Ca\,II]/[O\,I] are assigned unit weight. Mg\,I]\,$\lambda4571$, O\,I\,$\lambda8447$, and [Ni\,II]/[Fe\,II] are down-weighted to $w_i=0.2$, since these diagnostics are informative but more sensitive to line-formation uncertainties, blending, and model details \citep[e.g.,][]{Jerkstrand2015a}.  Combining the categorical scores $s_i$ with their weights $w_i$, we compute a normalized ECSN score

\[
\mathrm{ECSN\ score}
    = 
    \frac{\sum_i w_i s_i}{\sum_i w_i},
\]
which lies in the range
\[
-1 \leq \mathrm{ECSN\ score} \leq +1.
\]
Scores near $+1$ indicate that most diagnostics behave as expected for an ECSN, scores near $-1$ indicate the opposite, and intermediate values represent mixed or inconclusive evidence. To assess the robustness of the ECSN score to the adopted weights and thresholds, we performed 100 Monte Carlo realizations in which both were randomly perturbed by up to $\pm 50\%$ of their nominal values. For every realization we recomputed the full set of ECSN scores. The standard deviation of the ECSN score for each object was below $5\%$ of its nominal value. This shows that the ECSN score is stable against reasonable variations in the adopted weighting scheme and thresholds. \textcolor{black}{All the diagnostics and their color-coded interpretations for the LLIIP SNe and theoretical models are listed in Tables~\ref{tab:ecsn_diag} and \ref{tab:ecsn_theoretical_model_scores}, respectively.}

\input{ecsn_diagnostics_table_no_CI9100}

\subsection{ECSN scores of the LLIIP sample}
\label{subsec:ecsn_scores}

A small subset of objects attain mean ECSN scores higher than that of SN~2016bkv and SN~2018zd. The highest mean ECSN score is found for SN~2023bvj, based on its single available spectrum, followed by SN~2024btj and SN~2016bkv, which also remain positive across multiple epochs. In SN~2023bvj, the high score is driven primarily by the non-detection of He\,I\,$\lambda7065$ and O\,I\,$\lambda7774$, together with a detection of O\,I\,$\lambda8447$; Mg\,I]\,$\lambda4571$ and [C\,I]\,$\lambda8727$ remain ambiguous and the ratio diagnostics are inconclusive. We also note that [C\,I]\,$\lambda9850$ is not detected, further supporting an ECSN interpretation. SN~2024btj shows a robust absence of Mg\,I]\,$\lambda4571$ and O\,I\,$\lambda7774$ at both epochs, with [C\,I] and O\,I\,$\lambda8447$ either absent or marginal; its [Ni\,II]/[Fe\,II] ratio remains consistently low, which limits the overall ECSN score. SN~2016bkv likewise reaches high late-time scores because He\,I\,$\lambda7065$ and O\,I\,$\lambda7774$ are absent in the two later spectra while O\,I\,$\lambda8447$ is present at all epochs, although Mg\,I]\,$\lambda4571$ is clearly detected at late times. Among the remaining comparison objects, SN~2022zmb shows a modestly positive ECSN score, with non-detections of He\,I\,$\lambda7065$ and enhanced [Ni\,II]/[Fe\,II] partly offset by the absence of O\,I\,$\lambda8447$ and a detection of O\,I\,$\lambda7774$ in one spectrum. SN~2005cs has a lower overall score because [C\,I]\,$\lambda8727$\, is clearly present in one spectrum.

At the opposite extreme, SN~2024jxm has the most negative mean ECSN score in the sample, while SN~2024abfl, SN~2023kmk, and SN~2023vci also remain consistently negative at all observed epochs. SN~2024jxm shows He\,I\,$\lambda7065$ in the early spectra, [C\,I] emission that is present or at least not clearly absent, and persistently weak [Ni\,II]/[Fe\,II]. SN~2023kmk is driven negative by strong He\,I\,$\lambda7065$ at all epochs, repeated non-detections of [Ni\,II]/[Fe\,II], and at some epochs detected O\,I\,$\lambda7774$ or low [Ca\,II]/[O\,I]. SN~2023vci is likewise negative at both epochs because He\,I\,$\lambda7065$ is present while O\,I\,$\lambda8447$ and [Ni\,II]/[Fe\,II] are both weak. SN~2008bk also disfavors an ECSN interpretation, especially at 260\,d where He\,I\,$\lambda7065$, Mg\,I]\,$\lambda4571$, [C\,I]\,$\lambda8727$, and O\,I\,$\lambda8447$ are all clearly detected. \textcolor{black}{Figure~\ref{fig:ecsn_candidate_lightcurves} compares the $g$- and $r$-band
light curves of the three highest-scoring ECSN candidates with those of the ZTF
LLIIP nebular sample. The candidates are not photometrically distinct from the
broader LLIIP population.
}

Taken together, the ECSN scores show that only a handful of LLIIP SNe have nebular spectra that are at least as ECSN-like as SN~2016bkv and SN~2018zd in terms of line ratios and the presence or absence of key features. We regard SN~2023bvj, SN~2024btj, and SN~2016bkv as the strongest ECSN candidates in this diagnostic sense. However, we note that their nebular H$\alpha$ line widths and fractional [O\,I] fluxes $f_{\rm [O\,I]}$ are larger than predicted by the 9\,M$_\odot$ low-mass models of \citet{Jerkstrand2018}, which favor extremely narrow Balmer lines and very weak [O\,I] emission for ECSN-like explosions.

\subsection{Rate estimate of ECSNe from sAGB stars}
\label{subsec:ecsn_rate}

Our volume-limited ZTF CLU sample contains 28 LLIIP SNe within 100\,Mpc (Paper~I), of which 19 have nebular spectra of sufficient quality to measure the diagnostics described above. Among these 19, at most two events, SN~2023bvj and SN~2024btj, have ECSN scores higher than SN~2016bkv and SN~2018zd and therefore qualify as our strongest nebular ECSN candidates within the LLIIP sample. We can therefore use this systematic sample to place an approximate upper limit on the ECSN rate through the LLIIP channel. If both candidates are genuine ECSNe and the nebular subsample is representative of the parent LLIIP population, the implied ECSN fraction within the LLIIP class lies between $2/28 \simeq 0.07$ (normalizing to all 28 LLIIP SNe) and $2/19 \simeq 0.11$ (normalizing only to the nebular subsample). Combined with the volumetric LLIIP rate of $\sim\!7.3\times10^{3}\,\mathrm{Gpc^{-3}\,yr^{-1}}$ from Paper~I \citep{Das2025a}, this implies an approximate upper limit of $(5$--$8)\times10^{2}\,\mathrm{Gpc^{-3}\,yr^{-1}}$ on the rate of ECSNe that pass through the LLIIP channel.

In Paper~I, we showed that LLIIP SNe constitute $8^{+1}_{-2}\%$ of all CCSNe \citep{Das2025a}. Under the assumption that all ECSNe in the local Universe manifest as LLIIP SNe, the two ECSN candidates identified in our sample therefore imply an ECSN fraction of all CCSNe of approximately $f_{\mathrm{ECSN}} \sim$0.004--0.010 after accounting for the uncertainty in the LLIIP-to-CCSN fraction.

To translate this ECSN fraction into a constraint on the progenitor mass range, we adopt the simple picture in which ECSNe arise from sAGB stars at the low-mass end of the CCSN population. We assume that all stars with zero-age main-sequence masses between $M_{\mathrm{CC,min}}$ and $M_{\mathrm{CC,max}} = 120\,M_\odot$ can undergo core collapse, and that the lowest-mass CCSNe are ECSNe produced by sAGB stars. In this picture, the sAGB interval runs from $M_{\mathrm{sAGB,min}}$ to $M_{\mathrm{sAGB,max}}$, and we explicitly set $M_{\mathrm{sAGB,min}} = M_{\mathrm{CC,min}}$. For a Salpeter initial mass function (IMF), $\psi(M) \propto M^{-2.35}$, the fraction of CCSNe that are ECSNe is
\begin{equation*}
    f_{\mathrm{ECSN}} =
    \frac{\displaystyle \int_{M_{\mathrm{sAGB,min}}}^{M_{\mathrm{sAGB,max}}} M^{-2.35}\,dM}
         {\displaystyle \int_{M_{\mathrm{CC,min}}}^{M_{\mathrm{CC,max}}} M^{-2.35}\,dM},
\end{equation*}
where we parameterize the ECSN-producing interval by $\Delta M_{\mathrm{sAGB}} = M_{\mathrm{sAGB,max}} - M_{\mathrm{sAGB,min}}$, and adopt $M_{\mathrm{CC,max}} = 120\,M_\odot$.

For a fiducial choice of $M_{\mathrm{CC,min}} = M_{\mathrm{sAGB,min}} = 8.5\,M_\odot$, the nominal CCSN fractions above imply $\Delta M_{\mathrm{sAGB}} \simeq 0.035\,M_\odot$ to $0.052\,M_\odot$, corresponding to progenitor intervals of approximately $8.50$--$8.54\,M_\odot$ and $8.50$--$8.55\,M_\odot$. Varying the CCSN threshold over the plausible range $M_{\mathrm{CC,min}} \simeq 8$--$9\,M_\odot$ changes these values only slightly: the corresponding widths are $\sim$0.033--0.037\,M$_\odot$ for the $2/28$ normalization and $\sim$0.049--0.055\,M$_\odot$ for the $2/19$ normalization. Even allowing the full $6$--$9\%$ range for the LLIIP fraction of all CCSNe broadens the inferred window only to about $0.025$--$0.062\,M_\odot$. These trends are illustrated in Figure~\ref{fig:ecsnfraction}, which shows the required sAGB width $\Delta M_{\mathrm{sAGB}}$ as a function of the CCSN threshold mass $M_{\mathrm{CC,min}} = M_{\mathrm{sAGB,min}}$ for the two empirical ECSN fractions inferred above. The shaded region between the curves marks the range of $\Delta M_{\mathrm{sAGB}}$ values consistent with our measurements.

The evolution to sAGB stars is highly sensitive to dredge-up/dredge-out and mass-loss physics, yielding a broad range of predicted ECSN-producing mass windows. However, the progenitor interval inferred here is narrower than most theoretical expectations; for example, $\Delta M_{\rm sAGB} \approx 0.2$--$1.4\,M_\odot$ at solar metallicity \citep[e.g.,][]{Poelarends2008}. \textcolor{black}{Our inferred narrow progenitor interval is qualitatively consistent with  \citet{Podsiadlowski2004}, who argued that the mass range for single stars producing ECSNe may be extremely narrow or even absent. In that picture, binary interactions could instead provide the dominant ECSN channel by enabling some stars in the $\sim8-11\,M_\odot$ range, which would otherwise form ONeMg white dwarfs as single stars, to instead undergo ECSNe.}

We note that the LLIIP channel need not capture the entire ECSN population. Some ECSNe may instead appear as other low-luminosity or interaction-powered transients, such as intermediate-luminosity red transients \citep[e.g.,][]{Rose2024}, brighter Type IIP SNe (as proposed for SN~2018zd), or Type IIP-n events with stronger circumstellar interaction \citep{Hiramatsu2021,Smith2013}. A systematic, volume-limited survey of late-time nebular spectra across all nearby CCSN subtypes will be required to obtain a more comprehensive measurement of the ECSN rate and the associated sAGB progenitor mass range.

\begin{figure}
    \centering
    \includegraphics[width=1.0\columnwidth]{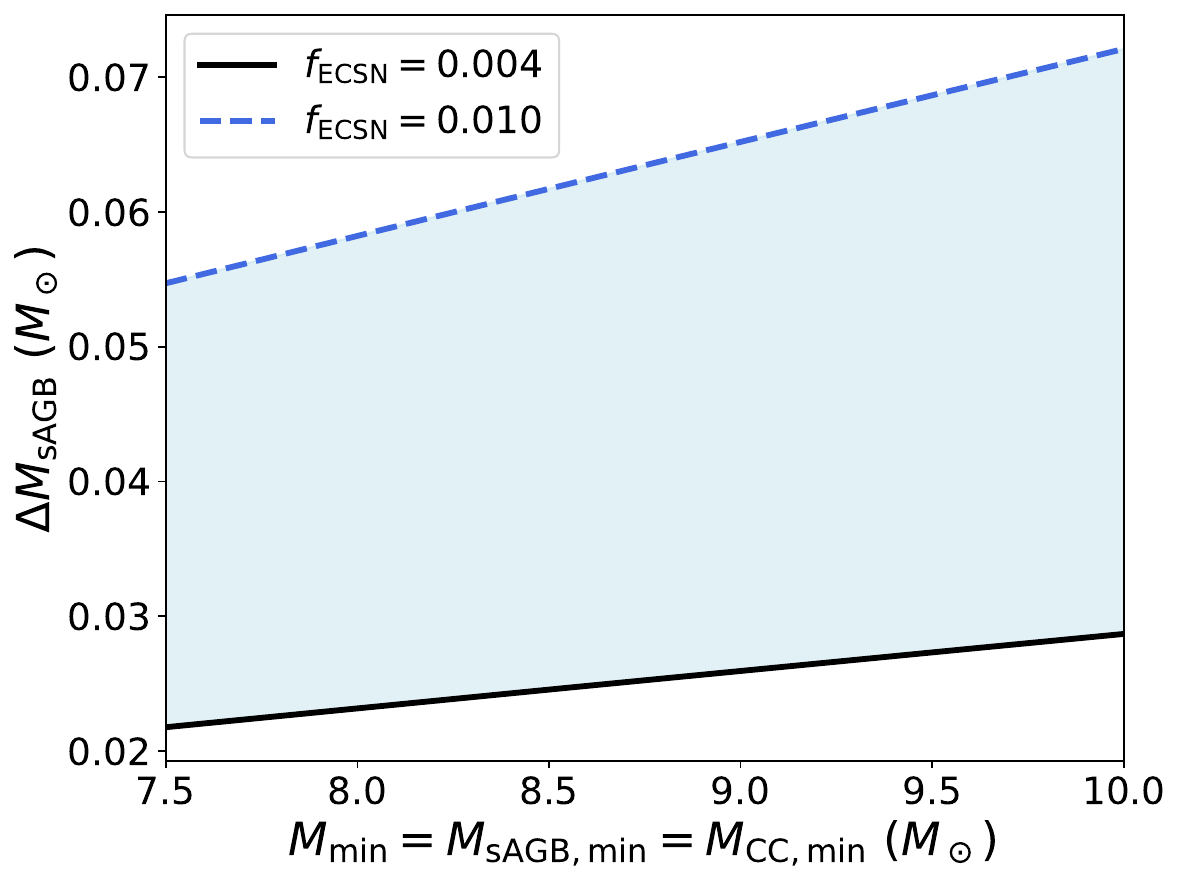}
\caption{Required width of the sAGB ECSN window $\Delta M_{\rm sAGB}$ as a function of the 
CCSN threshold mass $M_{\rm CC,min}$. The blue shaded region indicates the range of
$\Delta M_{\rm sAGB}$ values consistent with the empirical constraints.}
    \label{fig:ecsnfraction}
\end{figure}

\section{Conclusion}
\label{sec:conclusion}

We obtained usable nebular spectra for 19 (out of 28) LLIIP SNe within 100\,Mpc from the ZTF CLU survey discovered between 2022 November 1 and 2024 November 1 that satisfy the Paper~I quality criteria, at phases between $\sim$140 and 450\,d after explosion. Their low expansion velocities and explosion energies allow us to resolve narrow nebular emission lines that are typically blended in more energetic Type II SNe. We summarize the key takeaways here:

\begin{itemize}

    \item The LLIIP SNe in our sample exhibit systematically narrower nebular H\,\textsc{i}\,$\lambda6563$ emission lines than brighter comparison Type II SNe from the literature. A planar fit to the H\,\textsc{i}\,$\lambda6563$ FWHM as a function of phase and peak absolute magnitude,
    \[
    \begin{aligned}
    \mathrm{FWHM\ (\AA)} = &\, (-72.76 \pm 1.15) \\
    &+ (-0.05 \pm 0.01)\,\mathrm{Phase\ (d)} \\
    &+ (-9.05 \pm 0.07)\,M_{\mathrm{peak}},    \end{aligned}
    \]
    
    with a Pearson correlation coefficient $\mathrm{r} = 0.65$, shows that fainter SNe have lower nebular H\,\textsc{i} velocities at fixed phase, consistent with intrinsically low-energy explosions.

    \item When we compare the observed nebular H\,\textsc{i} widths to models for low-mass progenitors, we find that only SN~2024abfl in our LLIIP sample, together with SN~1997D and SN~2016bkv from the literature, approaches the extremely low H\,\textsc{i}\,$\lambda6563$ widths predicted for the weakest explosions of $\sim 9$\,M$_\odot$ RSG or sAGB progenitors \citep{Jerkstrand2018}. Most LLIIP SNe have nebular H\,\textsc{i} widths of at least $\sim 2000$\,km\,s$^{-1}$ despite their low explosion energies and $^{56}$Ni masses. Given that 8--10\,M$_\odot$ stars represent a substantial fraction of core-collapse progenitors for a Salpeter IMF, this scarcity suggests either that the effective lower mass limit for successful core collapse is higher than in some models or that real explosions near this threshold are not as uniformly weak as core-collapse models predict.

    \item We do not find any significant correlation between nebular-phase line widths or flux ratios and the lightcurve plateau duration. This indicates that the length of the recombination phase is largely decoupled from the explosion and other core properties probed by the nebular spectra.

    \item The fractional [O\,\textsc{i}] flux $f_{\rm [OI]}$ shows a modest but statistically significant correlation with phase. Across the full sample, the Pearson coefficient between $f_{\rm [OI]}$ and phase is $r \simeq 0.6$ (with $p < 10^{-5}$), and the best-fitting relation
    \[
    f_{\rm [OI]} \;=\; (2.0 \pm 0.1)\!\times\!10^{-4}\,t \;+\; (0.0112 \pm 0.0003),
    \]
    where $t$ is the phase in days, likely shows that $f_{\rm [OI]}$ increases slowly over time as the ejecta become more transparent and the oxygen-rich core contributes a larger fraction of the emergent flux.

    \item We estimated zero-age main sequence (ZAMS) progenitor masses using the nebular [O\,\textsc{i}]~$\lambda\lambda6300,6364$ emission as a proxy for oxygen-core mass. We find a moderate correlation between the inferred $M_{\rm ZAMS}$ and the optical peak magnitude, with Pearson coefficient $r = -0.58$ (and $p \simeq 6.8\times10^{-5}$) and best-fitting relation
    \[
M_{\rm ZAMS} \;\approx\; (-1.5 \pm 0.4)\,M_{r,\mathrm{peak}} - (11.7 \pm 1.2)\ \mathrm{M_\odot},    \]
    showing that fainter peaks correspond to systematically lower progenitor masses. We caution that circumstellar interaction or dust formation can modify the observed peak luminosity or suppress nebular [O\,\textsc{i}] emission, potentially biasing the inferred $M_{\rm ZAMS}$–$M_{r,\mathrm{peak}}$ relation.

    \item The distribution of ZAMS mass of LLIIP SNe is strongly weighted towards low-mass progenitors. The median is $M_{\rm ZAMS} \simeq 11.2$\,M$_\odot$ and the 16th to 84th percentiles lie at $\approx 9.9$ and 12.7\,M$_\odot$. Twelve out of nineteen SNe have $M_{\rm ZAMS,50} \leq 12$\,M$_\odot$, providing robust evidence that the LLIIP progenitor population is dominated by stars in the 8--12\,M$_\odot$ regime. From the volume-weighted ensemble we infer an IMF power-law index of $\alpha = 2.1 \pm 1.2$ over the $\sim 8$--22\,M$_\odot$ range, which is statistically consistent with a Salpeter-like progenitor distribution.

    \item To search for electron-capture supernovae, we introduced an ``ECSN score'' based on the absence of He- and O-shell emission lines and the Ni/Fe diagnostics in the nebular spectra. Only a handful of LLIIP SNe achieve scores that are at least as ECSN-like as SN~2016bkv and SN~2018zd. In our sample, we regard SN~2023bvj and SN~2024btj as the strongest ECSN candidates in this diagnostic sense. However, we note that their nebular H$\alpha$ line widths are larger than predicted by the 9\,M$_\odot$ low-mass models of \citet{Jerkstrand2018}, which favor extremely narrow Balmer lines for ECSN-like explosions.

\item Within our volume-limited sample of 28 LLIIP SNe, two events out of the full sample and out of the 19 SNe with usable nebular spectra satisfy our ECSN-score criteria. This corresponds to fractions of $2/28 \simeq 0.07$ and $2/19 \simeq 0.11$, which we treat as approximate upper limits on the fraction of LLIIP SNe arising from the ECSN channel. Assuming a volumetric LLIIP rate of $\sim 7.3\times10^{3}\,\mathrm{Gpc^{-3}\,yr^{-1}}$ \citep{Das2025a}, these fractions imply an approximate upper limit on the rate of ECSN candidates passing through the LLIIP channel of $(5$--$8)\times10^{2}\,\mathrm{Gpc^{-3}\,yr^{-1}}$. If ECSNe predominantly arise through the LLIIP SN channel, and if they originate from sAGB stars at the low-mass end of the CCSN population, then the fact that LLIIP SNe constitute $8^{+1}_{-2}\%$ of all CCSNe \citep{Das2025a} implies an ECSN fraction of all CCSNe of approximately $\sim0.004$--$0.010$. This corresponds to an sAGB progenitor mass window of order $0.02$--$0.06\,M_\odot$.

\end{itemize}

We emphasize that the LLIIP channel need not encompass the full ECSN population. Some ECSNe may instead manifest as other faint or interaction-powered transients, including intermediate-luminosity red transients, relatively brighter Type IIP SNe (as has been proposed for SN~2018zd), or Type IIP-n events with stronger circumstellar interaction. A truly robust measurement of the ECSN rate, and the associated sAGB progenitor mass window, will therefore require a systematic, volume-limited program of late-time nebular spectroscopy spanning all nearby core-collapse supernova subtypes.

We note that none of the nebular spectra match all predicted features of existing low-mass RSG or sAGB models in detail. The 9~M$_\odot$ RSG and sAGB nebular models of \citet{Jerkstrand2018} do not fully explore the effects of multi-dimensional mixing and core asymmetries, even though recent three-dimensional simulations suggest that the key one-dimensional ECSN versus Fe-core-collapse distinctions remain broadly robust at the lowest masses \citep{Bart2025}. An additional caveat is that the simplified sAGB nebular model is constructed by replacing the composition with H-zone material throughout, rather than using a fully evolved sAGB progenitor structure. A broader suite of nebular spectral calculations that spans both ONeMg-core and Fe-core explosions at the low-mass end, based on fully evolved sAGB stars, will allow one to establish more reliable and quantitative ECSN diagnostics. It will also be valuable to develop complementary mid-infrared diagnostics, for example with \textit{JWST}, to probe ejecta composition and cooling channels that may be less accessible at optical wavelengths.

Looking ahead, future wide-field time-domain surveys such as LSST will discover large numbers of faint transients. Expanding to a larger, volume-limited sample with uniform late-time spectroscopy across all nearby core-collapse subtypes will enable tighter constraints on the ECSN rate, the sAGB progenitor mass range, and the contribution of ECSNe at the low-mass end of the initial mass function, thereby improving our understanding of how the least massive core-collapse progenitors end their lives.

\section{Acknowldegement}

We thank Qiliang Fang, Ryan Chornock, Azalee Bostroem  and Luc Dessart for helpful discussions.

W.J.-G.\ is supported by NASA through Hubble Fellowship grant HSTHF2-51558.001-A awarded by the Space Telescope Science Institute, which is operated for NASA by the Association of Universities for Research in Astronomy, Inc., under contract NAS5-26555.

M.W.C. acknowledges support from the National Science Foundation with grant numbers PHY-2117997, PHY-2308862 and PHY-2409481.

Based on observations obtained with the Samuel Oschin Telescope 48-inch and the 60-inch Telescope at the Palomar Observatory as part of the Zwicky Transient Facility project. ZTF is supported by the National Science Foundation under Grants No. AST-1440341, AST-2034437, and currently Award 2407588. ZTF receives additional funding from the ZTF partnership. Current members include Caltech, USA; Caltech/IPAC, USA; University of Maryland, USA; University of California, Berkeley, USA; University of Wisconsin at Milwaukee, USA; Cornell University, USA; Drexel University, USA; University of North Carolina at Chapel Hill, USA; Institute of Science and Technology, Austria; National Central University, Taiwan, and OKC, University of Stockholm, Sweden. Operations are conducted by Caltech's Optical Observatory (COO), Caltech/IPAC, and the University of Washington at Seattle, USA.

Zwicky Transient Facility access for S.S. was supported by Northwestern University and the Center for Interdisciplinary Exploration and Research in Astrophysics (CIERA).


The ZTF forced-photometry service was funded under the Heising-Simons Foundation grant \#12540303 (PI: Graham).

The Gordon and Betty Moore Foundation, through both the Data-Driven Investigator Program and a dedicated grant, provided critical funding for SkyPortal .

This research has made use of the NASA/IPAC Extragalactic Database (NED), which is funded by the National Aeronautics and Space Administration and operated by the California Institute of Technology.

The Liverpool Telescope is operated on the island of La Palma by Liverpool John Moores University in the Spanish Observatorio del Roque de los Muchachos of the Instituto de Astrofisica de Canarias with financial support from the UK Science and Technology Facilities Council.

The W. M. Keck Observatory is operated as a scientific partnership among the California Institute of Technology, the University of California and the National Aeronautics and Space Administration. The Observatory was made possible by the generous financial support of the W. M. Keck Foundation. The authors wish to recognize and acknowledge the very significant cultural role and reverence that the summit of Maunakea has always had within the indigenous Hawaiian community.  We are most fortunate to have the opportunity to conduct observations from this mountain.

\textit{Software:} Global Relay of Observatories Watching Transients Happen Marshal \citep[GROWTH;][]{Kasliwal2019} and the Fritz SkyPortal Marshal  \citep[][]{Duev2019, skyportal, Coughlin2023}. Astropy \citep{Astropy-Collaboration13}, Matplotlib \citep{Hunter07}

\bibliography{main}
\bibliographystyle{aasjournal}

\newpage
\appendix
Table~\ref{tab:ecsn_weights_thresholds} lists the adopted ECSN diagnostic
weights and thresholds. Table~\ref{tab:ecsn_theoretical_model_scores} summarizes the
resulting ECSN scores measured from theoretical nebular spectral models. Table~\ref{table:bright_spectral_lo} lists the nebular spectral log for the brighter ZTF Type~IIP comparison sample. Figure~\ref{fig:analysis} illustrates the full nebular spectral analysis workflow on a representative object (ZTF22abtjefa/SN~2022aaad), including host-continuum fitting and subtraction, isolation and Gaussian fitting of the [O\,\textsc{i}]~$\lambda\lambda6300,6363$ feature, and multi-component fitting of the 7000--7500~\AA\ region to extract the [Ca\,\textsc{ii}]~$\lambda\lambda7291,7323$ profile. Table~\ref{table:literature_log} compiles the spectral log for the literature sample of 120 nebular spectra of non--ZTF Type~II SNe used for comparison. Table~\ref{tab:nebular_master} summarizes the full set of nebular measurements together with plateau properties, peak absolute magnitudes, and ZAMS mass estimates for the LLIIP sample. Table~\ref{tab:mzams_threecol} reports per-spectrum ZAMS mass estimates from the GP inversion.  The remainder of the appendix presents the nebular spectra for the LLIIP SNe analyzed in this paper (Figures~\ref{fig:ZTF22abtjefa_1_obs}--\ref{fig:ZTF24aaplfjd_obs}).

\begin{table*}[t]
\small
\centering
\caption{Diagnostic weights and quantitative thresholds used to compute the ECSN score. 
For each diagnostic, ``yes'' and ``no'' refer to whether the feature supports an ECSN interpretation. Intermediate cases are classified as 
``maybe'' and contribute zero to the score. Final classifications were also checked visually, 
especially for blended, noisy, or ambiguous spectral regions.}\begin{tabular}{l c c c}
\toprule
Diagnostic & Weight & Yes & No \\
\midrule

He I $\lambda7065$
 & 1.0
 &  $\mathrm{SNR} \ge 10$ 
 & $\mathrm{SNR} < 3$ or $\mathrm{PeakRatio} < 0.10$ \\[4pt]

O I $\lambda7774$
 & 1.0
 & $\mathrm{SNR} \ge 10$
 & $\mathrm{SNR} < 3$ \\[4pt]

[C I] $\lambda8727$
 & 1.0
 &  $\mathrm{SNR} \ge 10$ and $\mathrm{PeakRatio} \ge 0.25$
 & $\mathrm{PeakRatio} < 0.10$ \\[4pt]

Mg I] $\lambda4571$
 & 0.2
 &  $\mathrm{SNR} \ge 10$ and $\mathrm{PeakRatio} > 0.20$
 & $\mathrm{PeakRatio} < 0.15$ or $\mathrm{SNR} \le 3$ \\[4pt]

O I $\lambda8447$
 & 0.2
 & $\mathrm{SNR} \ge 10$ or $\mathrm{PeakRatio} \ge 0.30$
 & $\mathrm{SNR} < 3$ and $\mathrm{PeakRatio} < 0.30$ \\[4pt]

[Ni II]/[Fe II]
 & 0.2
 & $\mathrm{Ratio} > 0.7$
 & $\mathrm{Ratio} < 0.4$ \\[4pt]

[Ca II]/[O I]
 & 1.0
 & $--$
 & $\mathrm{Ratio} < 1.0$ \\[4pt]

\bottomrule
\end{tabular}
\label{tab:ecsn_weights_thresholds}
\end{table*}

\input{ecsn_theoretical_model_scores}

\input{ZTF_bright_table}

\input{nonZTF_spectral_log_threecol}

\input{master_obs_ZAMS_plateau_table}

\begin{figure*}[t]
    \centering
    \includegraphics[width=0.49\textwidth]{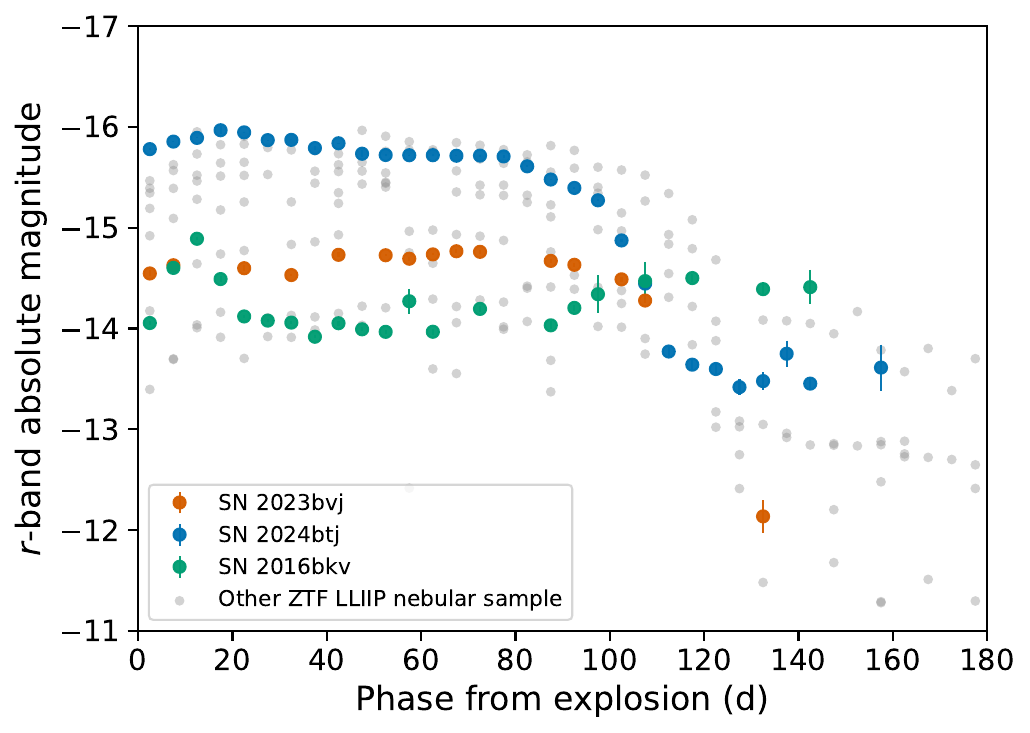}
    \includegraphics[width=0.49\textwidth]{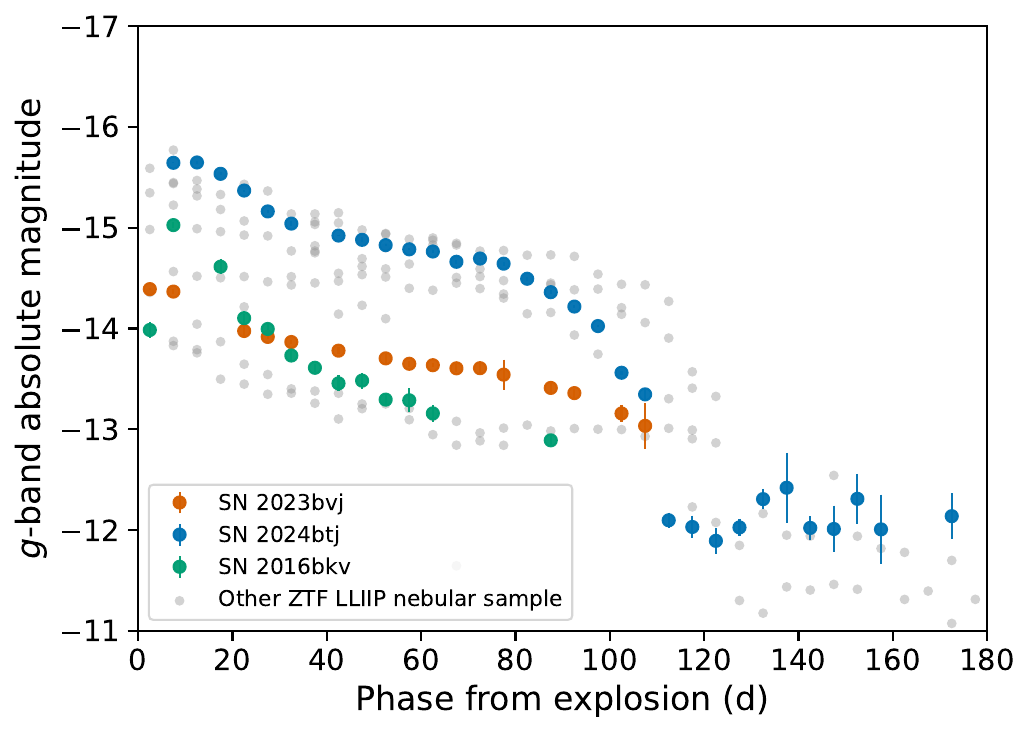}
    \caption{
    Absolute-magnitude light curves of the three highest-scoring ECSN candidates
    compared with the ZTF LLIIP nebular-spectroscopic sample.
    The left and right panels show the $r$- and $g$-band light curves, respectively.
    Gray points show the other ZTF LLIIP SNe with nebular spectra, while colored points
    mark SN~2023bvj, SN~2024btj, and SN~2016bkv.
    }
    \label{fig:ecsn_candidate_lightcurves}
\end{figure*}

\begin{figure*}
    \centering
     \includegraphics[width=0.99\textwidth]{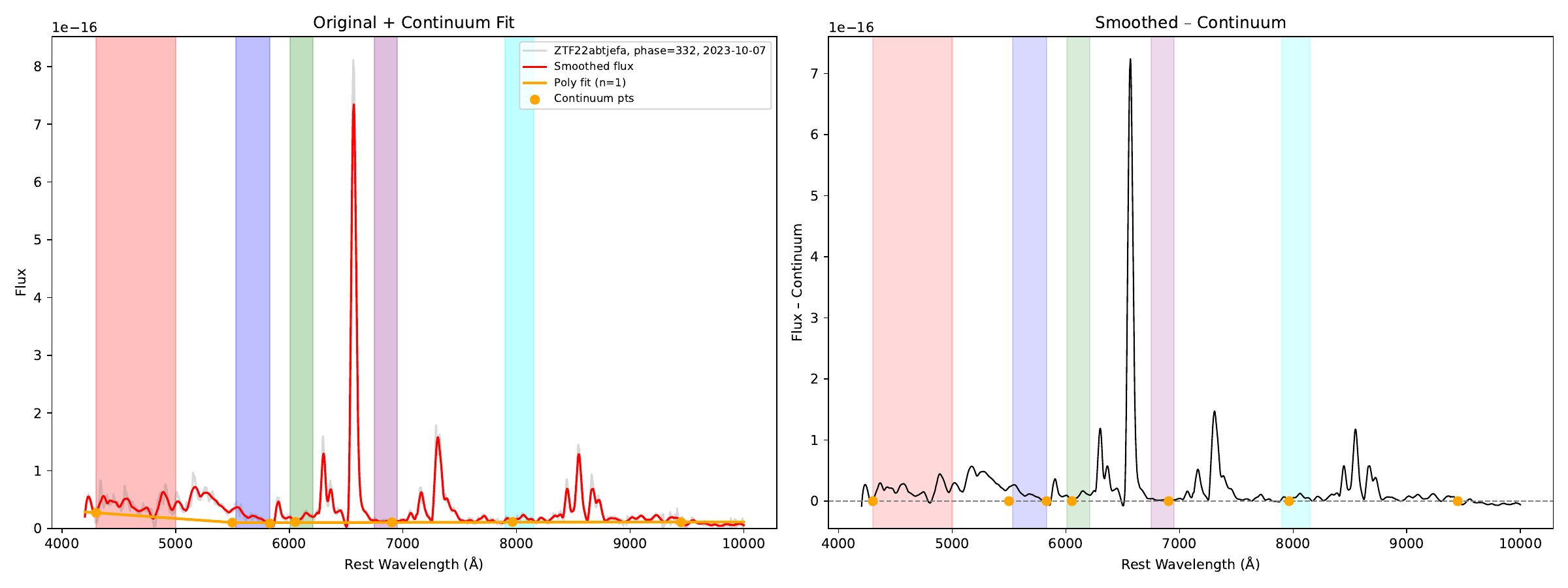}
     \includegraphics[width=0.5\textwidth]{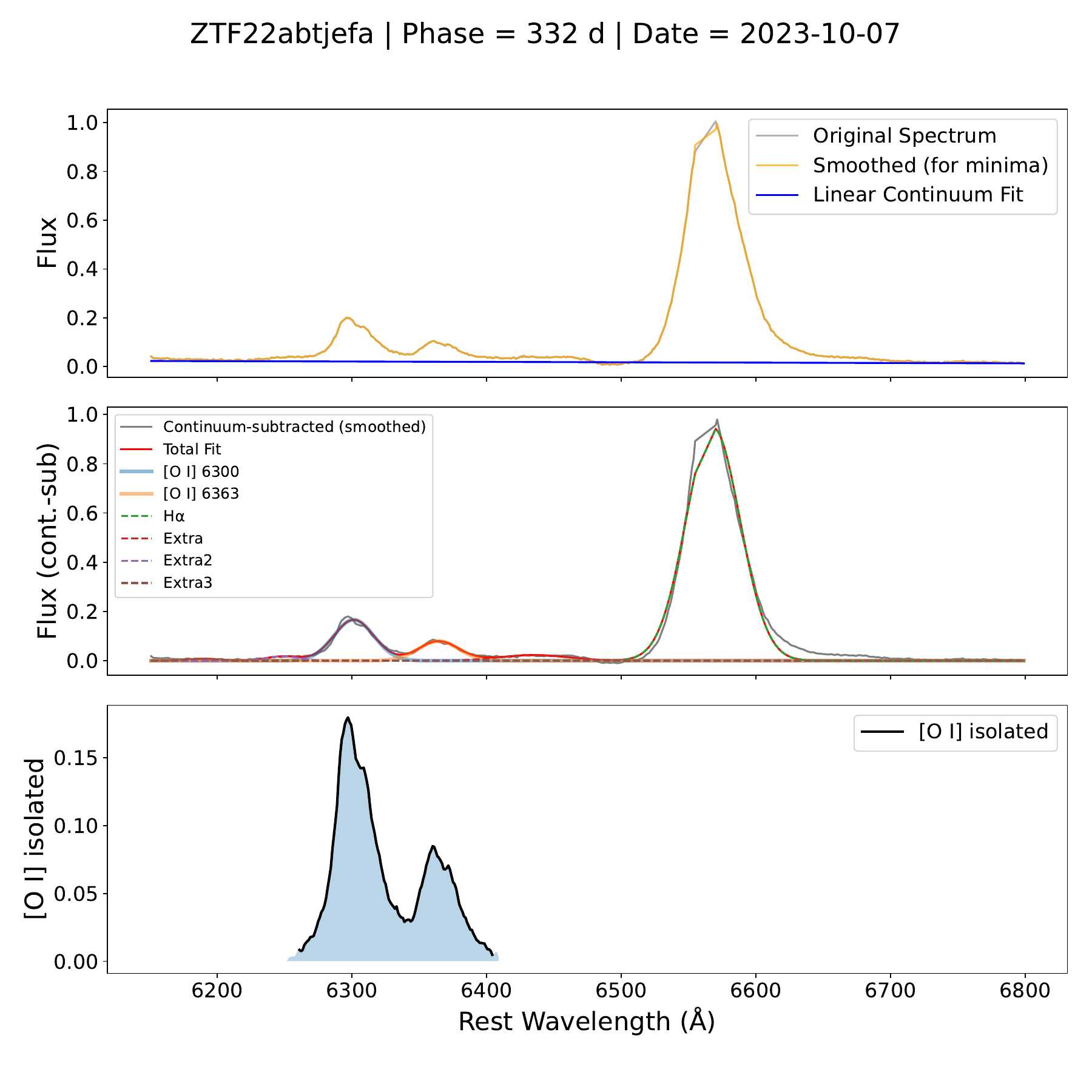}\includegraphics[width=0.5\textwidth]{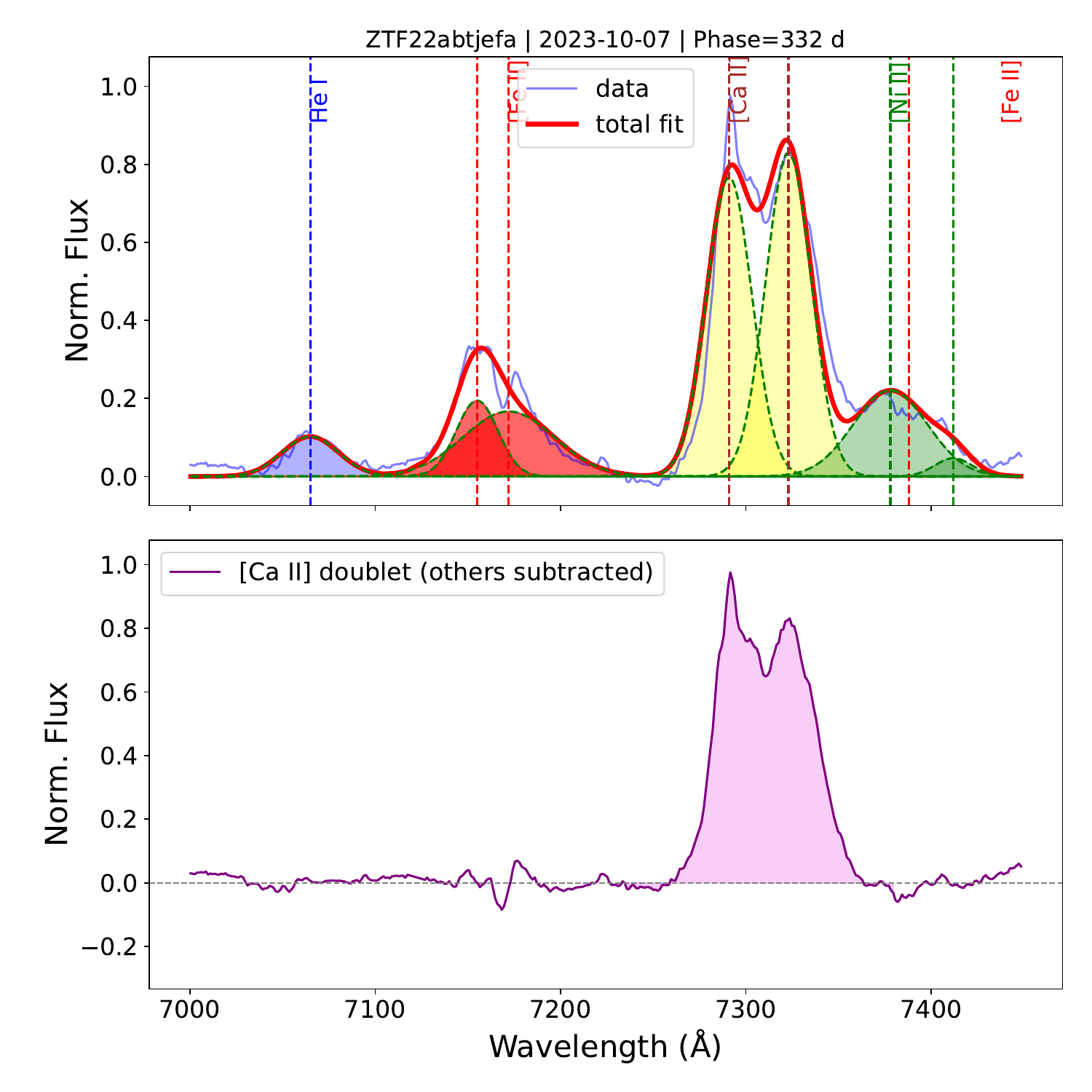}
    \caption{Example of the nebular spectral analysis procedure for ZTF22abtjefa/SN~2022aaad at phase 332\,d. Top panels show the original rest-frame, extinction-corrected spectrum with the fitted host-galaxy continuum (left) and the continuum-subtracted spectrum (right), with shaded regions indicating the line-poor windows used to define the continuum. Bottom left panels show the isolation of the [O\,\textsc{i}]~$\lambda\lambda6300,6363$ feature using a local pseudo-continuum defined by minima in the smoothed spectrum, followed by Gaussian fitting to measure the line flux and width. Bottom right panels show the multi-component Gaussian fit to the 7000--7500~\AA\ region and the resulting isolated [Ca\,\textsc{ii}]~$\lambda\lambda7291,7323$ profile after subtracting all other fitted lines.}

    \label{fig:analysis}
\end{figure*}

\input{mzams_threecol_table}

\begin{figure*}[htbp]
    \centering
    \includegraphics[width=1.0\textwidth]{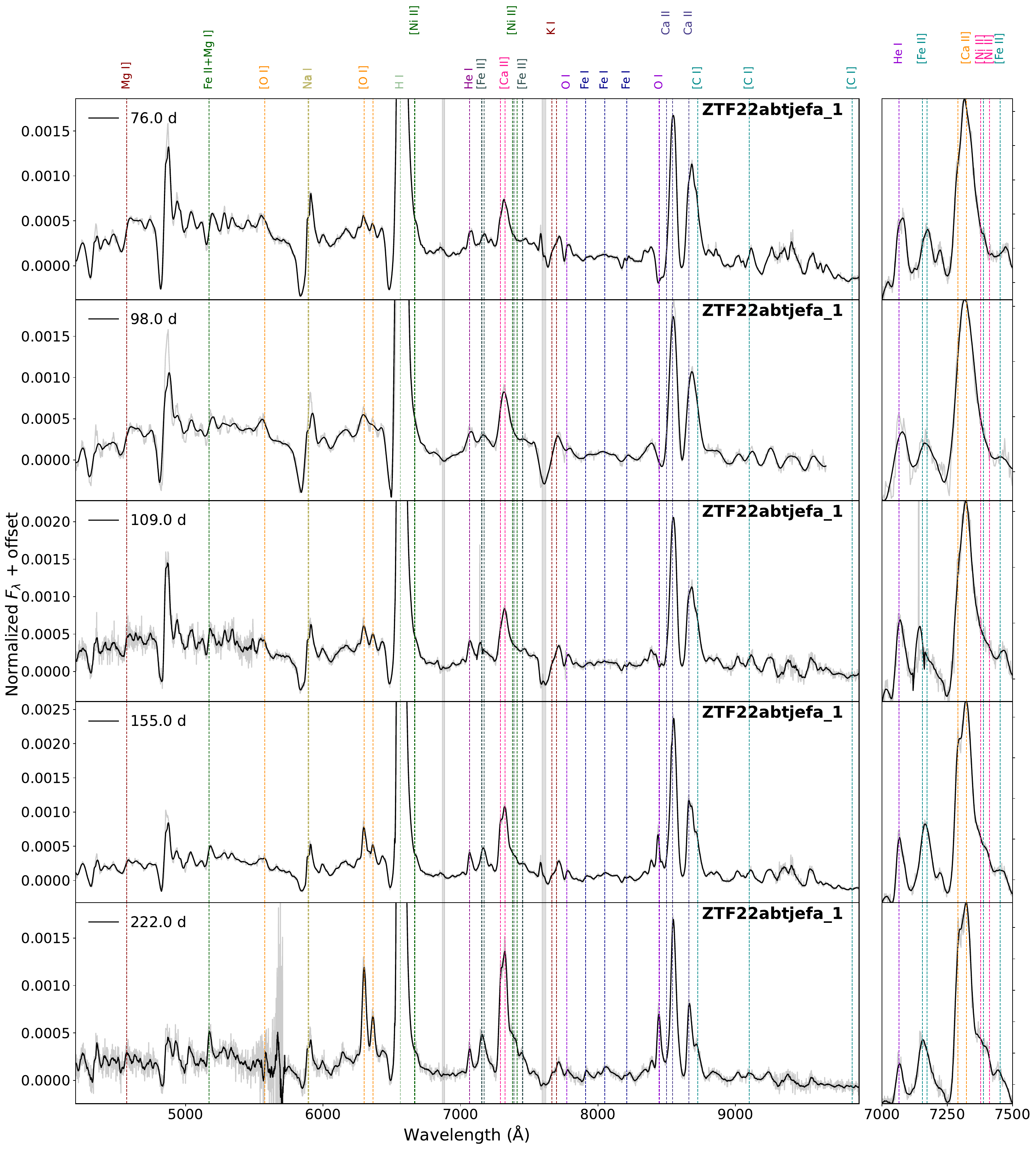}
    \caption{Nebular spectra of ZTF22abtjefa/SN~2022aaad.}
    \label{fig:ZTF22abtjefa_1_obs}
\end{figure*}

\begin{figure*}[htbp]
    \centering
    \includegraphics[width=1.0\textwidth]{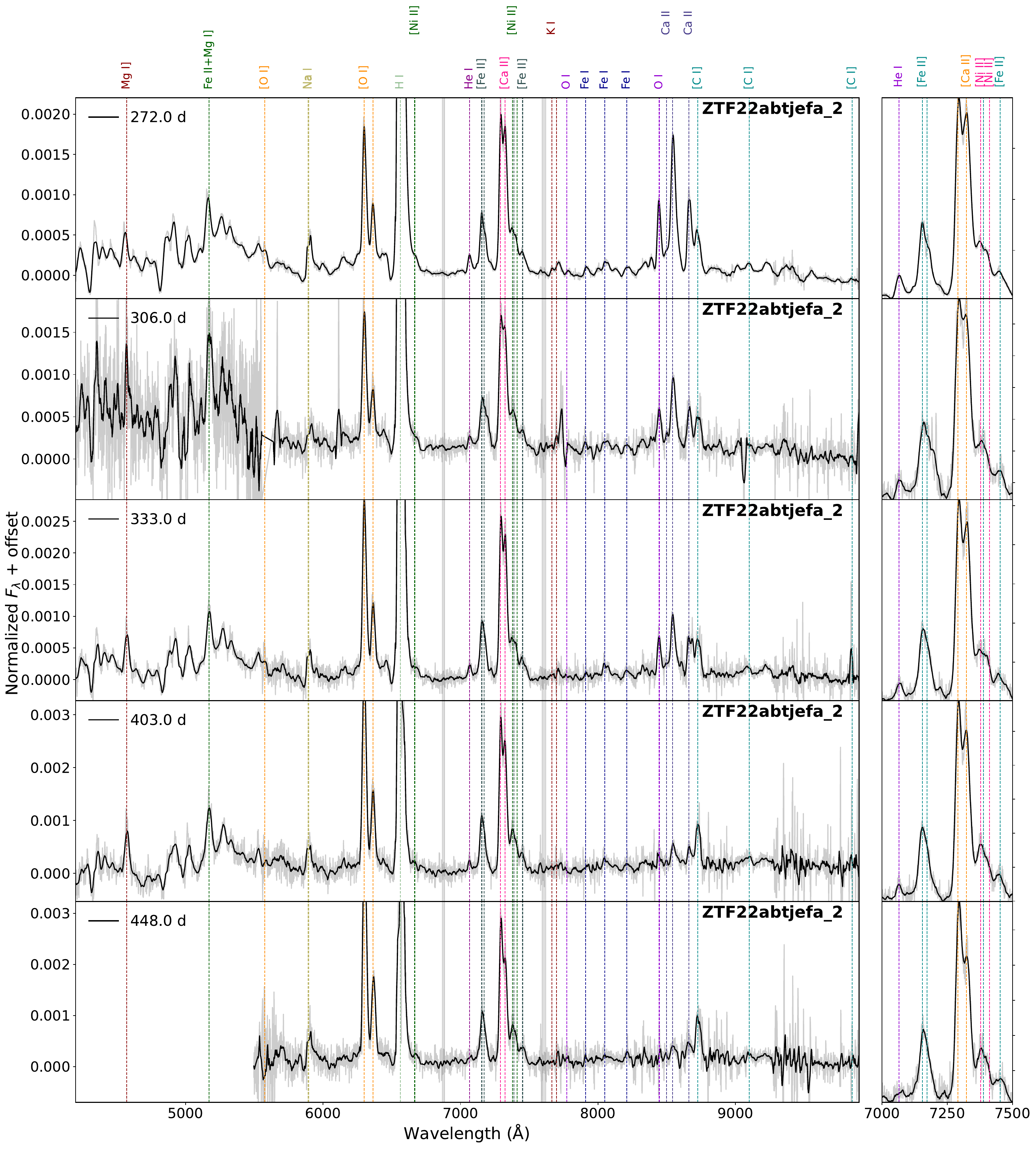}
    \caption{Nebular spectra of ZTF22abtjefa/SN~2022aaad.}
    \label{fig:ZTF22abtjefa_2_obs}
\end{figure*}

\begin{figure*}[htbp]
    \centering
    \includegraphics[width=1.0\textwidth]{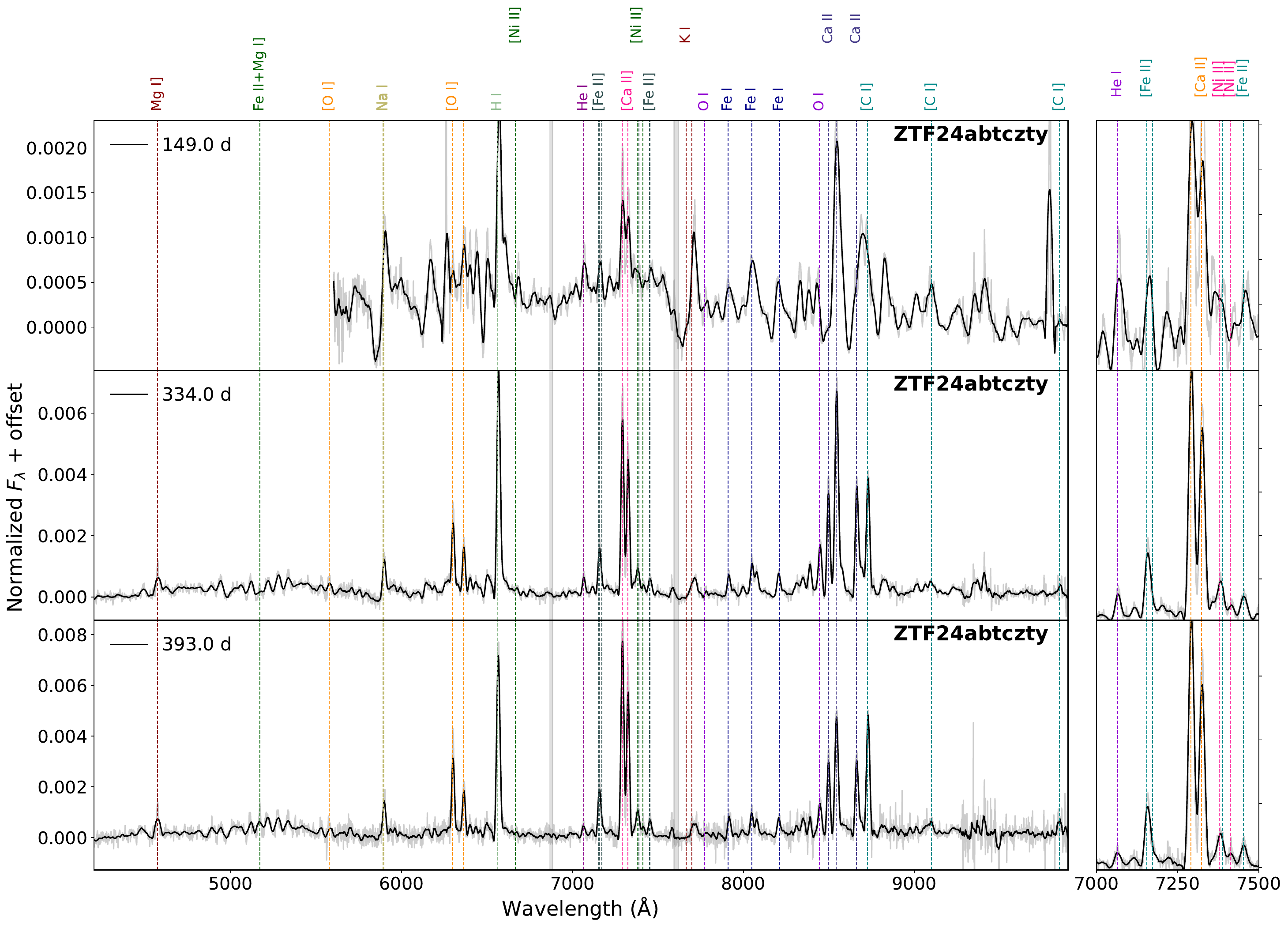}
    \caption{Nebular spectra of ZTF24abtczty/SN~2024abfl.}
    \label{fig:ZTF24abtczty_obs}
\end{figure*}

\begin{figure*}[htbp]
    \centering
    \includegraphics[width=1.0\textwidth]{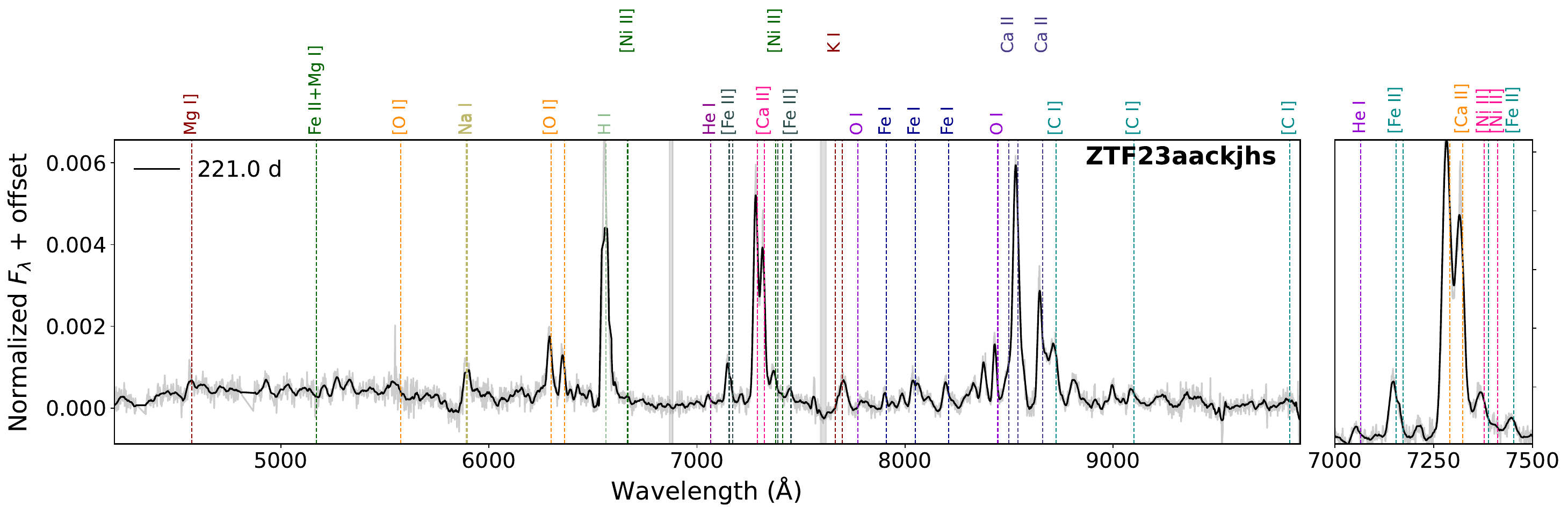}
    \caption{Nebular spectra of ZTF23aackjhs/SN~2023bvj.}
    \label{fig:ZTF23aackjhs_obs}
\end{figure*}

\begin{figure*}[htbp]
    \centering
    \includegraphics[width=1.0\textwidth]{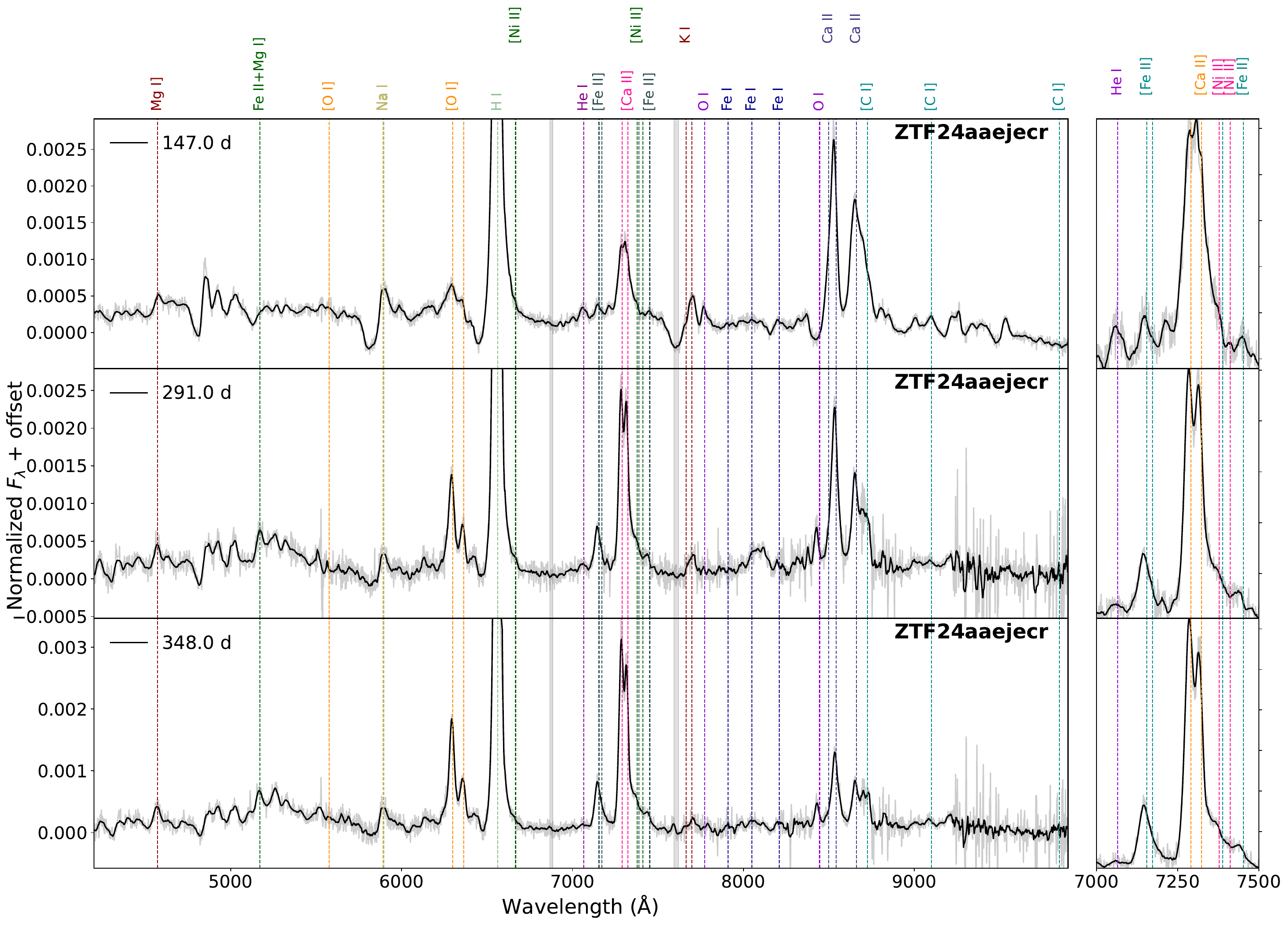}
    \caption{Nebular spectra of ZTF24aaejecr/SN~2024btj.}
    \label{fig:ZTF24aaejecr_obs}
\end{figure*}

\begin{figure*}[htbp]
    \centering
    \includegraphics[width=1.0\textwidth]{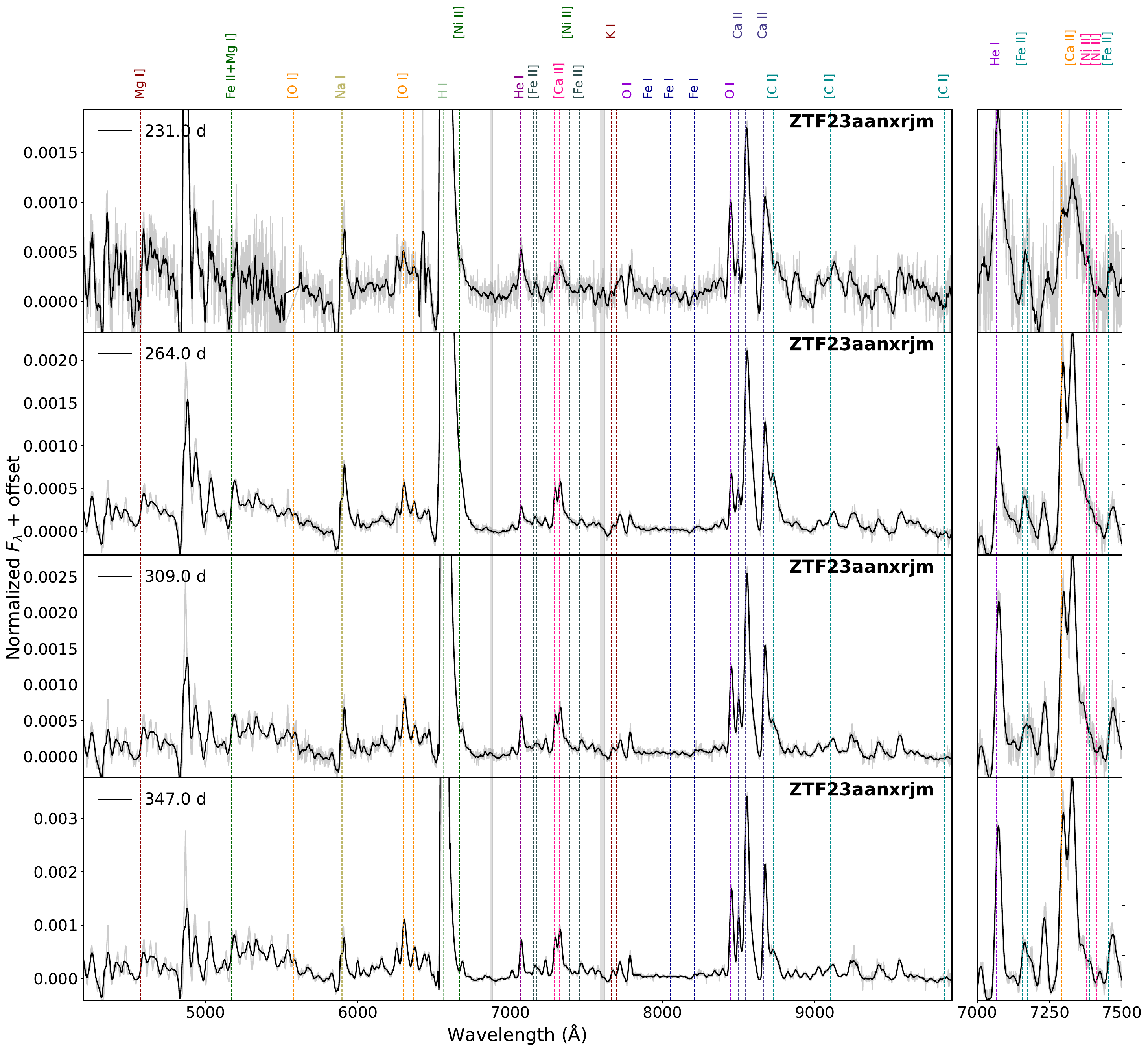}
    \caption{Nebular spectra of ZTF23aanxrjm/SN~2023kmk.}
    \label{fig:ZTF23aanxrjm_obs}
\end{figure*}

\begin{figure*}[htbp]
    \centering
    \includegraphics[width=1.0\textwidth]{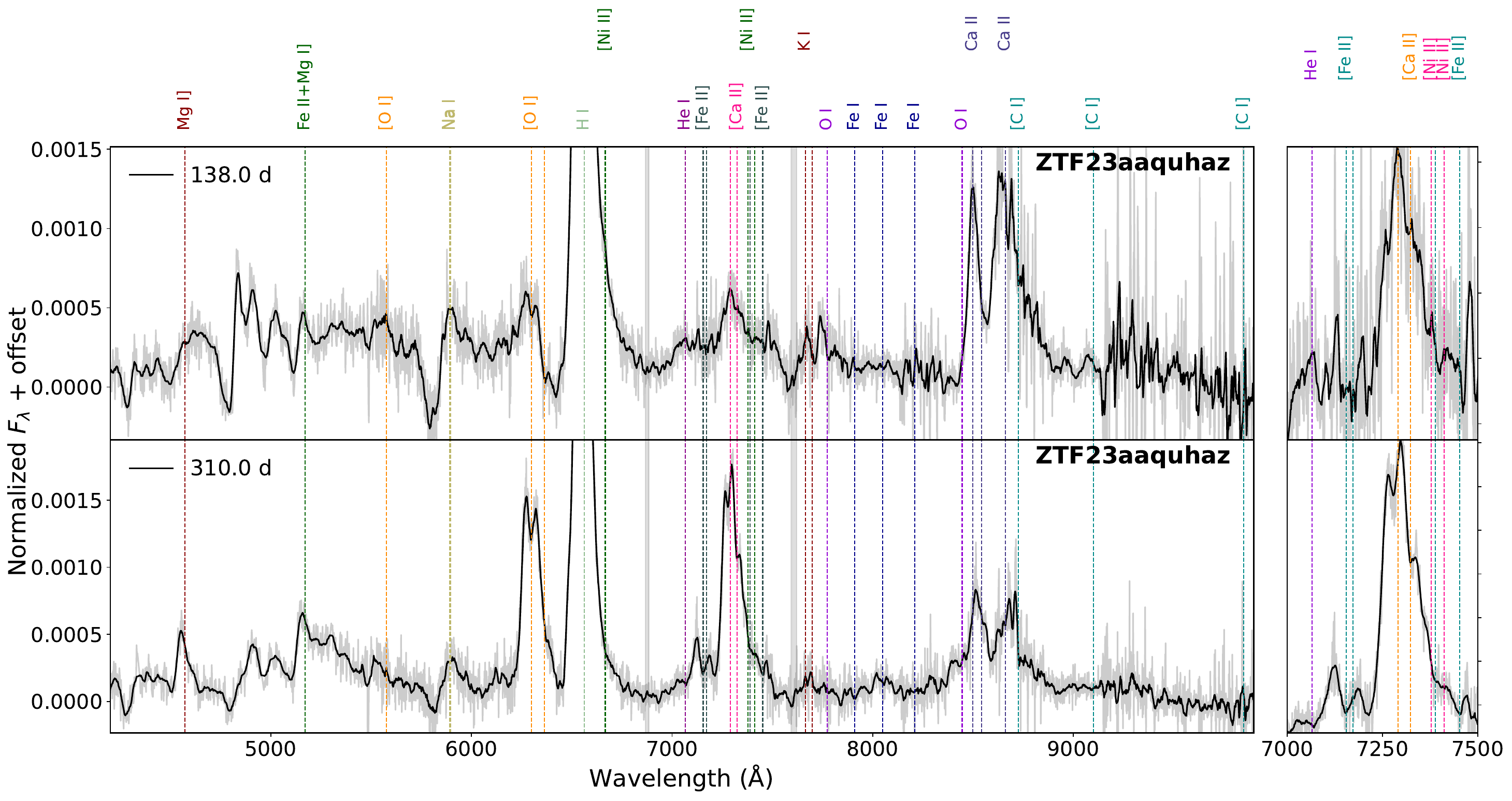}
    \caption{Nebular spectra of ZTF23aaquhaz/SN~2023mpz.}
    \label{fig:ZTF23aaquhaz_obs}
\end{figure*}

\begin{figure*}[htbp]
    \centering
    \includegraphics[width=1.0\textwidth]{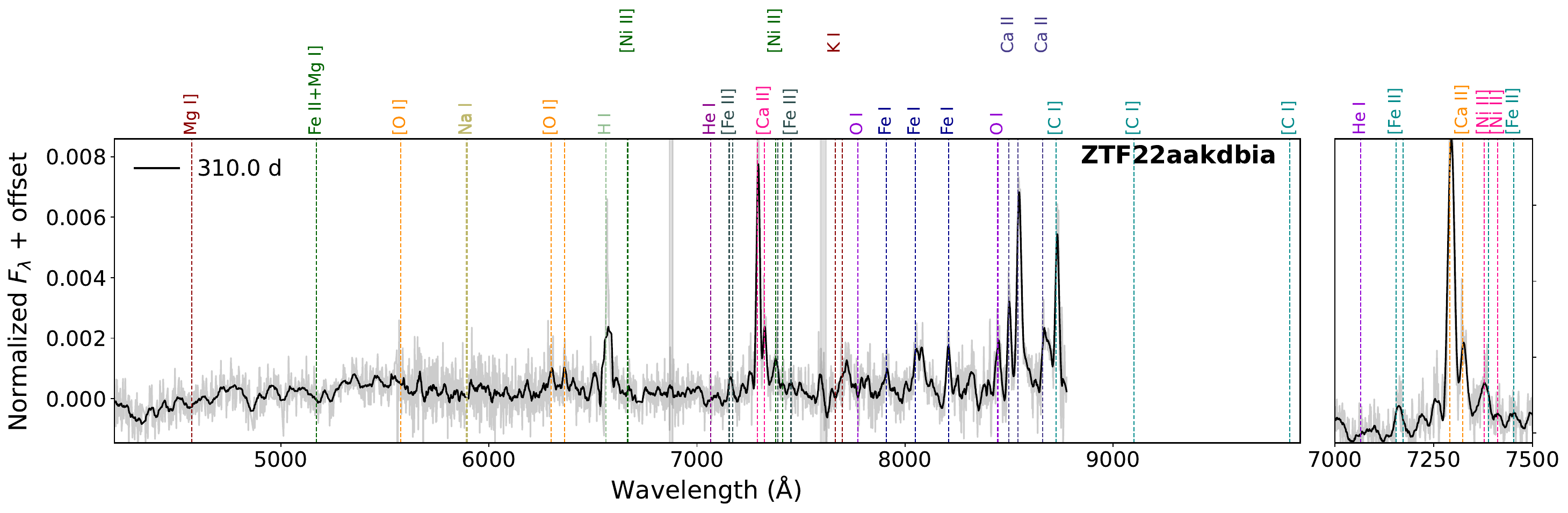}
    \caption{Nebular spectra of ZTF22aakdbia/SN~2022jzc.}
    \label{fig:ZTF22aakdbia_obs}
\end{figure*}

\begin{figure*}[htbp]
    \centering
    \includegraphics[width=1.0\textwidth]{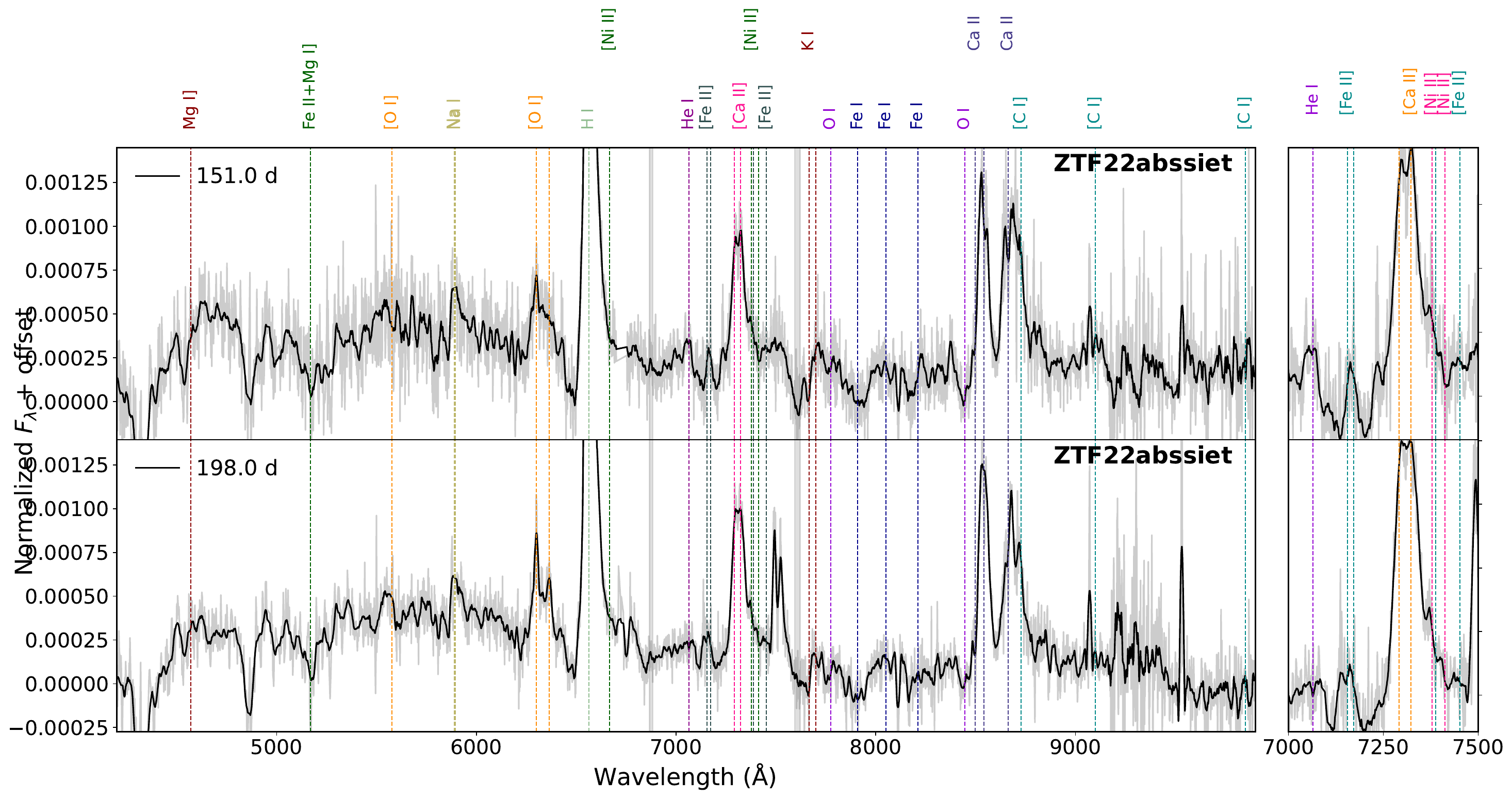}
    \caption{Nebular spectra of ZTF22abssiet/SN~2022zmb.}
    \label{fig:ZTF22abssiet_obs}
\end{figure*}

\begin{figure*}[htbp]
    \centering
    \includegraphics[width=1.0\textwidth]{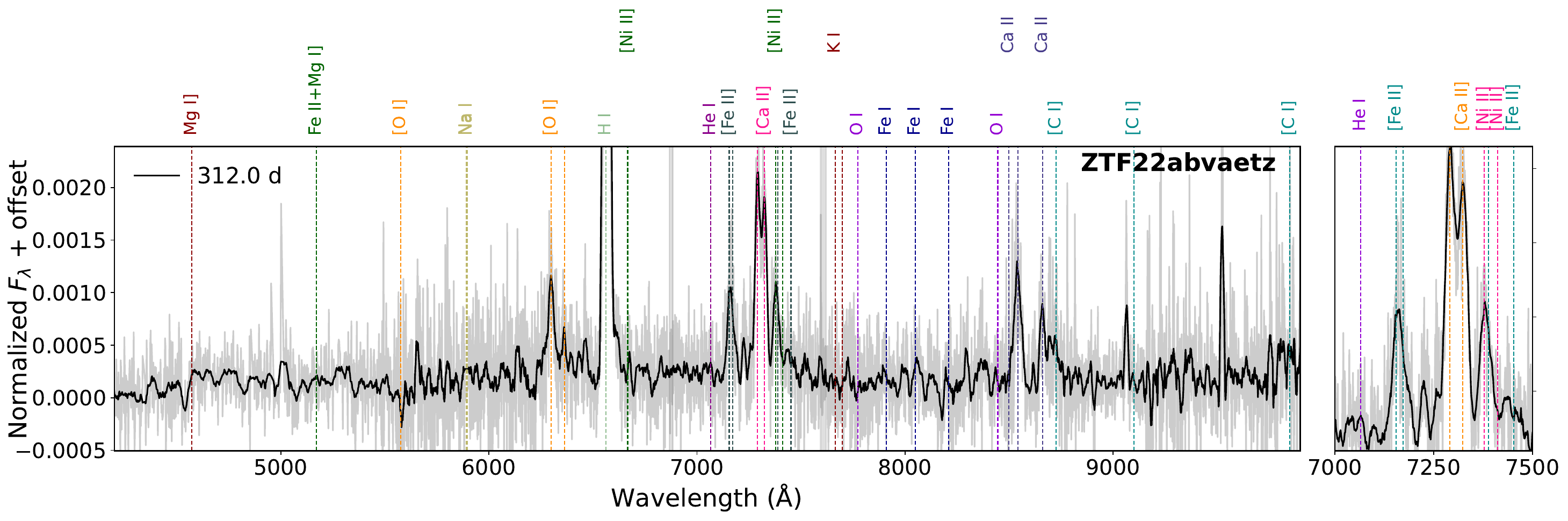}
    \caption{Nebular spectra of ZTF22abvaetz/SN~2022aang.}
    \label{fig:ZTF22abvaetz_obs}
\end{figure*}

\FloatBarrier

\begin{figure*}[htbp]
    \centering
    \includegraphics[width=1.0\textwidth]{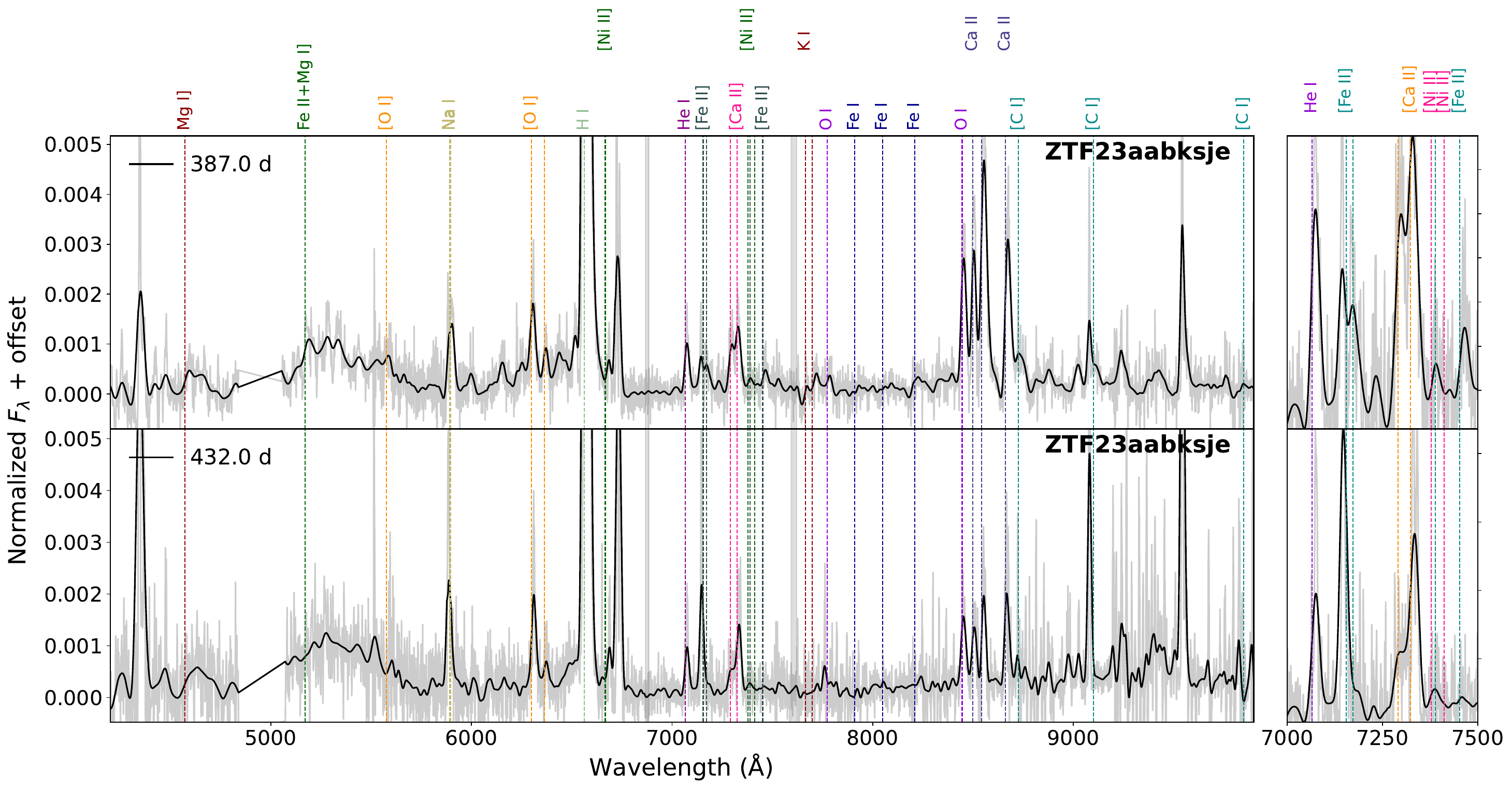}
    \caption{Nebular spectra of ZTF23aabksje/SN~2023azx.}
    \label{fig:ZTF23aabksje_obs}
\end{figure*}

\begin{figure*}[htbp]
    \centering
    \includegraphics[width=1.0\textwidth]{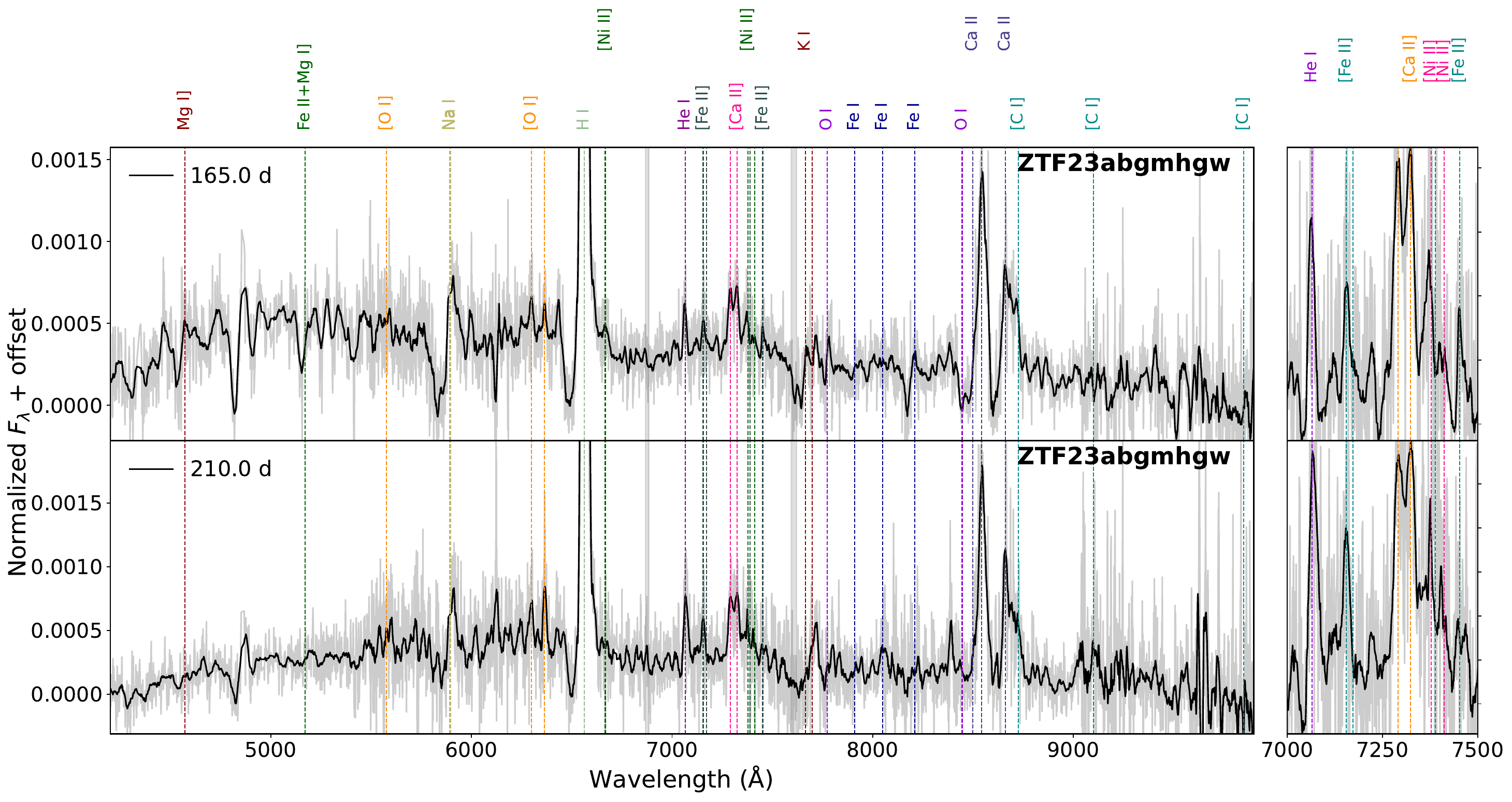}
    \caption{Nebular spectra of ZTF23abgmhgw/SN~2023vci.}
    \label{fig:ZTF23abgmhgw_obs}
\end{figure*}

\begin{figure*}[htbp]
    \centering
    \includegraphics[width=1.0\textwidth]{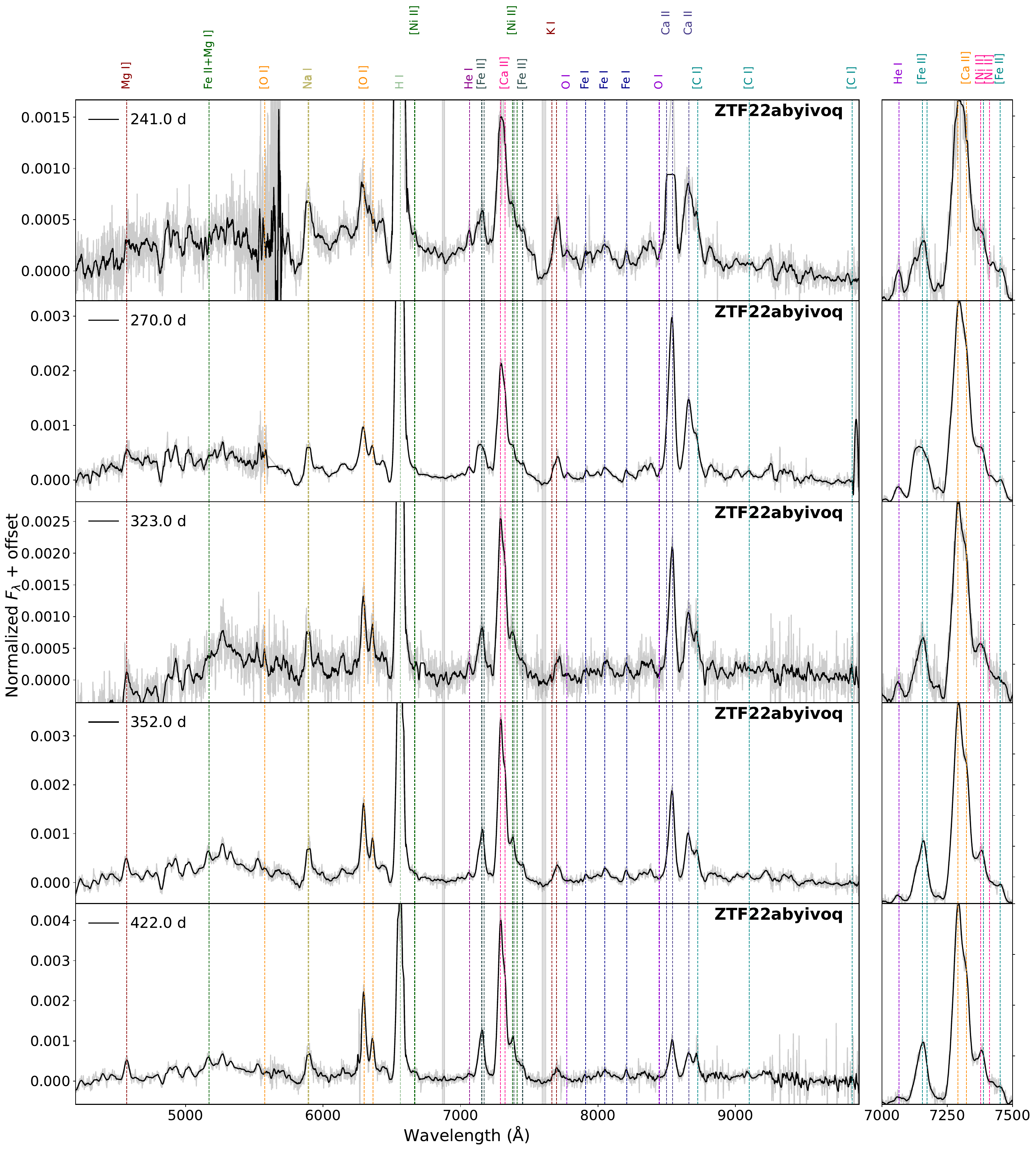}
    \caption{Nebular spectra of ZTF22abyivoq/SN~2022acko.}
    \label{fig:ZTF22abyivoq_obs}
\end{figure*}

\begin{figure*}[htbp]
    \centering
    \includegraphics[width=1.0\textwidth]{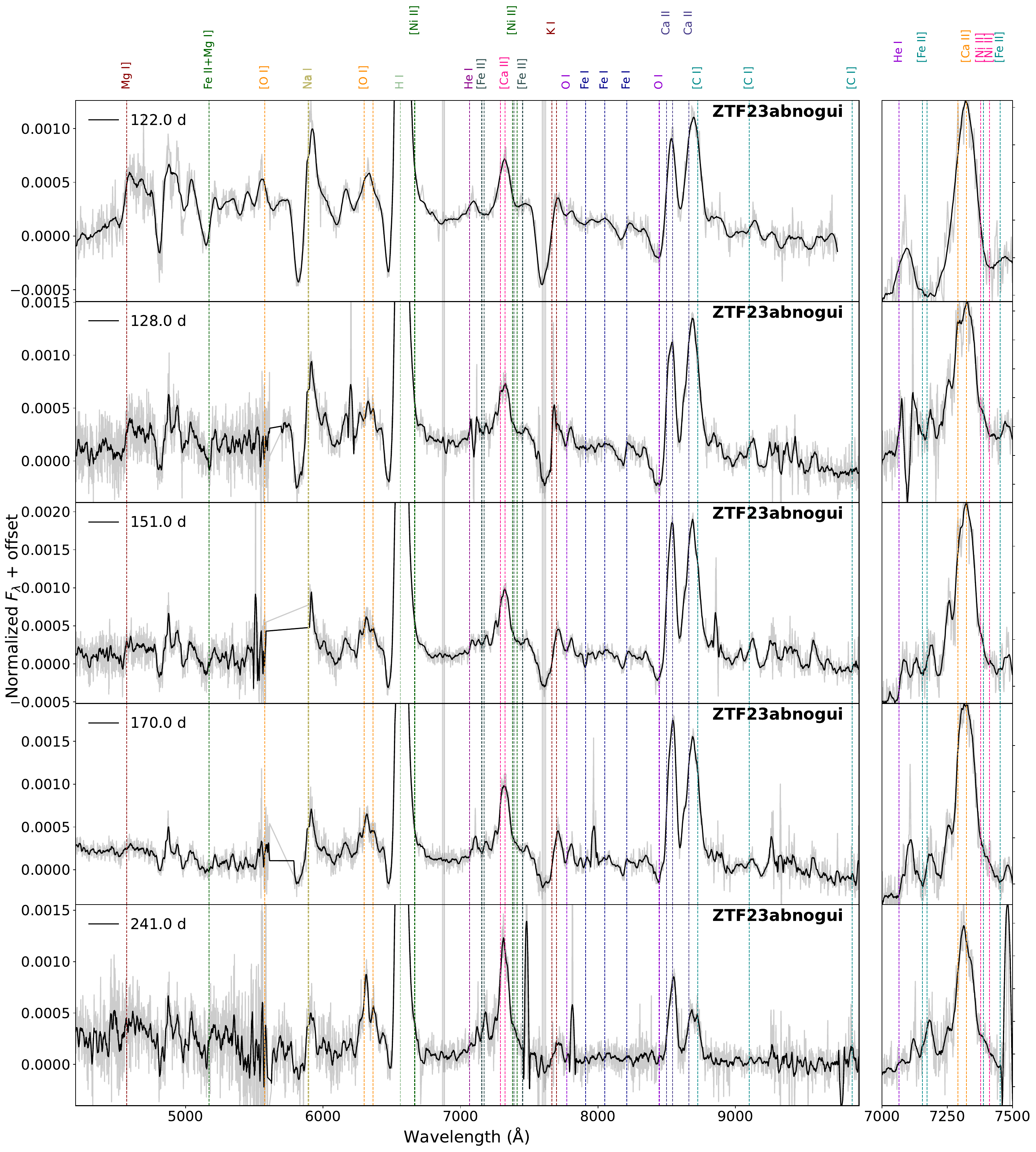}
    \caption{Nebular spectra of ZTF23abnogui/SN~2023wcr.}
    \label{fig:ZTF23abnogui_obs}
\end{figure*}

\begin{figure*}[htbp]
    \centering
    \includegraphics[width=1.0\textwidth]{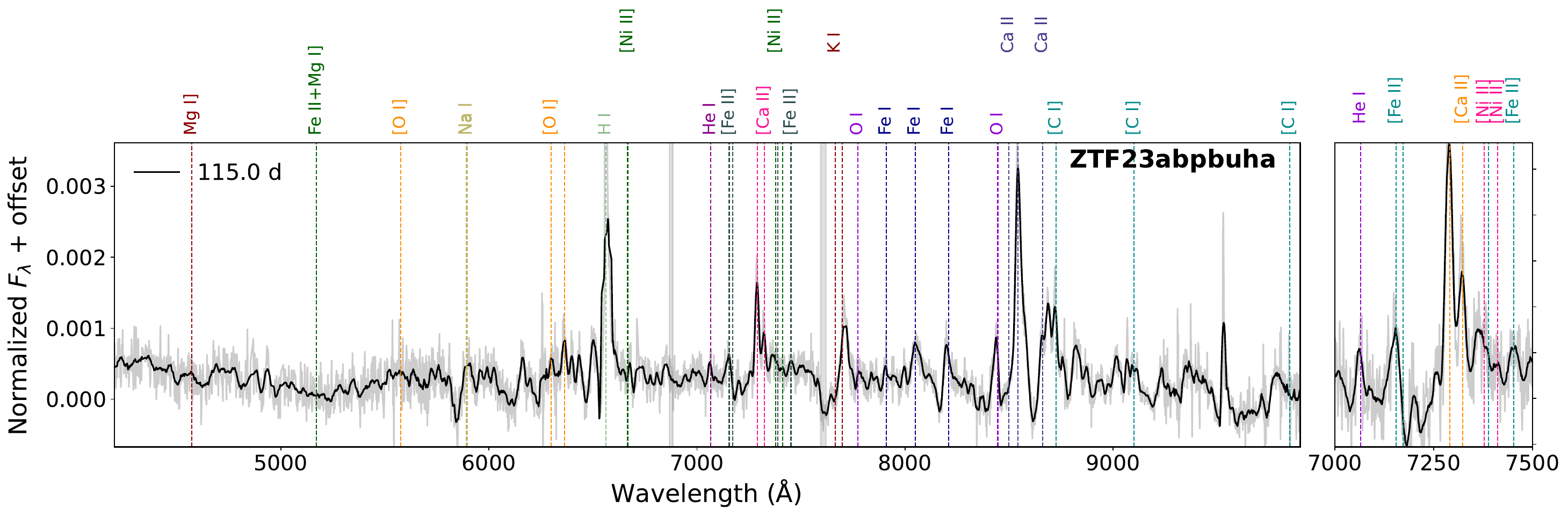}
    \caption{Nebular spectra of ZTF23abpbuha/SN~2023usp.}
    \label{fig:ZTF23abpbuha_obs}
\end{figure*}

\begin{figure*}[htbp]
    \centering
    \includegraphics[width=1.0\textwidth]{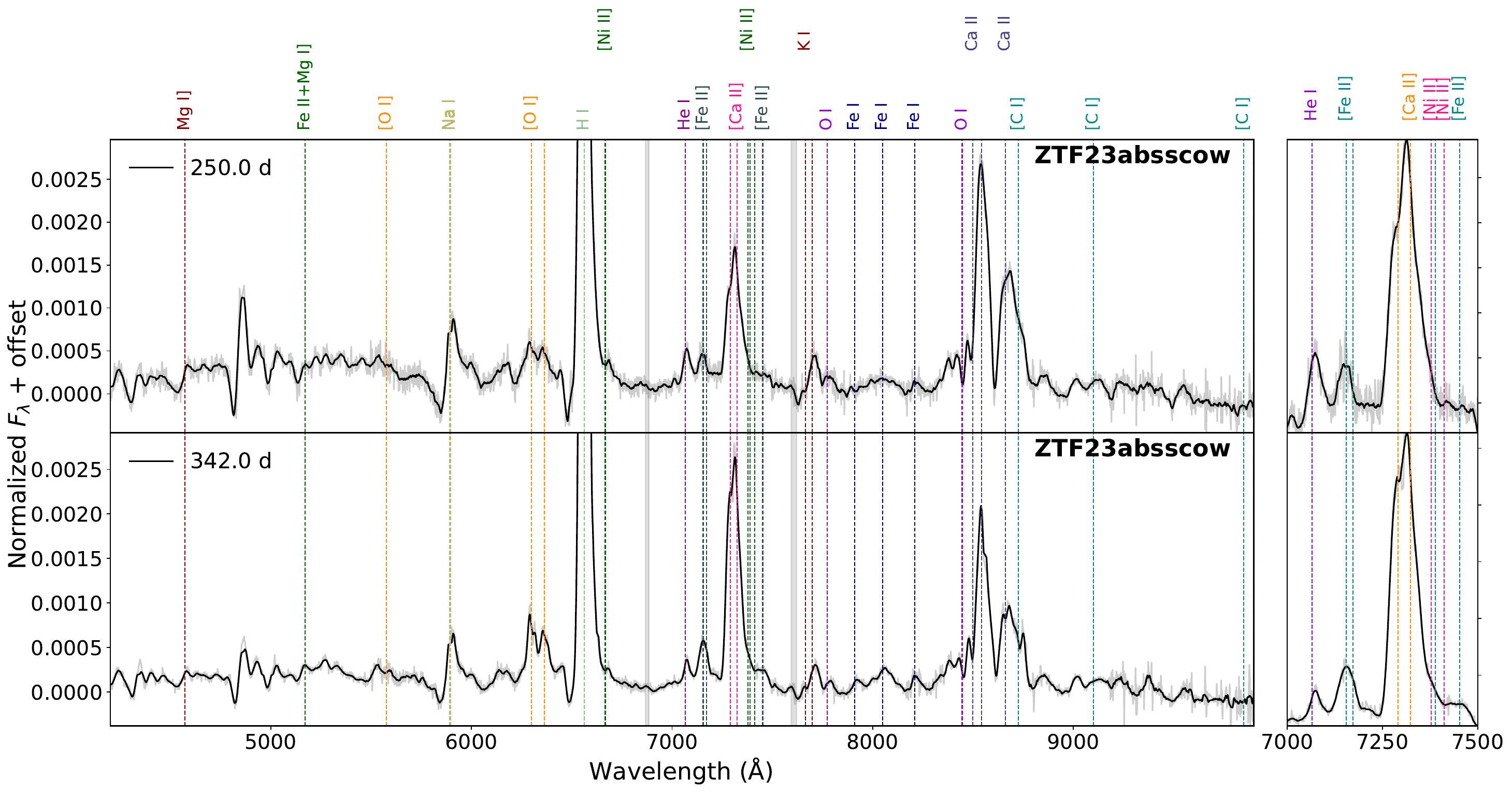}
    \caption{Nebular spectra of ZTF23absscow/SN~2023ywa.}
    \label{fig:ZTF23absscow_obs}
\end{figure*}

\begin{figure*}[htbp]
    \centering
    \includegraphics[width=1.0\textwidth]{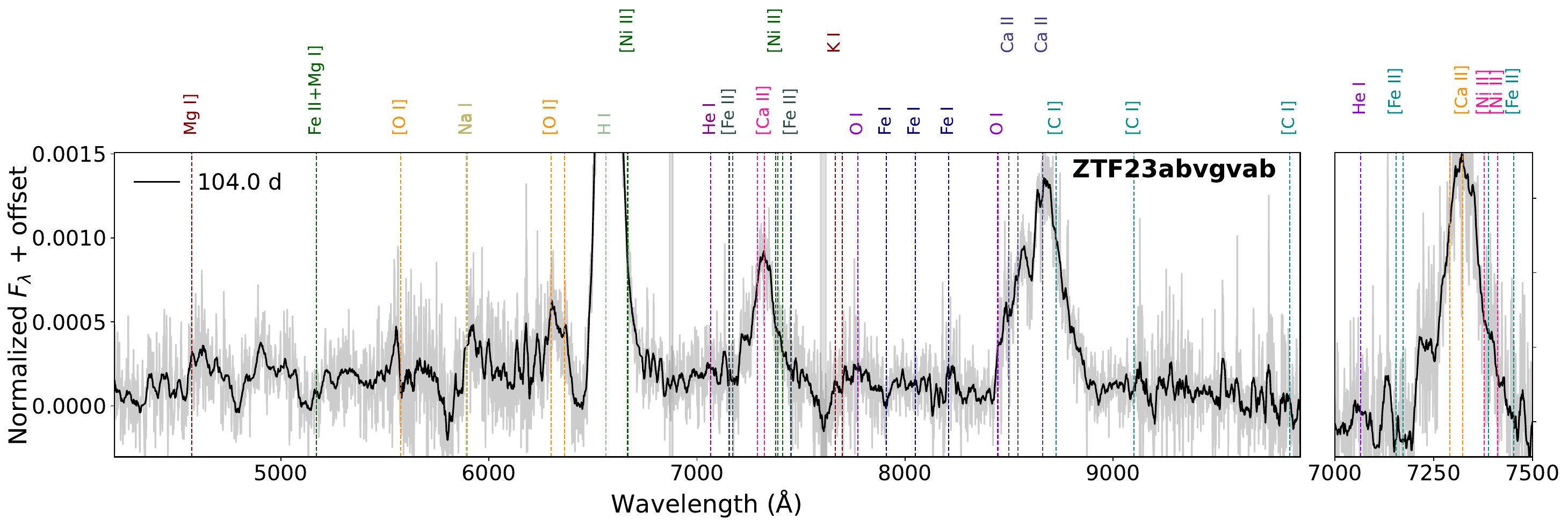}
    \caption{Nebular spectra of ZTF23abvgvab/SN~2023abim.}
    \label{fig:ZTF23abvgvab_obs}
\end{figure*}

\FloatBarrier

\begin{figure*}[htbp]
    \centering
    \includegraphics[width=1.0\textwidth]{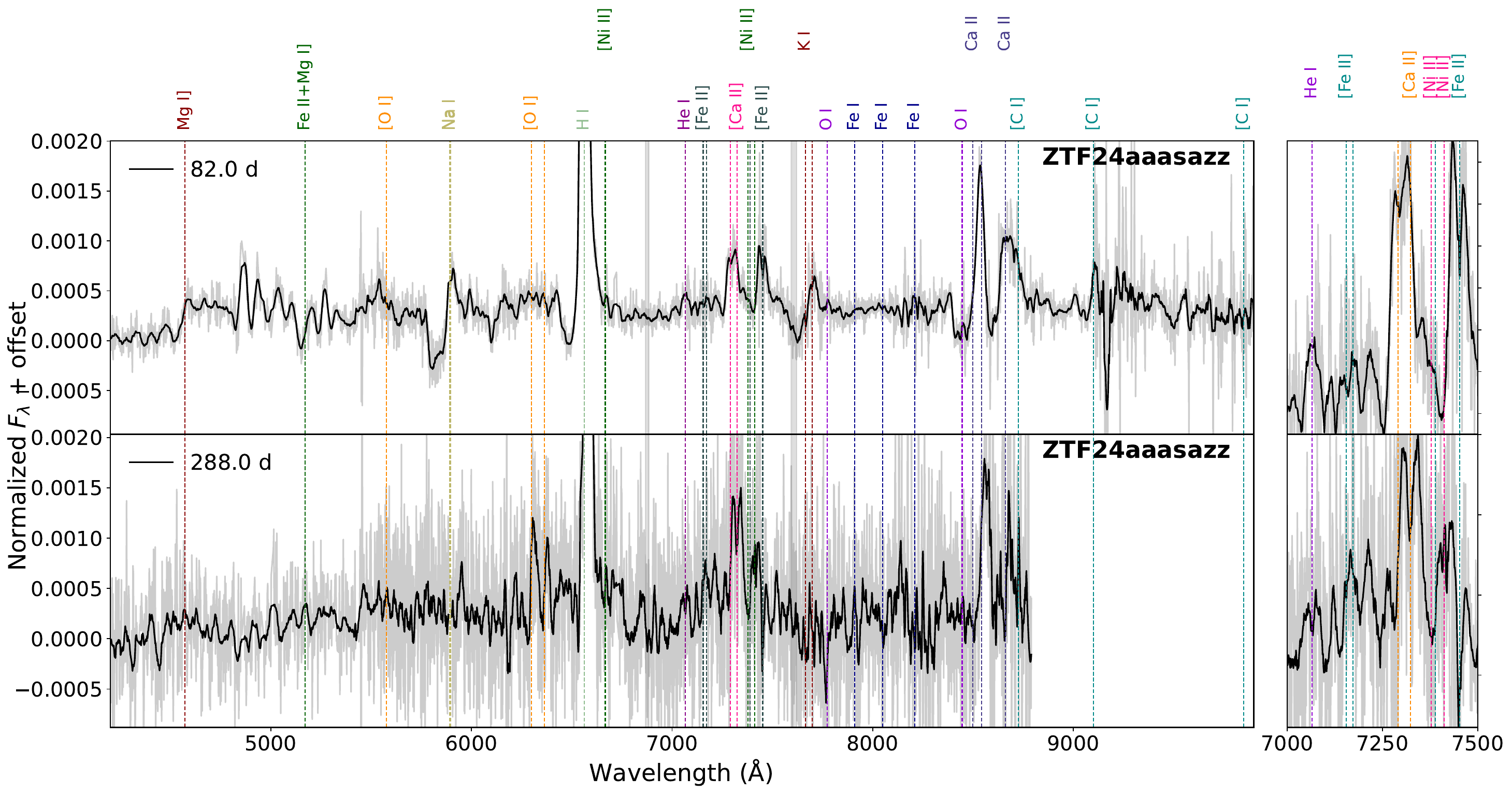}
    \caption{Nebular spectra of ZTF24aaasazz/SN~2024ov.}
    \label{fig:ZTF24aaasazz_obs}
\end{figure*}

\begin{figure*}[htbp]
    \centering
    \includegraphics[width=1.0\textwidth]{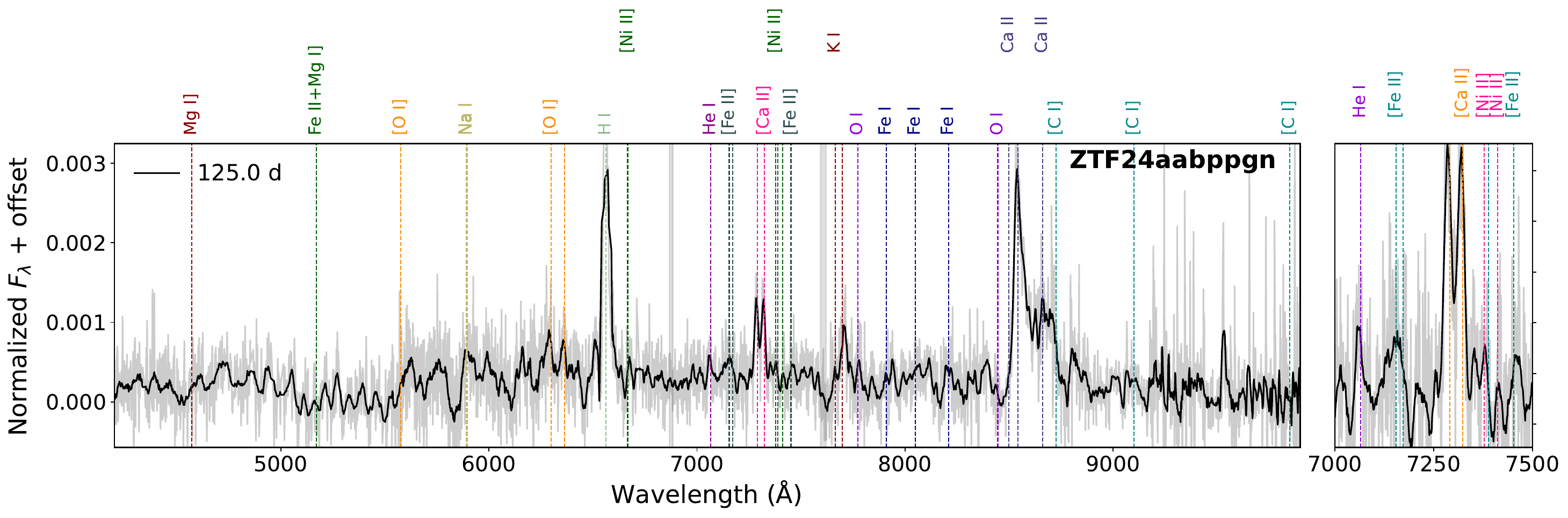}
    \caption{Nebular spectra of ZTF24aabppgn/SN~2024wp.}
    \label{fig:ZTF24aabppgn_obs}
\end{figure*}

\begin{figure*}[htbp]
    \centering
    \includegraphics[width=1.0\textwidth]{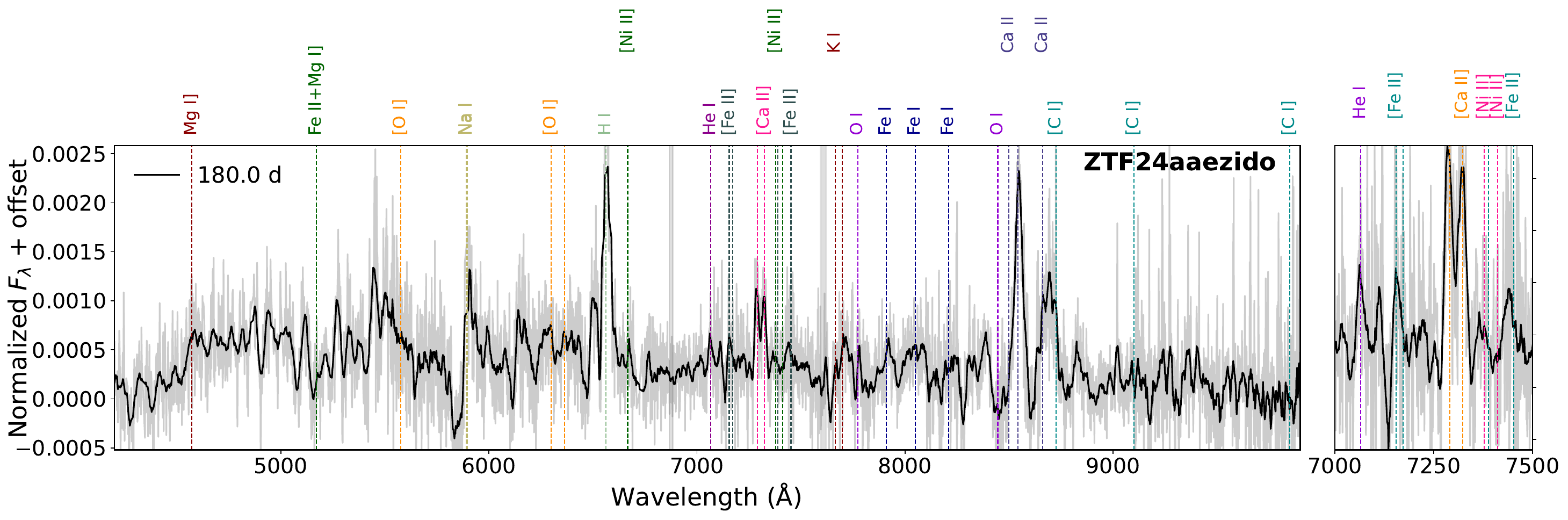}
    \caption{Nebular spectra of ZTF24aaezido/SN~2024cro.}
    \label{fig:ZTF24aaezido_obs}
\end{figure*}

\begin{figure*}[htbp]
    \centering
    \includegraphics[width=1.0\textwidth]{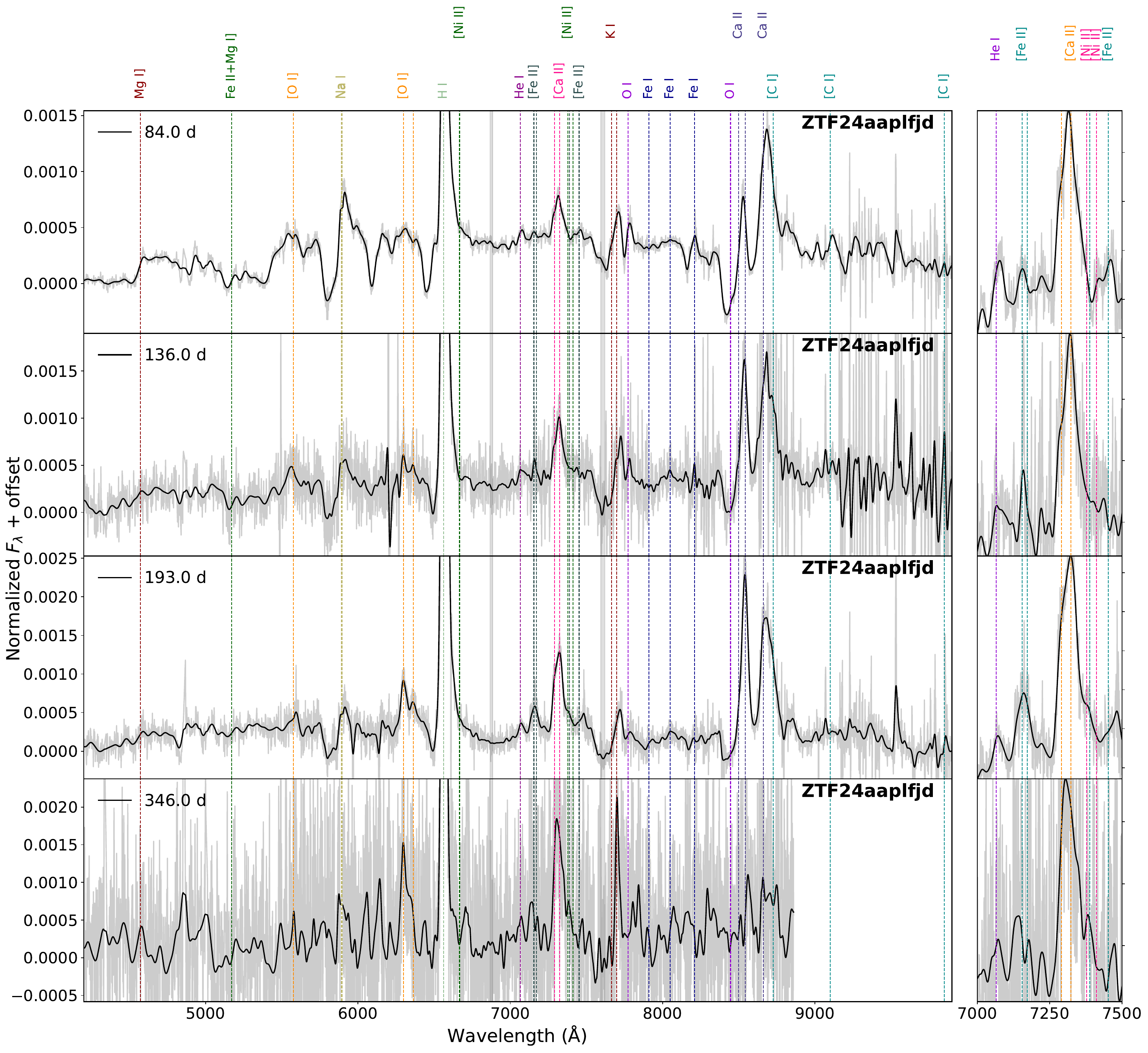}
    \caption{Nebular spectra of ZTF24aaplfjd/SN~2024jxm.}
    \label{fig:ZTF24aaplfjd_obs}
\end{figure*}

\FloatBarrier

\clearpage

\end{document}

%% file: ZTF_LLIIP_table.tex
\begin{table}
\label{table:LLIIP_spectral_log}
\centering
\scriptsize
\setlength{\tabcolsep}{2pt}
\caption{Spectral log of nebular spectra of LLIIP SNe. Throughout this paper, all quoted phases are measured relative to the estimated explosion epoch of each SN.}
\begin{tabular}{l l c c c c}
\hline
ZTF Name & IAU Name & Date & Phase (d) & Instrument \\
\hline
ZTF22aakdbia & SN2022jzc & 2023-04-26 & 310 & Keck/LRIS \\
ZTF22abssiet & SN2022zmb & 2023-04-26 & 151 & Keck/LRIS \\
ZTF22abssiet & SN2022zmb & 2023-06-12 & 198 & Keck/LRIS \\
ZTF22abtjefa & SN2022aaad & 2023-06-12 & 155 & Keck/LRIS \\
ZTF22abtjefa & SN2022aaad & 2023-08-18 & 222 & P200/DBSP \\
ZTF22abtjefa & SN2022aaad & 2023-10-07 & 272 & Keck/LRIS \\
ZTF22abtjefa & SN2022aaad & 2023-11-10 & 306 & P200/DBSP \\
ZTF22abtjefa & SN2022aaad & 2023-12-07 & 333 & Keck/LRIS \\
ZTF22abtjefa & SN2022aaad & 2024-02-15 & 403 & Keck/LRIS \\
ZTF22abtjefa & SN2022aaad & 2024-03-31 & 448 & Keck/LRIS \\
ZTF22abvaetz & SN2022aang & 2023-12-07 & 312 & Keck/LRIS \\
ZTF22abyivoq & SN2022acko & 2023-08-18 & 241 & P200/DBSP \\
ZTF22abyivoq & SN2022acko & 2023-09-16 & 270 & P200/DBSP \\
ZTF22abyivoq & SN2022acko & 2023-11-08 & 323 & Keck/LRIS \\
ZTF22abyivoq & SN2022acko & 2023-12-07 & 352 & Keck/LRIS \\
ZTF22abyivoq & SN2022acko & 2024-02-15 & 422 & Keck/LRIS \\
ZTF23aabksje & SN2023azx & 2024-02-15 & 387 & Keck/LRIS \\
ZTF23aabksje & SN2023azx & 2024-03-31 & 432 & Keck/LRIS \\
ZTF23aackjhs & SN2023bvj & 2023-12-07 & 220 & Keck/LRIS \\
ZTF23aanxrjm & SN2023kmk & 2024-01-13 & 231 & P200/DBSP \\
ZTF23aanxrjm & SN2023kmk & 2024-02-15 & 264 & Keck/LRIS \\
ZTF23aanxrjm & SN2023kmk & 2024-03-31 & 309 & Keck/LRIS \\
ZTF23aanxrjm & SN2023kmk & 2024-05-08 & 347 & Keck/LRIS \\
ZTF23aaquhaz & SN2023mpz & 2023-11-18 & 138 & Keck/LRIS \\
ZTF23aaquhaz & SN2023mpz & 2024-05-08 & 310 & Keck/LRIS \\
ZTF23abgmhgw & SN2023vci & 2024-02-15 & 165 & Keck/LRIS \\
ZTF23abgmhgw & SN2023vci & 2024-03-31 & 210 & Keck/LRIS \\
ZTF23abnogui & SN2023wcr & 2024-03-14 & 122 & Keck/LRIS \\
ZTF23abnogui & SN2023wcr & 2024-03-20 & 128 & P200/DBSP \\
ZTF23abnogui & SN2023wcr & 2024-04-12 & 151 & P200/DBSP \\
ZTF23abnogui & SN2023wcr & 2024-05-01 & 170 & P200/DBSP \\
ZTF23abnogui & SN2023wcr & 2024-07-11 & 241 & P200/DBSP \\
ZTF23abpbuha & SN2023usp & 2024-02-15 & 115 & Keck/LRIS \\
ZTF23absscow & SN2023ywa & 2024-07-09 & 250 & Keck/LRIS \\
ZTF23absscow & SN2023ywa & 2024-10-09 & 342 & Keck/LRIS \\
ZTF24aaasazz & SN2024ov & 2024-11-30 & 288 & Keck/LRIS \\
ZTF24aabppgn & SN2024wp & 2024-07-09 & 125 & Keck/LRIS \\
ZTF24aaejecr & SN2024btj & 2024-07-09 & 147 & Keck/LRIS \\
ZTF24aaejecr & SN2024btj & 2024-11-30 & 291 & Keck/LRIS \\
ZTF24aaejecr & SN2024btj & 2025-01-26 & 348 & Keck/LRIS \\
ZTF24aaezido & SN2024cro & 2024-07-09 & 180 & Keck/LRIS \\
ZTF24aaplfjd & SN2024jxm & 2024-11-30 & 136 & Keck/LRIS \\
ZTF24aaplfjd & SN2024jxm & 2025-01-26 & 193 & Keck/LRIS \\
ZTF24aaplfjd & SN2024jxm & 2025-06-28 & 346 & Keck/LRIS \\
ZTF24abtczty & SN2024abfl & 2025-04-20 & 149 & P200/NGPS \\
ZTF24abtczty & SN2024abfl & 2025-10-22 & 334 & Keck/LRIS \\
\hline
\end{tabular}
\end{table}

%% file: ecsn_diagnostics_table_no_CI9100.tex
\begin{table*}
\centering
\tiny
\setlength{\tabcolsep}{2pt}

\caption{Nebular spectral diagnostics for ZTF and selected comparison SNe, ordered from highest to lowest ECSN score. For each feature, the presence of an emission line is quantified using both the line SNR and peak ratio, with detection thresholds calibrated on the \citet{Jerkstrand2018} 9~M$_\odot$ nebular models (see Table~\ref{tab:ecsn_weights_thresholds}). Cells shaded in blue indicate measurements consistent with ECSN expectations, orange indicates measurements that contradict ECSN expectations, and grey indicates ambiguous cases; the text labels in each cell provide the corresponding line-detection outcome.}

\label{tab:ecsn_diag}
\resizebox{\linewidth}{!}{%
\begin{tabular}{lccccccccc}
\toprule
SN & Phase (d) & He I 7065 & Mg I] 4571 & [C I] 8727 & O I 7774 & O I 8447 & [Ca II]/[O I] & [Ni II]/[Fe II] & ECSN score \\
\midrule
SN~2023bvj & 296 & \cellcolor{ecsnAgree}no & \cellcolor{ecsnMaybe}maybe & \cellcolor{ecsnMaybe}maybe & \cellcolor{ecsnAgree}no & \cellcolor{ecsnAgree}yes & \cellcolor{ecsnMaybe}maybe & \cellcolor{ecsnMaybe}maybe & +0.48 \\
\hline
SN~2024btj  & 291 & \cellcolor{ecsnMaybe}maybe & \cellcolor{ecsnAgree}no & \cellcolor{ecsnMaybe}maybe & \cellcolor{ecsnAgree}no & \cellcolor{ecsnMaybe}maybe & \cellcolor{ecsnMaybe}maybe & \cellcolor{ecsnDisagree}no & +0.22 \\
SN~2024btj  & 348 & \cellcolor{ecsnAgree}no & \cellcolor{ecsnAgree}no & \cellcolor{ecsnMaybe}maybe & \cellcolor{ecsnAgree}no & \cellcolor{ecsnMaybe}maybe & \cellcolor{ecsnMaybe}maybe & \cellcolor{ecsnDisagree}no & +0.43 \\
\hline
SN~2016bkv & 257 & \cellcolor{ecsnMaybe}maybe & \cellcolor{ecsnMaybe}maybe & \cellcolor{ecsnMaybe}maybe & \cellcolor{ecsnMaybe}maybe & \cellcolor{ecsnAgree}yes & \cellcolor{ecsnMaybe}maybe & \cellcolor{ecsnMaybe}maybe & +0.04 \\
SN~2016bkv & 436 & \cellcolor{ecsnAgree}no & \cellcolor{ecsnDisagree}yes & \cellcolor{ecsnMaybe}maybe & \cellcolor{ecsnAgree}no & \cellcolor{ecsnAgree}yes & \cellcolor{ecsnMaybe}maybe & \cellcolor{ecsnMaybe}maybe & +0.43 \\
SN~2016bkv & 608 & \cellcolor{ecsnAgree}no & \cellcolor{ecsnDisagree}yes & \cellcolor{ecsnMaybe}maybe & \cellcolor{ecsnAgree}no & \cellcolor{ecsnAgree}yes & \cellcolor{ecsnMaybe}maybe & \cellcolor{ecsnMaybe}maybe & +0.43 \\
\hline
SN~2018zd & 228 & \cellcolor{ecsnMaybe}maybe & \cellcolor{ecsnMaybe}maybe & \cellcolor{ecsnMaybe}maybe & \cellcolor{ecsnMaybe}maybe & \cellcolor{ecsnMaybe}maybe & \cellcolor{ecsnMaybe}maybe & \cellcolor{ecsnAgree}yes & +0.04 \\
SN~2018zd & 291 & \cellcolor{ecsnMaybe}maybe & \cellcolor{ecsnMaybe}maybe & \cellcolor{ecsnMaybe}maybe & \cellcolor{ecsnMaybe}maybe & \cellcolor{ecsnAgree}yes & \cellcolor{ecsnMaybe}maybe & \cellcolor{ecsnAgree}yes & +0.09 \\
SN~2018zd & 306 & \cellcolor{ecsnAgree}no & \cellcolor{ecsnMaybe}maybe & \cellcolor{ecsnMaybe}maybe & \cellcolor{ecsnMaybe}maybe & \cellcolor{ecsnMaybe}maybe & \cellcolor{ecsnMaybe}maybe & \cellcolor{ecsnAgree}yes & +0.26 \\
SN~2018zd & 338 & \cellcolor{ecsnMaybe}maybe & \cellcolor{ecsnMaybe}maybe & \cellcolor{ecsnMaybe}maybe & \cellcolor{ecsnMaybe}maybe & \cellcolor{ecsnMaybe}maybe & \cellcolor{ecsnMaybe}maybe & \cellcolor{ecsnAgree}yes & +0.04 \\
SN~2018zd & 399 & \cellcolor{ecsnMaybe}maybe & \cellcolor{ecsnMaybe}maybe & \cellcolor{ecsnAgree}no & \cellcolor{ecsnMaybe}maybe & \cellcolor{ecsnMaybe}maybe & \cellcolor{ecsnMaybe}maybe & \cellcolor{ecsnAgree}yes & +0.26 \\
\hline
SN~2022zmb  & 175 & \cellcolor{ecsnAgree}no & \cellcolor{ecsnMaybe}maybe & \cellcolor{ecsnMaybe}maybe & \cellcolor{ecsnDisagree}yes & \cellcolor{ecsnDisagree}no & \cellcolor{ecsnMaybe}maybe & \cellcolor{ecsnAgree}yes & +0.00 \\
SN~2022zmb & 222 & \cellcolor{ecsnAgree}no & \cellcolor{ecsnMaybe}maybe & \cellcolor{ecsnMaybe}maybe & \cellcolor{ecsnMaybe}maybe & \cellcolor{ecsnDisagree}no & \cellcolor{ecsnMaybe}maybe & \cellcolor{ecsnAgree}yes & +0.22 \\
\hline
SN~2005cs & 304 & \cellcolor{ecsnMaybe}maybe & \cellcolor{ecsnAgree}no & \cellcolor{ecsnDisagree}yes & \cellcolor{ecsnMaybe}maybe & \cellcolor{ecsnMaybe}maybe & \cellcolor{ecsnMaybe}maybe & \cellcolor{ecsnAgree}yes & -0.13 \\
SN~2005cs & 334 & \cellcolor{ecsnAgree}no & \cellcolor{ecsnMaybe}maybe & \cellcolor{ecsnMaybe}maybe & \cellcolor{ecsnMaybe}maybe & \cellcolor{ecsnMaybe}maybe & \cellcolor{ecsnMaybe}maybe & \cellcolor{ecsnAgree}yes & +0.26 \\
\hline
SN~2022jzc  & 346 & \cellcolor{ecsnMaybe}maybe & \cellcolor{ecsnMaybe}maybe & \cellcolor{ecsnMaybe}maybe & \cellcolor{ecsnMaybe}maybe & \cellcolor{ecsnMaybe}maybe & \cellcolor{ecsnMaybe}maybe & \cellcolor{ecsnAgree}yes & +0.04 \\
\hline
SN~2023mpz  & 310 & \cellcolor{ecsnMaybe}maybe & \cellcolor{ecsnMaybe}maybe & \cellcolor{ecsnMaybe}maybe & \cellcolor{ecsnMaybe}maybe & \cellcolor{ecsnMaybe}maybe & \cellcolor{ecsnMaybe}maybe & \cellcolor{ecsnAgree}yes & +0.04 \\
\hline
SN~2024cro  & 180 & \cellcolor{ecsnMaybe}maybe & \cellcolor{ecsnMaybe}maybe & \cellcolor{ecsnMaybe}maybe & \cellcolor{ecsnMaybe}maybe & \cellcolor{ecsnDisagree}no & \cellcolor{ecsnMaybe}maybe & \cellcolor{ecsnMaybe}maybe & -0.04 \\
\hline
SN~2024ov  & 147 & \cellcolor{ecsnMaybe}maybe & \cellcolor{ecsnMaybe}maybe & \cellcolor{ecsnMaybe}maybe & \cellcolor{ecsnMaybe}maybe & \cellcolor{ecsnMaybe}maybe & \cellcolor{ecsnMaybe}maybe & \cellcolor{ecsnDisagree}no & -0.04 \\
\hline
SN~2023wcr  & 148 & \cellcolor{ecsnMaybe}maybe & \cellcolor{ecsnMaybe}maybe & \cellcolor{ecsnMaybe}maybe & \cellcolor{ecsnMaybe}maybe & \cellcolor{ecsnDisagree}no & \cellcolor{ecsnMaybe}maybe & \cellcolor{ecsnDisagree}no & -0.09 \\
SN~2023wcr  & 172 & \cellcolor{ecsnMaybe}maybe & \cellcolor{ecsnMaybe}maybe & \cellcolor{ecsnMaybe}maybe & \cellcolor{ecsnMaybe}maybe & \cellcolor{ecsnDisagree}no & \cellcolor{ecsnMaybe}maybe & \cellcolor{ecsnDisagree}no & -0.09 \\
SN~2023wcr  & 190 & \cellcolor{ecsnMaybe}maybe & \cellcolor{ecsnMaybe}maybe & \cellcolor{ecsnMaybe}maybe & \cellcolor{ecsnMaybe}maybe & \cellcolor{ecsnDisagree}no & \cellcolor{ecsnMaybe}maybe & \cellcolor{ecsnDisagree}no & -0.09 \\
SN~2023wcr  & 262 & \cellcolor{ecsnMaybe}maybe & \cellcolor{ecsnMaybe}maybe & \cellcolor{ecsnMaybe}maybe & \cellcolor{ecsnMaybe}maybe & \cellcolor{ecsnDisagree}no & \cellcolor{ecsnMaybe}maybe & \cellcolor{ecsnDisagree}no & -0.09 \\
\hline
SN~2023ywa & 250 & \cellcolor{ecsnDisagree}yes & \cellcolor{ecsnMaybe}maybe & \cellcolor{ecsnMaybe}maybe & \cellcolor{ecsnAgree}no & \cellcolor{ecsnMaybe}maybe & \cellcolor{ecsnDisagree}no & \cellcolor{ecsnDisagree}no & -0.26 \\
SN~2023ywa  & 342 & \cellcolor{ecsnDisagree}yes & \cellcolor{ecsnMaybe}maybe & \cellcolor{ecsnMaybe}maybe & \cellcolor{ecsnAgree}no & \cellcolor{ecsnDisagree}no & \cellcolor{ecsnMaybe}maybe & \cellcolor{ecsnMaybe}maybe & -0.04 \\
\hline
SN~2022aaad  & 136 & \cellcolor{ecsnDisagree}yes & \cellcolor{ecsnMaybe}maybe & \cellcolor{ecsnMaybe}maybe & \cellcolor{ecsnMaybe}maybe & \cellcolor{ecsnMaybe}maybe & \cellcolor{ecsnMaybe}maybe & \cellcolor{ecsnMaybe}maybe & -0.22 \\
SN~2022aaad  & 158 & \cellcolor{ecsnDisagree}yes & \cellcolor{ecsnMaybe}maybe & \cellcolor{ecsnMaybe}maybe & \cellcolor{ecsnMaybe}maybe & \cellcolor{ecsnMaybe}maybe & \cellcolor{ecsnMaybe}maybe & \cellcolor{ecsnMaybe}maybe & -0.22 \\
SN~2022aaad  & 169 & \cellcolor{ecsnDisagree}yes & \cellcolor{ecsnDisagree}yes & \cellcolor{ecsnMaybe}maybe & \cellcolor{ecsnMaybe}maybe & \cellcolor{ecsnDisagree}no & \cellcolor{ecsnMaybe}maybe & \cellcolor{ecsnMaybe}maybe & -0.30 \\
SN~2022aaad  & 215 & \cellcolor{ecsnDisagree}yes & \cellcolor{ecsnMaybe}maybe & \cellcolor{ecsnMaybe}maybe & \cellcolor{ecsnMaybe}maybe & \cellcolor{ecsnAgree}yes & \cellcolor{ecsnMaybe}maybe & \cellcolor{ecsnMaybe}maybe & -0.17 \\
SN~2022aaad  & 282 & \cellcolor{ecsnDisagree}yes & \cellcolor{ecsnDisagree}yes & \cellcolor{ecsnMaybe}maybe & \cellcolor{ecsnMaybe}maybe & \cellcolor{ecsnAgree}yes & \cellcolor{ecsnMaybe}maybe & \cellcolor{ecsnAgree}yes & -0.17 \\
SN~2022aaad  & 332 & \cellcolor{ecsnDisagree}yes & \cellcolor{ecsnMaybe}maybe & \cellcolor{ecsnMaybe}maybe & \cellcolor{ecsnMaybe}maybe & \cellcolor{ecsnAgree}yes & \cellcolor{ecsnMaybe}maybe & \cellcolor{ecsnAgree}yes & -0.13 \\
SN~2022aaad  & 366 & \cellcolor{ecsnDisagree}yes & \cellcolor{ecsnDisagree}yes & \cellcolor{ecsnMaybe}maybe & \cellcolor{ecsnMaybe}maybe & \cellcolor{ecsnAgree}yes & \cellcolor{ecsnMaybe}maybe & \cellcolor{ecsnMaybe}maybe & -0.22 \\
SN~2022aaad  & 393 & \cellcolor{ecsnDisagree}yes & \cellcolor{ecsnDisagree}yes & \cellcolor{ecsnMaybe}maybe & \cellcolor{ecsnMaybe}maybe & \cellcolor{ecsnAgree}yes & \cellcolor{ecsnMaybe}maybe & \cellcolor{ecsnMaybe}maybe & -0.22 \\
SN~2022aaad  & 463 & \cellcolor{ecsnDisagree}yes & \cellcolor{ecsnDisagree}yes & \cellcolor{ecsnMaybe}maybe & \cellcolor{ecsnMaybe}maybe & \cellcolor{ecsnAgree}yes & \cellcolor{ecsnMaybe}maybe & \cellcolor{ecsnMaybe}maybe & -0.22 \\
SN~2022aaad & 508 & \cellcolor{ecsnMaybe}maybe & \cellcolor{ecsnMaybe}maybe & \cellcolor{ecsnMaybe}maybe & \cellcolor{ecsnMaybe}maybe & \cellcolor{ecsnAgree}yes & \cellcolor{ecsnDisagree}no & \cellcolor{ecsnMaybe}maybe & -0.17 \\
\hline
SN~2022acko & 258 & \cellcolor{ecsnDisagree}yes & \cellcolor{ecsnDisagree}yes & \cellcolor{ecsnDisagree}yes & \cellcolor{ecsnAgree}no & \cellcolor{ecsnDisagree}no & \cellcolor{ecsnMaybe}maybe & \cellcolor{ecsnMaybe}maybe & -0.30 \\
SN~2022acko  & 287 & \cellcolor{ecsnDisagree}yes & \cellcolor{ecsnDisagree}yes & \cellcolor{ecsnDisagree}yes & \cellcolor{ecsnMaybe}maybe & \cellcolor{ecsnDisagree}no & \cellcolor{ecsnMaybe}maybe & \cellcolor{ecsnDisagree}no & -0.57 \\
SN~2022acko  & 340 & \cellcolor{ecsnDisagree}yes & \cellcolor{ecsnDisagree}yes & \cellcolor{ecsnAgree}no & \cellcolor{ecsnMaybe}maybe & \cellcolor{ecsnMaybe}maybe & \cellcolor{ecsnMaybe}maybe & \cellcolor{ecsnMaybe}maybe & -0.04 \\
SN~2022acko  & 369 & \cellcolor{ecsnDisagree}yes & \cellcolor{ecsnDisagree}yes & \cellcolor{ecsnDisagree}yes & \cellcolor{ecsnAgree}no & \cellcolor{ecsnDisagree}no & \cellcolor{ecsnMaybe}maybe & \cellcolor{ecsnMaybe}maybe & -0.30 \\
SN~2022acko  & 439 & \cellcolor{ecsnDisagree}yes & \cellcolor{ecsnDisagree}yes & \cellcolor{ecsnAgree}no & \cellcolor{ecsnAgree}no & \cellcolor{ecsnMaybe}maybe & \cellcolor{ecsnMaybe}maybe & \cellcolor{ecsnMaybe}maybe & +0.17 \\
\hline
SN~2022aang  & 391 & \cellcolor{ecsnDisagree}yes & \cellcolor{ecsnMaybe}maybe & \cellcolor{ecsnMaybe}maybe & \cellcolor{ecsnMaybe}maybe & \cellcolor{ecsnMaybe}maybe & \cellcolor{ecsnMaybe}maybe & \cellcolor{ecsnMaybe}maybe & -0.22 \\
\hline
SN~2023azx  & 387 & \cellcolor{ecsnDisagree}yes & \cellcolor{ecsnMaybe}maybe & \cellcolor{ecsnMaybe}maybe & \cellcolor{ecsnMaybe}maybe & \cellcolor{ecsnAgree}yes & \cellcolor{ecsnMaybe}maybe & \cellcolor{ecsnDisagree}no & -0.22 \\
SN~2023azx  & 432 & \cellcolor{ecsnDisagree}yes & \cellcolor{ecsnMaybe}maybe & \cellcolor{ecsnMaybe}maybe & \cellcolor{ecsnMaybe}maybe & \cellcolor{ecsnAgree}yes & \cellcolor{ecsnMaybe}maybe & \cellcolor{ecsnDisagree}no & -0.22 \\
\hline
SN~2024wp  & 180 & \cellcolor{ecsnDisagree}yes & \cellcolor{ecsnMaybe}maybe & \cellcolor{ecsnMaybe}maybe & \cellcolor{ecsnMaybe}maybe & \cellcolor{ecsnMaybe}maybe & \cellcolor{ecsnMaybe}maybe & \cellcolor{ecsnDisagree}no & -0.26 \\
\hline
SN~1997D & 207 & \cellcolor{ecsnDisagree}yes & \cellcolor{ecsnAgree}no & \cellcolor{ecsnMaybe}maybe & \cellcolor{ecsnMaybe}maybe & \cellcolor{ecsnDisagree}no & \cellcolor{ecsnMaybe}maybe & \cellcolor{ecsnDisagree}no & -0.26 \\
SN~1997D & 350 & \cellcolor{ecsnDisagree}yes & \cellcolor{ecsnDisagree}yes & \cellcolor{ecsnMaybe}maybe & \cellcolor{ecsnMaybe}maybe & \cellcolor{ecsnMaybe}maybe & \cellcolor{ecsnMaybe}maybe & \cellcolor{ecsnMaybe}maybe & -0.26 \\
\hline
SN~2008bk & 260 & \cellcolor{ecsnDisagree}yes & \cellcolor{ecsnDisagree}yes & \cellcolor{ecsnDisagree}yes & \cellcolor{ecsnMaybe}maybe & \cellcolor{ecsnAgree}yes & \cellcolor{ecsnMaybe}maybe & \cellcolor{ecsnDisagree}no & -0.48 \\
SN~2008bk & 529 & \cellcolor{ecsnMaybe}maybe & \cellcolor{ecsnDisagree}yes & \cellcolor{ecsnDisagree}yes & \cellcolor{ecsnAgree}no & \cellcolor{ecsnMaybe}maybe & \cellcolor{ecsnMaybe}maybe & \cellcolor{ecsnMaybe}maybe & -0.04 \\
\hline
SN~2023vci  & 165 & \cellcolor{ecsnDisagree}yes & \cellcolor{ecsnMaybe}maybe & \cellcolor{ecsnMaybe}maybe & \cellcolor{ecsnMaybe}maybe & \cellcolor{ecsnDisagree}no & \cellcolor{ecsnMaybe}maybe & \cellcolor{ecsnDisagree}no & -0.30 \\
SN~2023vci  & 210 & \cellcolor{ecsnDisagree}yes & \cellcolor{ecsnMaybe}maybe & \cellcolor{ecsnMaybe}maybe & \cellcolor{ecsnMaybe}maybe & \cellcolor{ecsnDisagree}no & \cellcolor{ecsnMaybe}maybe & \cellcolor{ecsnDisagree}no & -0.30 \\
\hline
SN~2023kmk & 231 & \cellcolor{ecsnDisagree}yes & \cellcolor{ecsnMaybe}maybe & \cellcolor{ecsnAgree}no & \cellcolor{ecsnDisagree}yes & \cellcolor{ecsnAgree}yes & \cellcolor{ecsnDisagree}no & \cellcolor{ecsnDisagree}no & -0.43 \\
SN~2023kmk  & 264 & \cellcolor{ecsnDisagree}yes & \cellcolor{ecsnMaybe}maybe & \cellcolor{ecsnMaybe}maybe & \cellcolor{ecsnMaybe}maybe & \cellcolor{ecsnAgree}yes & \cellcolor{ecsnMaybe}maybe & \cellcolor{ecsnDisagree}no & -0.22 \\
SN~2023kmk  & 309 & \cellcolor{ecsnDisagree}yes & \cellcolor{ecsnMaybe}maybe & \cellcolor{ecsnMaybe}maybe & \cellcolor{ecsnMaybe}maybe & \cellcolor{ecsnAgree}yes & \cellcolor{ecsnMaybe}maybe & \cellcolor{ecsnDisagree}no & -0.22 \\
SN~2023kmk  & 347 & \cellcolor{ecsnDisagree}yes & \cellcolor{ecsnMaybe}maybe & \cellcolor{ecsnMaybe}maybe & \cellcolor{ecsnMaybe}maybe & \cellcolor{ecsnAgree}yes & \cellcolor{ecsnDisagree}no & \cellcolor{ecsnDisagree}no & -0.43 \\
\hline
SN~2024abfl  & 149 & \cellcolor{ecsnDisagree}yes & \cellcolor{ecsnMaybe}maybe & \cellcolor{ecsnDisagree}yes & \cellcolor{ecsnMaybe}maybe & \cellcolor{ecsnAgree}yes & \cellcolor{ecsnDisagree}no & \cellcolor{ecsnAgree}yes & -0.57 \\
SN~2024abfl  & 334 & \cellcolor{ecsnDisagree}yes & \cellcolor{ecsnDisagree}yes & \cellcolor{ecsnDisagree}yes & \cellcolor{ecsnAgree}no & \cellcolor{ecsnAgree}yes & \cellcolor{ecsnMaybe}maybe & \cellcolor{ecsnMaybe}maybe & -0.22 \\
SN~2024abfl  & 393 & \cellcolor{ecsnDisagree}yes & \cellcolor{ecsnMaybe}maybe & \cellcolor{ecsnDisagree}yes & \cellcolor{ecsnAgree}no & \cellcolor{ecsnMaybe}maybe & \cellcolor{ecsnMaybe}maybe & \cellcolor{ecsnDisagree}no & -0.26 \\
\hline
SN~2023usp  & 115 & \cellcolor{ecsnDisagree}yes & \cellcolor{ecsnMaybe}maybe & \cellcolor{ecsnDisagree}yes & \cellcolor{ecsnMaybe}maybe & \cellcolor{ecsnAgree}yes & \cellcolor{ecsnMaybe}maybe & \cellcolor{ecsnAgree}yes & -0.35 \\
\hline
SN~2024jxm  & 132 & \cellcolor{ecsnDisagree}yes & \cellcolor{ecsnMaybe}maybe & \cellcolor{ecsnDisagree}yes & \cellcolor{ecsnDisagree}yes & \cellcolor{ecsnDisagree}no & \cellcolor{ecsnMaybe}maybe & \cellcolor{ecsnDisagree}no & -0.74 \\
SN~2024jxm  & 241 & \cellcolor{ecsnDisagree}yes & \cellcolor{ecsnMaybe}maybe & \cellcolor{ecsnMaybe}maybe & \cellcolor{ecsnMaybe}maybe & \cellcolor{ecsnDisagree}no & \cellcolor{ecsnMaybe}maybe & \cellcolor{ecsnDisagree}no & -0.30 \\
SN~2024jxm & 394 & \cellcolor{ecsnMaybe}maybe & \cellcolor{ecsnMaybe}maybe & \cellcolor{ecsnMaybe}maybe & \cellcolor{ecsnMaybe}maybe & \cellcolor{ecsnMaybe}maybe & \cellcolor{ecsnMaybe}maybe & \cellcolor{ecsnDisagree}no & -0.04 \\
\hline
\bottomrule
\end{tabular}
}
\end{table*}

%% file: ecsn_theoretical_model_scores.tex
\begin{table*}
\centering
\tiny
\setlength{\tabcolsep}{2pt}

\caption{sAGB ECSN vs RSG FeCCSN diagnostics for the 9~M$_\odot$ theoretical model nebular spectra.}
\label{tab:ecsn_theoretical_model_scores}
\resizebox{\linewidth}{!}{%
\begin{tabular}{lccccccccc}
\toprule
Model & Phase (d) & He I 7065 & Mg I] 4571 & [C I] 8727 & O I 7774 & O I 8447 & [Ca II]/[O I] & [Ni II]/[Fe II] & ECSN score \\
\midrule
9 Msun sAGB & 300 & \cellcolor{ecsnAgree}no & \cellcolor{ecsnMaybe}maybe & \cellcolor{ecsnAgree}no & \cellcolor{ecsnAgree}no & \cellcolor{ecsnAgree}yes & \cellcolor{ecsnMaybe}maybe & \cellcolor{ecsnAgree}yes & +0.74 \\
9 Msun sAGB & 400 & \cellcolor{ecsnAgree}no & \cellcolor{ecsnMaybe}maybe & \cellcolor{ecsnAgree}no & \cellcolor{ecsnAgree}no & \cellcolor{ecsnAgree}yes & \cellcolor{ecsnMaybe}maybe & \cellcolor{ecsnMaybe}maybe & +0.70 \\
9 Msun sAGB & 500 & \cellcolor{ecsnAgree}no & \cellcolor{ecsnAgree}no & \cellcolor{ecsnAgree}no & \cellcolor{ecsnAgree}no & \cellcolor{ecsnMaybe}maybe & \cellcolor{ecsnMaybe}maybe & \cellcolor{ecsnAgree}yes & +0.74 \\
9 Msun sAGB & 600 & \cellcolor{ecsnAgree}no & \cellcolor{ecsnAgree}no & \cellcolor{ecsnAgree}no & \cellcolor{ecsnAgree}no & \cellcolor{ecsnDisagree}no & \cellcolor{ecsnMaybe}maybe & \cellcolor{ecsnAgree}yes & +0.70 \\
\hline
9 Msun RSG & 200 & \cellcolor{ecsnDisagree}yes & \cellcolor{ecsnDisagree}yes & \cellcolor{ecsnMaybe}maybe & \cellcolor{ecsnDisagree}yes & \cellcolor{ecsnDisagree}no & \cellcolor{ecsnMaybe}maybe & \cellcolor{ecsnDisagree}no & -0.56 \\
9 Msun RSG & 300 & \cellcolor{ecsnDisagree}yes & \cellcolor{ecsnDisagree}yes & \cellcolor{ecsnDisagree}yes & \cellcolor{ecsnDisagree}yes & \cellcolor{ecsnDisagree}no & \cellcolor{ecsnMaybe}maybe & \cellcolor{ecsnDisagree}no & -0.78 \\
9 Msun RSG & 400 & \cellcolor{ecsnDisagree}yes & \cellcolor{ecsnDisagree}yes & \cellcolor{ecsnDisagree}yes & \cellcolor{ecsnDisagree}yes & \cellcolor{ecsnDisagree}no & \cellcolor{ecsnMaybe}maybe & \cellcolor{ecsnMaybe}maybe & -0.74 \\
9 Msun RSG & 500 & \cellcolor{ecsnDisagree}yes & \cellcolor{ecsnDisagree}yes & \cellcolor{ecsnDisagree}yes & \cellcolor{ecsnMaybe}maybe & \cellcolor{ecsnDisagree}no & \cellcolor{ecsnMaybe}maybe & \cellcolor{ecsnAgree}yes & -0.48 \\
9 Msun RSG & 600 & \cellcolor{ecsnMaybe}maybe & \cellcolor{ecsnDisagree}yes & \cellcolor{ecsnDisagree}yes & \cellcolor{ecsnMaybe}maybe & \cellcolor{ecsnDisagree}no & \cellcolor{ecsnMaybe}maybe & \cellcolor{ecsnAgree}yes & -0.26 \\
\hline
\bottomrule
\end{tabular}
}
\end{table*}

%% file: ZTF_bright_table.tex
\begin{table*}
\centering
\scriptsize
\setlength{\tabcolsep}{2pt}
\caption{Spectral log of nebular spectra of brighter ZTF Type IIP SNe.}
\begin{tabular}{l l c c c}
\hline
ZTF Name & IAU Name & Date & Phase (d) & Instrument \\
\hline
ZTF19abqhobb & SN~2019nvm & 2020-01-29 & 152 & NOT/ALFOSC \\
ZTF20aaynrrh & SN~2020jfo\textsuperscript{\tiny a} & 2020-12-06 & 207 & NOT/ALFOSC \\
ZTF20aaynrrh & SN~2020jfo\textsuperscript{\tiny a} & 2021-01-15 & 247 & NOT/ALFOSC \\
ZTF20aaynrrh & SN~2020jfo\textsuperscript{\tiny a} & 2021-02-08 & 271 & NOT/ALFOSC \\
ZTF20aaynrrh & SN~2020jfo\textsuperscript{\tiny a} & 2021-03-08 & 299 & NOT/ALFOSC \\
ZTF20aaynrrh & SN~2020jfo\textsuperscript{\tiny a} & 2021-04-20 & 342 & NOT/ALFOSC \\
ZTF20abotkfn & SN~2020qmp & 2021-02-20 & 111 & P200/DBSP \\
ZTF21abouuat & SN~2021ucg & 2022-04-27 & 257 & Keck/LRIS \\
ZTF23absdcgi & SN~2023zcu & 2024-11-11 & 329 & P200/DBSP \\
ZTF23abvgvab & SN~2023abim & 2024-03-31 & 104 & Keck/LRIS \\
\hline
\end{tabular}

\tablenotetext{\scriptsize $a$}{\footnotesize Nebular spectra of SN~2020jfo are taken from \citet{Sollerman2021}.}

\label{table:bright_spectral_lo}
\end{table*}

%% file: nonZTF_spectral_log_threecol.tex
\begin{table*}[ht]
\centering
\scriptsize
\caption{Spectral Log for 120 nebular spectra of 54 non--ZTF Type~II SNe}
\label{table:literature_log}
\begin{minipage}{0.4\linewidth}
\centering
\begin{tabular}{lccc}
\hline
SN & MJD & Phase (d) & Ref. \\
\hline
1990E & 48191 & 253 & (1) \\
1990E & 48207 & 269 & (1) \\
1990E & 48242 & 304 & (1) \\
1990E & 48267 & 330 & (1) \\
1990E & 48268 & 330 & (1) \\
1990Q & 48362 & 320 & (1) \\
1991G & 48635 & 355 & (2)(3) \\
1992H & 49042 & 386 & (2)(4) \\
1992ad & 49030 & 225 & (2)(5) \\
1992ad & 49091 & 286 & (2)(5) \\
1992ad & 49092 & 287 & (2)(5) \\
1993K & 49369 & 295 & (2) \\
1996W & 50432 & 252 & (6) \\
1996W & 50475 & 295 & (6) \\
1997D & 50569 & 207 &  (64)(65)\\
1997D & 50712 & 350 &  (64)(65)\\
1998S & 51017 & 143 &  (66)\\
1998S & 51188 & 314 &  (66)\\
1999em & 51793 & 317 & (7)(8) \\
2002hh & 52972 & 396 & (2)(9) \\
2003B & 52897 & 275 & (10)(11) \\
2003gd & 52847 & 130 & (8)(12) \\
2003gd & 52967 & 250 & (8)(12) \\
2004A & 53296 & 286 & (2)(13)(14) \\
2004dj & 53323 & 136 & (2)(15) \\
2004dj & 53351 & 164 & (2)(15) \\
2004dj & 53386 & 199 & (2)(15) \\
2004dj & 53413 & 226 & (2)(15) \\
2004dj & 53440 & 253 & (2)(15) \\
2004dj & 53442 & 274 & (2)(15) \\
2004dj & 53477 & 290 & (2)(15) \\
2004et & 53432 & 162 & (8)(16)(17) \\
2004et & 53471 & 201 & (8)(16)(17) \\
2004et & 53496 & 226 & (8)(16)(17) \\
2004et & 53528 & 258 & (8)(16)(17) \\
2004et & 53552 & 282 & (8)(16)(17) \\
2004et & 53583 & 313 & (8)(16)(17) \\
2004et & 53623 & 353 & (8)(16)(17) \\
2004et & 53624 & 354 & (8)(16)(17) \\
2005ay & 53741 & 285 & (8)(18)(19) \\
2005cs & 53706 & 157 & (8)(20)(21) \\
2005cs & 53852 & 304 & (8)(20)(21) \\
2005cs & 53883 & 334 & (8)(20)(21) \\
2007aa & 54515 & 384 & (22)(23) \\
2007it & 54616 & 268 & (11)(24) \\
2008bk & 54810 & 260 & (11)(25) \\
2008bk & 55079 & 529 & (11)(25) \\
2008cn & 54938 & 340 & (23)(26) \\
2008ex & 54979 & 285 & (2)(27) \\
2009N & 55259 & 411 & (28) \\
2009dd & 55157 & 232 & (6) \\
2009dd & 55333 & 408 & (6) \\
2009ib & 55260 & 219 & (29) \\
2009ib & 55303 & 262 & (29) \\
2012A & 56019 & 86 & (30) \\
2012A & 56341 & 408 & (30) \\
2012aw & 56371 & 369 & (31)(32) \\
2012ch & 56402 & 357 & (2)(33) \\
2012ec & 56546 & 403 & (34) 
\\\hline
\end{tabular}
\end{minipage}
\hspace{0.005\linewidth}
\begin{minipage}{0.4\linewidth}
\centering
\begin{tabular}{lccc}
\hline
SN & MJD & Phase (d) & Ref. \\
\hline
2012ho & 56422 & 168 & (2)(35) \\
2012ho & 56484 & 229 & (2)(35) \\
2012ho & 56504 & 249 & (2)(35) \\
2012ho & 56506 & 252 & (2)(35) \\
2012ho & 56520 & 265 & (2)(35) \\
2012ho & 56573 & 318 & (2)(35) \\
2013am & 56652 & 281 & (2)(36) \\
2013by & 56564 & 160 & (37)(38) \\
2013by & 56691 & 287 & (37)(38) \\
2013ej & 56932 & 435 & (2)(39)(40) \\
2013ej & 57305 & 808 & (2)(39)(40) \\
2013ej & 56834 & 337 & (2)(39)(40) \\
2013fs & 56841 & 270 & (41) \\
2014G & 56806 & 137 & (42) \\
2014G & 56856 & 187 & (42) \\
2014G & 57010 & 341 & (42) \\
2014cx & 57251 & 350 & (43) \\
2015bs & 57340 & 420 & (44) \\
2015bs & 57341 & 421 & (44) \\
ASASSN15oz & 57604 & 342 & (45) \\
2016X & 57745 & 340 & (46)(47) \\
2016aqf & 57642 & 202 & (48) \\
2016aqf & 57699 & 259 & (48) \\
2016aqf & 57725 & 285 & (48) \\
2016aqf & 57743 & 303 & (48) \\
2016aqf & 57770 & 330 & (48) \\
2016aqf & 57791 & 351 & (48) \\
2016bkv & 57724 & 257 &  (67)\\
2016bkv & 57903 & 436 &  (67)\\
2016bkv & 58075 & 608 &  (67)\\
2017eaw & 58136 & 250 & (49) \\
2017ivv & 58202 & 110 & (50) \\
2017ivv & 58228 & 136 & (50) \\
2017ivv & 58250 & 158 & (50) \\
2017ivv & 58335 & 243 & (50) \\
2017ivv & 58371 & 279 & (50) \\
2017ivv & 58417 & 325 & (50) \\
2017ivv & 58424 & 332 & (50) \\
2018cuf & 58627 & 335 & (52) \\
2018gj & 58397 & 270 & (69) \\
2018hwm & 58810 & 385 & (53) \\
2018is & 58519 & 386 & (51) \\
2018zd & 58360 & 182 &  (68)\\
2018zd & 58406 & 228 &  (68)\\
2018zd & 58462 & 284 &  (68)\\
2018zd & 58469 & 291 &  (68)\\
2018zd & 58484 & 306 &  (68)\\
2018zd & 58493 & 315 &  (68)\\
2018zd & 58516 & 338 &  (68)\\
2018zd & 58577 & 399 &  (68)\\
2020jfo & 59190 & 217 & (54)(55)(56)(57) \\
2020jfo & 59230 & 256 & (54)(55)(56)(57) \\
2020jfo & 59254 & 280 & (54)(55)(56)(57) \\
2020jfo & 59281 & 308 & (54)(55)(56)(57) \\
2020jfo & 59324 & 351 & (54)(55)(56)(57) \\
2021dbg & 59610 & 352 & (58) \\
2021gmj & 59678 & 385 & (59)(60) \\
2022jox & 59947 & 240 & (61) \\
2023ixf & 60343 & 260 & (62)(63) \\
\hline
\end{tabular}
\end{minipage}
\vspace{1.5cm}
\begin{minipage}{0.95\linewidth}
\tiny
\noindent {\bf References:} 
(1)~\citet{gomez00};
(2)~\citet{SNDB};
(3)~\citet{1991G};
(4)~\citet{1992H};
(5)~\citet{1992ad};
(6)~\citet{inserra13};
(7)~\citet{1999em};
(8)~\citet{faran14};
(9)~\citet{2002hh};
(10)~\citet{anderson14};
(11)~\citet{gutierrez17};
(12)~\citet{2003gd};
(13)~\citet{2004A1};
(14)~\citet{2004A2};
(15)~\citet{2004dj};
(16)~\citet{2004et1};
(17)~\citet{2004et2};
(18)~\citet{2005ay1};
(19)~\citet{2005ay2};
(20)~\citet{2005cs1};
(21)~\citet{2005cs2};
(22)~\citet{2007aa};
(23)~\citet{maguire12};
(24)~\citet{2007it};
(25)~\citet{2008bk};
(26)~\citet{2008cn};
(27)~\citet{2008ex};
(28)~\citet{2009N};
(29)~\citet{2009ib};
(30)~\citet{2012A};
(31)~\citet{2012aw};
(32)~\citet{jerk14};
(33)~\citet{2012ch};
(34)~\citet{jerk15};
(35)~\citet{2012ho};
(36)~\citet{2013am};
(37)~\citet{2013by};
(38)~\citet{black17};
(39)~\citet{2013ej};
(40)~\citet{2013ej2};
(41)~\citet{2013fs};
(42)~\citet{2014G};
(43)~\citet{2014cx};
(44)~\citet{Anderson2018};
(45)~\citet{15oz};
(46)~\citet{2016X1};
(47)~\citet{2016X2};
(48)~\citet{Muller2020};
(49)~\citet{2017eaw};
(50)~\citet{2017ivv};
(51)~\citet{Dastidar2025};
(52)~\citet{Dong2021};
(53)~\citet{Reguitti2021};
(54)~\citet{Sollerman2021};
(55)~\citet{Teja2022};
(56)~\citet{Ailawadhi2023};
(57)~\citet{kilpatrick23};
(58)~\citet{Zhao2024};
(59)~\citet{murai24};
(60)~\citet{MezaRetamal2024};
(61)~\citet{Andrews2024};
(62)~\citet{Singh2024};
(63)~\citet{Ferrari2024};
(64)~\citet{Turatto1998};
(65)~\citet{Benetti2001};
(66)~\citet{Pozzo2004};
(67)~\citet{Hosseinzadeh2018};
(68)~\citet{Hiramatsu2021};
(69)~\citet{Teja2023}.

\end{minipage}
\end{table*}

%% file: master_obs_ZAMS_plateau_table.tex
\begin{table*}[ht]
\centering
\tiny
\caption{Nebular measurements, plateau properties, peak absolute magnitudes, and ZAMS estimates for the LLIIP sample.}
\label{tab:nebular_master}
\begin{tabular}{lrllllllll}
\hline
SN & Phase & $M_{r,\mathrm{peak}}$ & H I 6563 FWHM & [Ca II]/[O I] & [Ni II]/[Fe II] & $f_{\rm [O\,I]}$ & Plat. dur. & Slope & $M_{\rm ZAMS}$ \\
 & (days) & (mag) & (\AA) & & & & (days) & ($\times 0.01$ mag d$^{-1}$) & (M$_\odot$) \\
\hline

ZTF22abssiet & 175 & $-15.72$ & $55.71 \pm 0.67$ & $3.83 \pm 0.53$ & $1.47$ & $0.022 \pm 0.002$ & 66 & 0.182 & $13.62^{+2.84}_{-3.14}$ \\
ZTF22abssiet & 222 & $-15.72$ & $52.70 \pm 0.89$ & $1.41 \pm 0.09$ & $1.04$ & $0.056 \pm 0.001$ & 66 & 0.182 & $13.62^{+2.84}_{-3.14}$ \\
ZTF22abtjefa & 136 & $-15.06$ & $65.21 \pm 0.52$ & $--$ & $--$ & $0.023 \pm 0.001$ & 66 & -0.015 & $12.85^{+2.68}_{-3.05}$ \\
ZTF22abtjefa & 158 & $-15.06$ & $62.72 \pm 0.73$ & $1.55 \pm 0.13$ & $0.55$ & $0.032 \pm 0.001$ & 66 & -0.015 & $12.85^{+2.68}_{-3.05}$ \\
ZTF22abtjefa & 169 & $-15.06$ & $61.42 \pm 0.46$ & $1.17 \pm 0.13$ & $0.47$ & $0.032 \pm 0.001$ & 66 & -0.015 & $12.85^{+2.68}_{-3.05}$ \\
ZTF22abtjefa & 215 & $-15.06$ & $56.43 \pm 0.30$ & $1.45 \pm 0.09$ & $0.52$ & $0.041 \pm 0.001$ & 66 & -0.015 & $12.85^{+2.68}_{-3.05}$ \\
ZTF22abtjefa & 282 & $-15.06$ & $50.70 \pm 0.34$ & $1.32 \pm 0.10$ & $0.81$ & $0.063 \pm 0.001$ & 66 & -0.015 & $12.85^{+2.68}_{-3.05}$ \\
ZTF22abtjefa & 332 & $-15.06$ & $47.08 \pm 0.26$ & $1.15 \pm 0.09$ & $0.71$ & $0.094 \pm 0.001$ & 66 & -0.015 & $12.85^{+2.68}_{-3.05}$ \\
ZTF22abtjefa & 366 & $-15.06$ & $47.30 \pm 0.48$ & $1.26 \pm 0.15$ & $0.53$ & $0.095 \pm 0.005$ & 66 & -0.015 & $12.85^{+2.68}_{-3.05}$ \\
ZTF22abtjefa & 393 & $-15.06$ & $44.48 \pm 0.33$ & $1.29 \pm 0.27$ & $0.64$ & $0.122 \pm 0.023$ & 66 & -0.015 & $12.85^{+2.68}_{-3.05}$ \\
ZTF22abvaetz & 391 & $-15.44$ & $39.50 \pm 1.24$ & $5.84 \pm 1.13$ & $0.46$ & $0.035 \pm 0.006$ & -- & -- & $--$ \\
ZTF22abyivoq & 258 & $-15.83$ & $56.56 \pm 1.56$ & $3.65 \pm 0.42$ & $0.42$ & $0.039 \pm 0.001$ & 70 & 0.229 & $11.19^{+2.02}_{-2.11}$ \\
ZTF22abyivoq & 287 & $-15.83$ & $54.80 \pm 0.36$ & $2.52 \pm 0.13$ & $0.33$ & $0.057 \pm 0.002$ & 70 & 0.229 & $11.19^{+2.02}_{-2.11}$ \\
ZTF22abyivoq & 340 & $-15.83$ & $51.76 \pm 0.60$ & $4.34 \pm 0.58$ & $0.40$ & $0.061 \pm 0.006$ & 70 & 0.229 & $11.19^{+2.02}_{-2.11}$ \\
ZTF22abyivoq & 369 & $-15.83$ & $49.25 \pm 0.27$ & $2.75 \pm 0.19$ & $0.54$ & $0.078 \pm 0.004$ & 70 & 0.229 & $11.19^{+2.02}_{-2.11}$ \\
ZTF22abyivoq & 439 & $-15.83$ & $48.18 \pm 0.70$ & $2.32 \pm 0.12$ & $0.52$ & $0.104 \pm 0.001$ & 70 & 0.229 & $11.19^{+2.02}_{-2.11}$ \\
ZTF23aabksje & 387 & $-15.56$ & $50.05 \pm 1.26$ & $3.07 \pm 0.51$ & $0.07$ & $0.056 \pm 0.001$ & -- & -- & $9.85^{+1.81}_{-1.53}$ \\
ZTF23aabksje & 432 & $-15.56$ & $35.45 \pm 2.06$ & $2.69 \pm 1.19$ & $0.07$ & $0.045 \pm 0.006$ & -- & -- & $9.85^{+1.81}_{-1.53}$ \\
ZTF23aackjhs & 296 & $-14.88$ & $26.17 \pm 0.55$ & $4.32 \pm 0.59$ & $0.56$ & $0.059 \pm 0.007$ & 97 & 0.144 & $11.89^{+2.77}_{-2.52}$ \\
ZTF23aanxrjm & 231 & $-15.02$ & $50.98 \pm 0.81$ & $0.79 \pm 0.18$ & $0.18$ & $0.050 \pm 0.001$ & -- & -- & $11.24^{+2.37}_{-2.23}$ \\
ZTF23aanxrjm & 264 & $-15.02$ & $56.54 \pm 0.46$ & $1.13 \pm 0.11$ & $0.21$ & $0.036 \pm 0.001$ & -- & -- & $11.24^{+2.37}_{-2.23}$ \\
ZTF23aanxrjm & 309 & $-15.02$ & $43.36 \pm 0.34$ & $1.18 \pm 0.16$ & $0.17$ & $0.051 \pm 0.001$ & -- & -- & $11.24^{+2.37}_{-2.23}$ \\
ZTF23aanxrjm & 347 & $-15.02$ & $36.01 \pm 0.23$ & $0.85 \pm 0.04$ & $0.19$ & $0.064 \pm 0.001$ & -- & -- & $11.24^{+2.37}_{-2.23}$ \\
ZTF23abgmhgw & 165 & $-15.69$ & $39.21 \pm 0.49$ & $3.70 \pm 0.43$ & $0.33$ & $0.032 \pm 0.001$ & -- & -- & $12.13^{+2.27}_{-1.89}$ \\
ZTF23abgmhgw & 210 & $-15.69$ & $37.07 \pm 0.39$ & $2.33 \pm 0.35$ & $0.24$ & $0.040 \pm 0.001$ & -- & -- & $12.13^{+2.27}_{-1.89}$ \\
ZTF23abnogui & 142 & $-15.58$ & $76.82 \pm 1.44$ & $1.46 \pm 0.15$ & $16.68$ & $0.032 \pm 0.001$ & 87 & 0.584 & $11.60^{+2.31}_{-1.75}$ \\
ZTF23abnogui & 148 & $-15.58$ & $77.43 \pm 0.84$ & $1.52 \pm 0.17$ & $0.36$ & $0.039 \pm 0.001$ & 87 & 0.584 & $11.60^{+2.31}_{-1.75}$ \\
ZTF23abnogui & 172 & $-15.58$ & $70.72 \pm 0.73$ & $1.71 \pm 0.13$ & $0.33$ & $0.053 \pm 0.001$ & 87 & 0.584 & $11.60^{+2.31}_{-1.75}$ \\
ZTF23abnogui & 190 & $-15.58$ & $66.70 \pm 14.44$ & $2.12 \pm 0.17$ & $0.35$ & $0.048 \pm 0.001$ & 87 & 0.584 & $11.60^{+2.31}_{-1.75}$ \\
ZTF23abnogui & 262 & $-15.58$ & $63.03 \pm 0.94$ & $2.82 \pm 0.28$ & $0.21$ & $0.045 \pm 0.002$ & 87 & 0.584 & $11.60^{+2.31}_{-1.75}$ \\
ZTF23abpbuha & 115 & $-14.26$ & $35.32 \pm 7.47$ & $1.24 \pm 0.16$ & $1.21$ & $0.064 \pm 0.001$ & -- & -- & $--$ \\
ZTF23absscow & 342 & $-15.91$ & $56.28 \pm 0.39$ & $2.84 \pm 0.14$ & $0.44$ & $0.061 \pm 0.002$ & -- & -- & $10.88^{+2.39}_{-1.71}$ \\
ZTF24aaasazz & 147 & $-16.03$ & $52.38 \pm 0.86$ & $1.11 \pm 0.21$ & $0.03$ & $0.039 \pm 0.001$ & -- & -- & $10.28^{+1.78}_{-1.76}$ \\
ZTF24aaasazz & 353 & $-16.03$ & $40.24 \pm 1.38$ & $4.34 \pm 0.82$ & $0.14$ & $0.053 \pm 0.007$ & -- & -- & $10.28^{+1.78}_{-1.76}$ \\
ZTF24aabppgn & 180 & $-15.52$ & $35.14 \pm 1.55$ & $4.03 \pm 2.10$ & $0.18$ & $0.028 \pm 0.014$ & 95 & 0.105 & $--$ \\
ZTF24aaejecr & 147 & $-16.02$ & $57.39 \pm 0.71$ & $3.21 \pm 0.18$ & $0.17$ & $0.030 \pm 0.001$ & 78 & 0.549 & $14.05^{+3.15}_{-3.84}$ \\
ZTF24aaejecr & 291 & $-16.02$ & $50.32 \pm 0.37$ & $2.14 \pm 0.13$ & $0.35$ & $0.084 \pm 0.001$ & 78 & 0.549 & $14.05^{+3.15}_{-3.84}$ \\
ZTF24aaejecr & 348 & $-16.02$ & $48.12 \pm 0.45$ & $1.93 \pm 0.12$ & $0.36$ & $0.106 \pm 0.001$ & 78 & 0.549 & $14.05^{+3.15}_{-3.84}$ \\
ZTF24aaezido & 180 & $-14.98$ & $99.98$ & $3.18 \pm 0.53$ & $0.47$ & $0.050 \pm 0.004$ & -- & -- & $--$ \\
ZTF24aaplfjd & 132 & $-16.00$ & $56.30 \pm 1.02$ & $1.64 \pm 0.16$ & $0.05$ & $0.022 \pm 0.001$ & 75 & 0.241 & $10.50^{+2.00}_{-1.95}$ \\
ZTF24aaplfjd & 184 & $-16.00$ & $50.79 \pm 1.13$ & $5.07 \pm 1.34$ & $0.19$ & $0.024 \pm 0.006$ & 75 & 0.241 & $10.50^{+2.00}_{-1.95}$ \\
ZTF24aaplfjd & 394 & $-16.00$ & $47.74 \pm 2.55$ & $6.98 \pm 0.69$ & $0.17$ & $0.072 \pm 0.002$ & 75 & 0.241 & $10.50^{+2.00}_{-1.95}$ \\
ZTF24abtczty & 149 & $-14.13$ & $35.33 \pm 11.88$ & $0.87 \pm 0.10$ & $1.05$ & $0.064 \pm 0.001$ & 99 & -0.040 & $10.91^{+1.94}_{-2.11}$ \\
ZTF24abtczty & 334 & $-14.13$ & $21.06 \pm 0.32$ & $3.99 \pm 0.33$ & $0.49$ & $0.055 \pm 0.002$ & 99 & -0.040 & $10.91^{+1.94}_{-2.11}$ \\
ZTF24abtczty & 393 & $-14.13$ & $20.36 \pm 0.37$ & $4.11 \pm 0.28$ & $0.34$ & $0.075 \pm 0.003$ & 99 & -0.040 & $10.91^{+1.94}_{-2.11}$ \\
1987A & 198 & $--$ & $--$ & $--$ & $2.76$ & $--$ & -- & -- & $--$ \\
1998S & 143 & $--$ & $182.06 \pm 2.60$ & $1.95 \pm 0.11$ & $2.21$ & $0.025 \pm 0.001$ & -- & -- & $8.92^{+0.78}_{-0.72}$ \\
1998S & 314 & $--$ & $182.53 \pm 8.97$ & $1.28 \pm 0.06$ & $0.98$ & $0.023 \pm 0.001$ & -- & -- & $8.92^{+0.78}_{-0.72}$ \\
2004dj & 274 & $-17.45$ & $59.91 \pm 1.24$ & $1.58 \pm 0.54$ & $0.24$ & $0.078 \pm 0.027$ & 50 & 1.258 & $14.78^{+3.65}_{-3.83}$ \\
2004dj & 274 & $-17.45$ & $59.91 \pm 1.24$ & $1.58 \pm 0.54$ & $0.24$ & $0.078 \pm 0.027$ & 50 & 1.258 & $14.78^{+3.65}_{-3.83}$ \\
2004et & 253 & $-19.04$ & $--$ & $--$ & $1.09$ & $--$ & 100 & 0.657 & $12.95^{+2.90}_{-3.07}$ \\
2004et & 253 & $-19.04$ & $--$ & $--$ & $1.09$ & $--$ & 100 & 0.657 & $12.95^{+2.90}_{-3.07}$ \\
2004et & 253 & $-19.04$ & $--$ & $--$ & $1.09$ & $--$ & 100 & 0.657 & $13.02^{+2.98}_{-3.20}$ \\
2004et & 253 & $-19.04$ & $--$ & $--$ & $1.09$ & $--$ & 100 & 0.657 & $13.02^{+2.98}_{-3.20}$ \\
2004et & 355 & $-19.04$ & $63.53 \pm 0.52$ & $3.34 \pm 0.13$ & $1.02$ & $0.092 \pm 0.001$ & 100 & 0.657 & $12.95^{+2.90}_{-3.07}$ \\
2004et & 355 & $-19.04$ & $63.53 \pm 0.52$ & $3.34 \pm 0.13$ & $1.02$ & $0.092 \pm 0.001$ & 100 & 0.657 & $12.95^{+2.90}_{-3.07}$ \\
2004et & 355 & $-19.04$ & $63.53 \pm 0.52$ & $3.34 \pm 0.13$ & $1.02$ & $0.092 \pm 0.001$ & 100 & 0.657 & $13.02^{+2.98}_{-3.20}$ \\
2004et & 355 & $-19.04$ & $63.53 \pm 0.52$ & $3.34 \pm 0.13$ & $1.02$ & $0.092 \pm 0.001$ & 100 & 0.657 & $13.02^{+2.98}_{-3.20}$ \\
2013by & 287 & $-18.76$ & $117.25 \pm 1.38$ & $0.77 \pm 0.05$ & $0.86$ & $0.116 \pm 0.001$ & 67 & 2.053 & $17.86^{+2.14}_{-5.50}$ \\
2013by & 287 & $-18.76$ & $117.25 \pm 1.38$ & $0.77 \pm 0.05$ & $0.86$ & $0.116 \pm 0.001$ & 67 & 2.053 & $17.86^{+2.14}_{-5.50}$ \\
2013by & 287 & $-18.76$ & $117.25 \pm 1.38$ & $0.77 \pm 0.05$ & $0.86$ & $0.116 \pm 0.001$ & 67 & 2.053 & $17.32^{+2.68}_{-4.97}$ \\
2013by & 287 & $-18.76$ & $117.25 \pm 1.38$ & $0.77 \pm 0.05$ & $0.86$ & $0.116 \pm 0.001$ & 67 & 2.053 & $17.32^{+2.68}_{-4.97}$ \\
2013ej & 435 & $-17.72$ & $82.36 \pm 0.66$ & $1.89 \pm 0.34$ & $0.67$ & $0.142 \pm 0.024$ & 75 & 1.742 & $--$ \\
2013ej & 435 & $-17.72$ & $82.36 \pm 0.66$ & $1.89 \pm 0.34$ & $0.67$ & $0.142 \pm 0.024$ & 75 & 1.742 & $--$ \\
2015bs & 421 & $--$ & $139.45 \pm 3.26$ & $0.94 \pm 0.06$ & $0.82$ & $0.256 \pm 0.001$ & 61 & 0.492 & $20.00^{+0.00}_{-6.82}$ \\
2015bs & 421 & $--$ & $139.45 \pm 3.26$ & $0.94 \pm 0.06$ & $0.82$ & $0.256 \pm 0.001$ & 61 & 0.492 & $20.00^{+0.00}_{-6.82}$ \\
2017ivv & 332 & $--$ & $195.82 \pm 4.88$ & $1.05 \pm 0.07$ & $3.79$ & $0.215 \pm 0.004$ & -- & -- & $20.00^{+0.00}_{-5.47}$ \\
2017ivv & 332 & $--$ & $195.82 \pm 4.88$ & $1.05 \pm 0.07$ & $3.79$ & $0.215 \pm 0.004$ & -- & -- & $20.00^{+0.00}_{-5.47}$ \\
2017ivv & 332 & $--$ & $195.82 \pm 4.88$ & $1.05 \pm 0.07$ & $3.79$ & $0.215 \pm 0.004$ & -- & -- & $19.69^{+0.31}_{-5.31}$ \\
\hline
\end{tabular}
\end{table*}

\begin{table*}[ht]
\ContinuedFloat
\centering
\tiny
\caption{Continued.}
\begin{tabular}{lrllllllll}
\hline
SN & Phase & $M_{r,\mathrm{peak}}$ & H I 6563 FWHM & [Ca II]/[O I] & [Ni II]/[Fe II] & $f_{\rm [O\,I]}$ & Plat. dur. & Slope & $M_{\rm ZAMS}$ \\
 & (days) & (mag) & (\AA) & & & & (days) & ($\times 0.01$ mag d$^{-1}$) & (M$_\odot$) \\
\hline

2017ivv & 332 & $--$ & $195.82 \pm 4.88$ & $1.05 \pm 0.07$ & $3.79$ & $0.215 \pm 0.004$ & -- & -- & $19.69^{+0.31}_{-5.31}$ \\
2018zd & 182 & $--$ & $71.66 \pm 0.85$ & $1.58 \pm 0.18$ & $0.86$ & $0.044 \pm 0.001$ & -- & -- & $11.47^{+2.47}_{-2.06}$ \\
2018zd & 228 & $--$ & $65.43 \pm 0.68$ & $1.59 \pm 0.32$ & $1.31$ & $0.042 \pm 0.001$ & -- & -- & $11.47^{+2.47}_{-2.06}$ \\
2018zd & 291 & $--$ & $59.60 \pm 0.52$ & $1.18 \pm 0.15$ & $1.07$ & $0.055 \pm 0.001$ & -- & -- & $11.47^{+2.47}_{-2.06}$ \\
2018zd & 306 & $--$ & $61.82 \pm 0.87$ & $1.17 \pm 0.29$ & $0.76$ & $0.088 \pm 0.015$ & -- & -- & $11.47^{+2.47}_{-2.06}$ \\
2018zd & 315 & $--$ & $59.75 \pm 0.73$ & $1.26 \pm 0.24$ & $0.59$ & $0.085 \pm 0.010$ & -- & -- & $11.47^{+2.47}_{-2.06}$ \\
2018zd & 338 & $--$ & $57.61 \pm 0.66$ & $1.32 \pm 0.19$ & $0.83$ & $0.064 \pm 0.005$ & -- & -- & $11.47^{+2.47}_{-2.06}$ \\
2018zd & 399 & $--$ & $56.21 \pm 0.65$ & $1.45 \pm 0.23$ & $0.70$ & $0.108 \pm 0.003$ & -- & -- & $11.47^{+2.47}_{-2.06}$ \\
ASASSN15oz & 342 & $--$ & $87.36 \pm 1.64$ & $0.90 \pm 0.03$ & $0.87$ & $0.130 \pm 0.001$ & -- & -- & $15.74^{+3.34}_{-4.26}$ \\
1990E & 253 & $--$ & $88.43 \pm 0.87$ & $1.80 \pm 0.08$ & $0.47$ & $0.080 \pm 0.001$ & -- & -- & $16.34^{+3.32}_{-5.14}$ \\
1990E & 269 & $--$ & $90.79 \pm 2.90$ & $1.58 \pm 0.07$ & $0.27$ & $0.092 \pm 0.001$ & -- & -- & $16.34^{+3.32}_{-5.14}$ \\
1990E & 304 & $--$ & $76.83 \pm 0.88$ & $1.89 \pm 0.07$ & $0.58$ & $0.100 \pm 0.001$ & -- & -- & $16.34^{+3.32}_{-5.14}$ \\
1990E & 330 & $--$ & $55.08 \pm 3.92$ & $1.38 \pm 0.05$ & $0.70$ & $0.128 \pm 0.001$ & -- & -- & $16.34^{+3.32}_{-5.14}$ \\
1990E & 330 & $--$ & $65.19 \pm 1.16$ & $1.77 \pm 0.04$ & $0.36$ & $0.122 \pm 0.002$ & -- & -- & $16.34^{+3.32}_{-5.14}$ \\
1990Q & 320 & $--$ & $60.32 \pm 3.50$ & $2.98 \pm 0.37$ & $2.16$ & $0.085 \pm 0.010$ & -- & -- & $13.38^{+3.30}_{-3.15}$ \\
1991G & 355 & $-15.52$ & $64.07 \pm 0.95$ & $1.58 \pm 0.11$ & $0.85$ & $0.070 \pm 0.001$ & 83 & 0.289 & $11.40^{+2.52}_{-2.22}$ \\
1992H & 386 & $-17.92$ & $82.92 \pm 0.79$ & $1.49 \pm 0.29$ & $0.34$ & $0.126 \pm 0.023$ & -- & -- & $--$ \\
1992ad & 225 & $--$ & $114.56 \pm 1.13$ & $2.83 \pm 0.85$ & $0.52$ & $0.056 \pm 0.017$ & -- & -- & $--$ \\
1992ad & 286 & $--$ & $112.05 \pm 1.26$ & $3.72 \pm 2.34$ & $0.64$ & $0.063 \pm 0.040$ & -- & -- & $--$ \\
1992ad & 287 & $--$ & $111.76 \pm 1.26$ & $2.70 \pm 0.90$ & $0.64$ & $0.083 \pm 0.027$ & -- & -- & $--$ \\
1993K & 295 & $--$ & $97.57 \pm 1.62$ & $1.72 \pm 0.11$ & $0.21$ & $0.128 \pm 0.005$ & -- & -- & $17.60^{+2.40}_{-5.34}$ \\
1996W & 252 & $--$ & $69.43 \pm 1.02$ & $2.28 \pm 0.07$ & $1.95$ & $0.063 \pm 0.001$ & -- & -- & $13.66^{+3.69}_{-3.40}$ \\
1996W & 295 & $--$ & $65.30 \pm 1.37$ & $2.41 \pm 0.10$ & $1.09$ & $0.080 \pm 0.002$ & -- & -- & $13.66^{+3.69}_{-3.40}$ \\
1997D & 207 & $-14.75$ & $28.29 \pm 1.71$ & $2.81 \pm 0.68$ & $0.07$ & $0.073 \pm 0.001$ & -- & -- & $15.77^{+3.09}_{-4.95}$ \\
1997D & 350 & $-14.75$ & $22.72 \pm 0.46$ & $1.57 \pm 0.11$ & $0.46$ & $0.131 \pm 0.001$ & -- & -- & $15.77^{+3.09}_{-4.95}$ \\
1999em & 317 & $-17.11$ & $60.71 \pm 0.42$ & $3.05 \pm 0.13$ & $0.76$ & $0.075 \pm 0.001$ & 94 & 0.169 & $12.88^{+2.97}_{-3.03}$ \\
2002hh & 396 & $--$ & $106.28 \pm 1.06$ & $3.58 \pm 0.17$ & $0.55$ & $0.056 \pm 0.001$ & 200 & 0.978 & $9.71^{+1.62}_{-1.52}$ \\
2003B & 275 & $-15.60$ & $43.68 \pm 1.14$ & $1.45 \pm 0.05$ & $0.47$ & $0.119 \pm 0.001$ & 52 & 0.710 & $17.36^{+2.64}_{-4.99}$ \\
2003gd & 130 & $-16.48$ & $59.52 \pm 0.83$ & $1.75 \pm 0.10$ & $2.69$ & $0.040 \pm 0.001$ & -- & -- & $14.04^{+3.31}_{-3.48}$ \\
2003gd & 250 & $-16.48$ & $45.65 \pm 0.44$ & $2.06 \pm 0.11$ & $0.97$ & $0.066 \pm 0.001$ & -- & -- & $14.04^{+3.31}_{-3.48}$ \\
2004A & 286 & $-16.68$ & $46.96 \pm 0.39$ & $1.36 \pm 0.06$ & $0.43$ & $0.075 \pm 0.001$ & 86 & 0.519 & $13.81^{+3.50}_{-3.42}$ \\
2004dj & 136 & $-17.45$ & $69.00 \pm 1.83$ & $1.10 \pm 0.04$ & $0.27$ & $0.045 \pm 0.001$ & 50 & 1.258 & $14.78^{+3.65}_{-3.83}$ \\
2004dj & 136 & $-17.45$ & $69.00 \pm 1.83$ & $1.10 \pm 0.04$ & $0.27$ & $0.045 \pm 0.001$ & 50 & 1.258 & $14.78^{+3.65}_{-3.83}$ \\
2004dj & 164 & $-17.45$ & $62.83 \pm 1.38$ & $1.79 \pm 0.21$ & $0.24$ & $0.033 \pm 0.004$ & 50 & 1.258 & $14.78^{+3.65}_{-3.83}$ \\
2004dj & 164 & $-17.45$ & $62.83 \pm 1.38$ & $1.79 \pm 0.21$ & $0.24$ & $0.033 \pm 0.004$ & 50 & 1.258 & $14.78^{+3.65}_{-3.83}$ \\
2004dj & 199 & $-17.45$ & $62.23 \pm 1.08$ & $1.32 \pm 0.03$ & $0.16$ & $0.055 \pm 0.001$ & 50 & 1.258 & $14.78^{+3.65}_{-3.83}$ \\
2004dj & 199 & $-17.45$ & $62.23 \pm 1.08$ & $1.32 \pm 0.03$ & $0.16$ & $0.055 \pm 0.001$ & 50 & 1.258 & $14.78^{+3.65}_{-3.83}$ \\
2004dj & 226 & $-17.45$ & $58.96 \pm 0.95$ & $1.37 \pm 0.04$ & $0.19$ & $0.067 \pm 0.001$ & 50 & 1.258 & $14.78^{+3.65}_{-3.83}$ \\
2004dj & 226 & $-17.45$ & $58.96 \pm 0.95$ & $1.37 \pm 0.04$ & $0.19$ & $0.067 \pm 0.001$ & 50 & 1.258 & $14.78^{+3.65}_{-3.83}$ \\
2004dj & 253 & $-17.45$ & $58.15 \pm 0.81$ & $1.67 \pm 0.24$ & $0.22$ & $0.064 \pm 0.009$ & 50 & 1.258 & $14.78^{+3.65}_{-3.83}$ \\
2004dj & 253 & $-17.45$ & $58.15 \pm 0.81$ & $1.67 \pm 0.24$ & $0.22$ & $0.064 \pm 0.009$ & 50 & 1.258 & $14.78^{+3.65}_{-3.83}$ \\
2004dj & 290 & $-17.45$ & $58.90 \pm 0.65$ & $1.51 \pm 0.04$ & $0.29$ & $0.090 \pm 0.001$ & 50 & 1.258 & $14.78^{+3.65}_{-3.83}$ \\
2004dj & 290 & $-17.45$ & $58.90 \pm 0.65$ & $1.51 \pm 0.04$ & $0.29$ & $0.090 \pm 0.001$ & 50 & 1.258 & $14.78^{+3.65}_{-3.83}$ \\
2004et & 162 & $-17.34$ & $78.16 \pm 15.54$ & $3.24 \pm 0.08$ & $1.27$ & $0.030 \pm 0.001$ & 100 & 0.657 & $12.95^{+2.90}_{-3.07}$ \\
2004et & 162 & $-17.34$ & $78.16 \pm 15.54$ & $3.24 \pm 0.08$ & $1.27$ & $0.030 \pm 0.001$ & 100 & 0.657 & $12.95^{+2.90}_{-3.07}$ \\
2004et & 162 & $-17.34$ & $78.16 \pm 15.54$ & $3.24 \pm 0.08$ & $1.27$ & $0.030 \pm 0.001$ & 100 & 0.657 & $13.02^{+2.98}_{-3.20}$ \\
2004et & 162 & $-17.34$ & $78.16 \pm 15.54$ & $3.24 \pm 0.08$ & $1.27$ & $0.030 \pm 0.001$ & 100 & 0.657 & $13.02^{+2.98}_{-3.20}$ \\
2004et & 201 & $-17.34$ & $74.40 \pm 23.11$ & $3.38 \pm 0.08$ & $1.70$ & $0.045 \pm 0.001$ & 100 & 0.657 & $12.95^{+2.90}_{-3.07}$ \\
2004et & 201 & $-17.34$ & $74.40 \pm 23.11$ & $3.38 \pm 0.08$ & $1.70$ & $0.045 \pm 0.001$ & 100 & 0.657 & $12.95^{+2.90}_{-3.07}$ \\
2004et & 201 & $-17.34$ & $74.40 \pm 23.11$ & $3.38 \pm 0.08$ & $1.70$ & $0.045 \pm 0.001$ & 100 & 0.657 & $13.02^{+2.98}_{-3.20}$ \\
2004et & 201 & $-17.34$ & $74.40 \pm 23.11$ & $3.38 \pm 0.08$ & $1.70$ & $0.045 \pm 0.001$ & 100 & 0.657 & $13.02^{+2.98}_{-3.20}$ \\
2004et & 226 & $-17.34$ & $69.59 \pm 0.53$ & $3.01 \pm 0.07$ & $1.07$ & $0.051 \pm 0.001$ & 100 & 0.657 & $12.95^{+2.90}_{-3.07}$ \\
2004et & 226 & $-17.34$ & $69.59 \pm 0.53$ & $3.01 \pm 0.07$ & $1.07$ & $0.051 \pm 0.001$ & 100 & 0.657 & $12.95^{+2.90}_{-3.07}$ \\
2004et & 226 & $-17.34$ & $69.59 \pm 0.53$ & $3.01 \pm 0.07$ & $1.07$ & $0.051 \pm 0.001$ & 100 & 0.657 & $13.02^{+2.98}_{-3.20}$ \\
2004et & 226 & $-17.34$ & $69.59 \pm 0.53$ & $3.01 \pm 0.07$ & $1.07$ & $0.051 \pm 0.001$ & 100 & 0.657 & $13.02^{+2.98}_{-3.20}$ \\
2004et & 258 & $-17.34$ & $66.56 \pm 0.49$ & $3.11 \pm 0.09$ & $1.17$ & $0.061 \pm 0.001$ & 100 & 0.657 & $12.95^{+2.90}_{-3.07}$ \\
2004et & 258 & $-17.34$ & $66.56 \pm 0.49$ & $3.11 \pm 0.09$ & $1.17$ & $0.061 \pm 0.001$ & 100 & 0.657 & $12.95^{+2.90}_{-3.07}$ \\
2004et & 258 & $-17.34$ & $66.56 \pm 0.49$ & $3.11 \pm 0.09$ & $1.17$ & $0.061 \pm 0.001$ & 100 & 0.657 & $13.02^{+2.98}_{-3.20}$ \\
2004et & 258 & $-17.34$ & $66.56 \pm 0.49$ & $3.11 \pm 0.09$ & $1.17$ & $0.061 \pm 0.001$ & 100 & 0.657 & $13.02^{+2.98}_{-3.20}$ \\
2004et & 282 & $-17.34$ & $66.02 \pm 0.59$ & $3.24 \pm 0.11$ & $1.22$ & $0.073 \pm 0.001$ & 100 & 0.657 & $12.95^{+2.90}_{-3.07}$ \\
2004et & 282 & $-17.34$ & $66.02 \pm 0.59$ & $3.24 \pm 0.11$ & $1.22$ & $0.073 \pm 0.001$ & 100 & 0.657 & $12.95^{+2.90}_{-3.07}$ \\
2004et & 282 & $-17.34$ & $66.02 \pm 0.59$ & $3.24 \pm 0.11$ & $1.22$ & $0.073 \pm 0.001$ & 100 & 0.657 & $13.02^{+2.98}_{-3.20}$ \\
2004et & 282 & $-17.34$ & $66.02 \pm 0.59$ & $3.24 \pm 0.11$ & $1.22$ & $0.073 \pm 0.001$ & 100 & 0.657 & $13.02^{+2.98}_{-3.20}$ \\
2004et & 313 & $-17.34$ & $64.68 \pm 0.50$ & $1.73 \pm 0.04$ & $0.79$ & $0.099 \pm 0.001$ & 100 & 0.657 & $12.95^{+2.90}_{-3.07}$ \\
2004et & 313 & $-17.34$ & $64.68 \pm 0.50$ & $1.73 \pm 0.04$ & $0.79$ & $0.099 \pm 0.001$ & 100 & 0.657 & $12.95^{+2.90}_{-3.07}$ \\
2004et & 313 & $-17.34$ & $64.68 \pm 0.50$ & $1.73 \pm 0.04$ & $0.79$ & $0.099 \pm 0.001$ & 100 & 0.657 & $13.02^{+2.98}_{-3.20}$ \\
2004et & 313 & $-17.34$ & $64.68 \pm 0.50$ & $1.73 \pm 0.04$ & $0.79$ & $0.099 \pm 0.001$ & 100 & 0.657 & $13.02^{+2.98}_{-3.20}$ \\
2004et & 353 & $-17.34$ & $63.53 \pm 0.52$ & $3.33 \pm 0.09$ & $1.01$ & $0.092 \pm 0.001$ & 100 & 0.657 & $12.95^{+2.90}_{-3.07}$ \\
2004et & 353 & $-17.34$ & $63.53 \pm 0.52$ & $3.33 \pm 0.09$ & $1.01$ & $0.092 \pm 0.001$ & 100 & 0.657 & $12.95^{+2.90}_{-3.07}$ \\
2004et & 353 & $-17.34$ & $63.53 \pm 0.52$ & $3.33 \pm 0.09$ & $1.01$ & $0.092 \pm 0.001$ & 100 & 0.657 & $13.02^{+2.98}_{-3.20}$ \\
\hline
\end{tabular}
\end{table*}

\begin{table*}[ht]
\ContinuedFloat
\centering
\tiny
\caption{Continued.}
\begin{tabular}{lrllllllll}
\hline
SN & Phase & $M_{r,\mathrm{peak}}$ & H I 6563 FWHM & [Ca II]/[O I] & [Ni II]/[Fe II] & $f_{\rm [O\,I]}$ & Plat. dur. & Slope & $M_{\rm ZAMS}$ \\
 & (days) & (mag) & (\AA) & & & & (days) & ($\times 0.01$ mag d$^{-1}$) & (M$_\odot$) \\
\hline

2004et & 353 & $-17.34$ & $63.53 \pm 0.52$ & $3.33 \pm 0.09$ & $1.01$ & $0.092 \pm 0.001$ & 100 & 0.657 & $13.02^{+2.98}_{-3.20}$ \\
2004et & 354 & $-17.34$ & $63.53 \pm 0.52$ & $3.33 \pm 0.13$ & $1.01$ & $0.092 \pm 0.003$ & 100 & 0.657 & $12.95^{+2.90}_{-3.07}$ \\
2004et & 354 & $-17.34$ & $63.53 \pm 0.52$ & $3.33 \pm 0.13$ & $1.01$ & $0.092 \pm 0.003$ & 100 & 0.657 & $12.95^{+2.90}_{-3.07}$ \\
2004et & 354 & $-17.34$ & $63.53 \pm 0.52$ & $3.33 \pm 0.13$ & $1.01$ & $0.092 \pm 0.003$ & 100 & 0.657 & $13.02^{+2.98}_{-3.20}$ \\
2004et & 354 & $-17.34$ & $63.53 \pm 0.52$ & $3.33 \pm 0.13$ & $1.01$ & $0.092 \pm 0.003$ & 100 & 0.657 & $13.02^{+2.98}_{-3.20}$ \\
2005ay & 285 & $-16.27$ & $49.57 \pm 0.58$ & $1.24 \pm 0.08$ & $0.94$ & $0.078 \pm 0.001$ & 91 & 0.154 & $14.15^{+3.41}_{-3.09}$ \\
2005cs & 304 & $-15.66$ & $38.69 \pm 0.94$ & $2.76 \pm 0.72$ & $1.50$ & $0.083 \pm 0.020$ & 113 & 0.512 & $11.21^{+2.17}_{-1.98}$ \\
2005cs & 334 & $-15.66$ & $37.38 \pm 0.55$ & $4.52 \pm 0.23$ & $1.34$ & $0.058 \pm 0.001$ & 113 & 0.512 & $11.21^{+2.17}_{-1.98}$ \\
2007aa & 384 & $--$ & $53.87 \pm 0.57$ & $1.35 \pm 0.05$ & $0.88$ & $0.153 \pm 0.001$ & -- & -- & $15.40^{+4.18}_{-4.03}$ \\
2007it & 268 & $--$ & $--$ & $--$ & $--$ & $--$ & -- & -- & $--$ \\
2007it & 268 & $-15.00$ & $--$ & $--$ & $--$ & $--$ & -- & -- & $--$ \\
2007it & 268 & $--$ & $73.08 \pm 1.08$ & $0.76 \pm 0.04$ & $1.39$ & $--$ & -- & -- & $--$ \\
2007it & 268 & $-15.00$ & $73.08 \pm 1.08$ & $0.76 \pm 0.04$ & $1.39$ & $--$ & -- & -- & $--$ \\
2008cn & 340 & $--$ & $77.53 \pm 0.90$ & $1.78 \pm 0.06$ & $0.71$ & $0.115 \pm 0.002$ & -- & -- & $15.00^{+3.37}_{-4.83}$ \\
2008ex & 285 & $--$ & $76.37 \pm 1.12$ & $1.18 \pm 0.16$ & $0.59$ & $0.081 \pm 0.010$ & -- & -- & $14.64^{+3.16}_{-3.81}$ \\
2009N & 411 & $-16.25$ & $30.40 \pm 0.44$ & $3.60 \pm 0.42$ & $1.16$ & $0.076 \pm 0.009$ & 83 & 0.407 & $10.41^{+1.89}_{-2.00}$ \\
2009dd & 232 & $-17.63$ & $74.10 \pm 1.11$ & $1.52 \pm 0.04$ & $0.59$ & $0.055 \pm 0.001$ & 80 & 0.231 & $13.54^{+2.82}_{-2.72}$ \\
2009dd & 408 & $-17.63$ & $34.30 \pm 1.64$ & $2.05 \pm 0.39$ & $0.64$ & $0.046 \pm 0.008$ & 80 & 0.231 & $13.54^{+2.82}_{-2.72}$ \\
2009ib & 219 & $-16.49$ & $53.75 \pm 0.80$ & $1.19 \pm 0.06$ & $0.60$ & $0.045 \pm 0.001$ & 116 & 0.377 & $17.85^{+2.15}_{-5.31}$ \\
2009ib & 262 & $-16.49$ & $55.94 \pm 0.97$ & $0.53 \pm 0.03$ & $1.11$ & $0.111 \pm 0.001$ & 116 & 0.377 & $17.85^{+2.15}_{-5.31}$ \\
2012A & 86 & $-16.37$ & $85.49 \pm 5.57$ & $--$ & $--$ & $0.019 \pm 0.001$ & 84 & 1.088 & $12.46^{+2.81}_{-3.00}$ \\
2012A & 86 & $-16.37$ & $85.37 \pm 9.64$ & $--$ & $--$ & $0.023 \pm 0.001$ & 84 & 1.088 & $12.46^{+2.81}_{-3.00}$ \\
2012A & 408 & $-16.37$ & $44.84 \pm 0.22$ & $1.82 \pm 0.21$ & $0.35$ & $0.118 \pm 0.013$ & 84 & 1.088 & $12.46^{+2.81}_{-3.00}$ \\
2012aw & 369 & $-17.03$ & $59.56 \pm 0.69$ & $1.24 \pm 0.06$ & $0.71$ & $0.124 \pm 0.001$ & 102 & 0.361 & $14.32^{+3.28}_{-3.29}$ \\
2012ch & 357 & $--$ & $77.45 \pm 1.71$ & $3.38 \pm 0.87$ & $1.48$ & $0.085 \pm 0.022$ & -- & -- & $--$ \\
2012ec & 403 & $-16.93$ & $58.09 \pm 0.60$ & $0.62 \pm 0.08$ & $1.27$ & $0.182 \pm 0.010$ & 85 & 0.629 & $16.03^{+3.97}_{-4.59}$ \\
2012ho & 168 & $--$ & $90.14 \pm 15.75$ & $3.98 \pm 0.16$ & $0.95$ & $0.028 \pm 0.001$ & -- & -- & $14.33^{+3.16}_{-3.88}$ \\
2012ho & 229 & $--$ & $76.47 \pm 1.21$ & $3.50 \pm 0.11$ & $0.65$ & $0.043 \pm 0.001$ & -- & -- & $14.33^{+3.16}_{-3.88}$ \\
2012ho & 249 & $--$ & $74.21 \pm 1.17$ & $2.44 \pm 0.08$ & $0.76$ & $0.069 \pm 0.001$ & -- & -- & $14.33^{+3.16}_{-3.88}$ \\
2012ho & 252 & $--$ & $74.98 \pm 0.48$ & $2.62 \pm 0.74$ & $0.60$ & $0.067 \pm 0.019$ & -- & -- & $14.33^{+3.16}_{-3.88}$ \\
2012ho & 265 & $--$ & $71.73 \pm 1.11$ & $2.44 \pm 0.07$ & $0.59$ & $0.071 \pm 0.001$ & -- & -- & $14.33^{+3.16}_{-3.88}$ \\
2012ho & 318 & $--$ & $77.70 \pm 1.01$ & $3.24 \pm 0.48$ & $0.51$ & $0.091 \pm 0.012$ & -- & -- & $14.33^{+3.16}_{-3.88}$ \\
2012ho & 318 & $--$ & $77.70 \pm 1.01$ & $3.17 \pm 0.47$ & $0.51$ & $0.093 \pm 0.012$ & -- & -- & $14.33^{+3.16}_{-3.88}$ \\
2013am & 281 & $-16.17$ & $35.86 \pm 0.93$ & $--$ & $0.73$ & $--$ & 81 & 0.111 & $--$ \\
2013by & 160 & $-18.76$ & $135.53 \pm 8.84$ & $0.60 \pm 0.02$ & $157.61$ & $0.061 \pm 0.001$ & 67 & 2.053 & $17.86^{+2.14}_{-5.50}$ \\
2013by & 160 & $-18.76$ & $135.53 \pm 8.84$ & $0.60 \pm 0.02$ & $157.61$ & $0.061 \pm 0.001$ & 67 & 2.053 & $17.86^{+2.14}_{-5.50}$ \\
2013by & 160 & $-18.76$ & $135.53 \pm 8.84$ & $0.60 \pm 0.02$ & $157.61$ & $0.061 \pm 0.001$ & 67 & 2.053 & $17.32^{+2.68}_{-4.97}$ \\
2013by & 160 & $-18.76$ & $135.53 \pm 8.84$ & $0.60 \pm 0.02$ & $157.61$ & $0.061 \pm 0.001$ & 67 & 2.053 & $17.32^{+2.68}_{-4.97}$ \\
2013by & 292 & $-18.76$ & $110.23 \pm 1.43$ & $0.90 \pm 0.03$ & $199.45$ & $0.124 \pm 0.003$ & 67 & 2.053 & $17.86^{+2.14}_{-5.50}$ \\
2013by & 292 & $-18.76$ & $110.23 \pm 1.43$ & $0.90 \pm 0.03$ & $199.45$ & $0.124 \pm 0.003$ & 67 & 2.053 & $17.86^{+2.14}_{-5.50}$ \\
2013by & 292 & $-18.76$ & $110.23 \pm 1.43$ & $0.90 \pm 0.03$ & $199.45$ & $0.124 \pm 0.003$ & 67 & 2.053 & $17.32^{+2.68}_{-4.97}$ \\
2013by & 292 & $-18.76$ & $110.23 \pm 1.43$ & $0.90 \pm 0.03$ & $199.45$ & $0.124 \pm 0.003$ & 67 & 2.053 & $17.32^{+2.68}_{-4.97}$ \\
2013fs & 270 & $-17.63$ & $107.45 \pm 0.78$ & $1.46 \pm 0.05$ & $1.56$ & $0.140 \pm 0.001$ & 69 & 1.174 & $19.93^{+0.07}_{-6.06}$ \\
2014G & 137 & $-18.64$ & $133.74 \pm 1.57$ & $1.80 \pm 0.10$ & $1.32$ & $0.034 \pm 0.001$ & 62 & 2.194 & $18.28^{+1.72}_{-5.15}$ \\
2014G & 187 & $-18.64$ & $115.16 \pm 1.47$ & $1.46 \pm 0.08$ & $2.30$ & $0.067 \pm 0.001$ & 62 & 2.194 & $18.28^{+1.72}_{-5.15}$ \\
2014G & 341 & $-18.64$ & $123.77 \pm 2.91$ & $1.12 \pm 0.07$ & $2.52$ & $0.177 \pm 0.001$ & 62 & 2.194 & $18.28^{+1.72}_{-5.15}$ \\
2014cx & 350 & $-17.33$ & $78.76 \pm 2.82$ & $0.76 \pm 0.15$ & $0.68$ & $0.196 \pm 0.035$ & 82 & 0.586 & $--$ \\
2015bs & 420 & $--$ & $139.59 \pm 3.26$ & $0.93 \pm 0.18$ & $0.82$ & $0.238 \pm 0.043$ & 61 & 0.492 & $20.00^{+0.00}_{-6.82}$ \\
2015bs & 420 & $--$ & $139.59 \pm 3.26$ & $0.93 \pm 0.18$ & $0.82$ & $0.238 \pm 0.043$ & 61 & 0.492 & $20.00^{+0.00}_{-6.82}$ \\
2016X & 340 & $-17.10$ & $114.37 \pm 3.39$ & $--$ & $--$ & $--$ & 67 & 1.311 & $--$ \\
2016aqf & 202 & $--$ & $48.05 \pm 0.75$ & $1.62 \pm 0.10$ & $0.40$ & $0.034 \pm 0.001$ & 69 & 1.601 & $12.79^{+2.87}_{-3.08}$ \\
2016aqf & 259 & $--$ & $44.80 \pm 0.58$ & $1.46 \pm 0.07$ & $0.31$ & $0.043 \pm 0.001$ & 69 & 1.601 & $12.79^{+2.87}_{-3.08}$ \\
2016aqf & 285 & $--$ & $41.46 \pm 1.29$ & $1.75 \pm 0.20$ & $0.58$ & $0.063 \pm 0.001$ & 69 & 1.601 & $12.79^{+2.87}_{-3.08}$ \\
2016aqf & 303 & $--$ & $45.15 \pm 0.83$ & $1.26 \pm 0.08$ & $0.65$ & $0.082 \pm 0.001$ & 69 & 1.601 & $12.79^{+2.87}_{-3.08}$ \\
2016aqf & 330 & $--$ & $43.26 \pm 0.54$ & $1.18 \pm 0.08$ & $0.40$ & $0.069 \pm 0.001$ & 69 & 1.601 & $12.79^{+2.87}_{-3.08}$ \\
2016aqf & 351 & $--$ & $45.22 \pm 1.01$ & $1.23 \pm 0.11$ & $0.30$ & $0.088 \pm 0.001$ & 69 & 1.601 & $12.79^{+2.87}_{-3.08}$ \\
2016bkv & 257 & $-16.00$ & $25.34 \pm 0.43$ & $1.49 \pm 0.08$ & $0.44$ & $0.094 \pm 0.001$ & -- & -- & $9.08^{+1.26}_{-1.08}$ \\
2016bkv & 436 & $-16.00$ & $21.35 \pm 0.18$ & $5.27 \pm 0.32$ & $0.62$ & $0.059 \pm 0.001$ & -- & -- & $9.08^{+1.26}_{-1.08}$ \\
2017eaw & 250 & $-17.91$ & $56.08 \pm 0.57$ & $2.93 \pm 0.10$ & $0.84$ & $0.065 \pm 0.002$ & 95 & 0.645 & $13.83^{+3.20}_{-3.23}$ \\
2017ivv & 110 & $--$ & $147.60 \pm 16.67$ & $1.66 \pm 0.07$ & $32.78$ & $0.058 \pm 0.001$ & -- & -- & $20.00^{+0.00}_{-5.47}$ \\
2017ivv & 110 & $--$ & $147.60 \pm 16.67$ & $1.66 \pm 0.07$ & $32.78$ & $0.058 \pm 0.001$ & -- & -- & $20.00^{+0.00}_{-5.47}$ \\
2017ivv & 110 & $--$ & $147.60 \pm 16.67$ & $1.66 \pm 0.07$ & $32.78$ & $0.058 \pm 0.001$ & -- & -- & $19.69^{+0.31}_{-5.31}$ \\
2017ivv & 110 & $--$ & $147.60 \pm 16.67$ & $1.66 \pm 0.07$ & $32.78$ & $0.058 \pm 0.001$ & -- & -- & $19.69^{+0.31}_{-5.31}$ \\
2017ivv & 136 & $--$ & $139.78 \pm 66.47$ & $1.36 \pm 0.04$ & $36.96$ & $0.086 \pm 0.001$ & -- & -- & $20.00^{+0.00}_{-5.47}$ \\
2017ivv & 136 & $--$ & $139.78 \pm 66.47$ & $1.36 \pm 0.04$ & $36.96$ & $0.086 \pm 0.001$ & -- & -- & $20.00^{+0.00}_{-5.47}$ \\
2017ivv & 136 & $--$ & $139.78 \pm 66.47$ & $1.36 \pm 0.04$ & $36.96$ & $0.086 \pm 0.001$ & -- & -- & $19.69^{+0.31}_{-5.31}$ \\
2017ivv & 136 & $--$ & $139.78 \pm 66.47$ & $1.36 \pm 0.04$ & $36.96$ & $0.086 \pm 0.001$ & -- & -- & $19.69^{+0.31}_{-5.31}$ \\
2017ivv & 158 & $--$ & $141.56 \pm 42.49$ & $1.35 \pm 0.05$ & $753.63$ & $0.109 \pm 0.001$ & -- & -- & $20.00^{+0.00}_{-5.47}$ \\
2017ivv & 158 & $--$ & $141.56 \pm 42.49$ & $1.35 \pm 0.05$ & $753.63$ & $0.109 \pm 0.001$ & -- & -- & $20.00^{+0.00}_{-5.47}$ \\
2017ivv & 158 & $--$ & $141.56 \pm 42.49$ & $1.35 \pm 0.05$ & $753.63$ & $0.109 \pm 0.001$ & -- & -- & $19.69^{+0.31}_{-5.31}$ \\
\hline
\end{tabular}
\end{table*}

\begin{table*}[ht]
\ContinuedFloat
\centering
\tiny
\caption{Continued.}
\begin{tabular}{lrllllllll}
\hline
SN & Phase & $M_{r,\mathrm{peak}}$ & H I 6563 FWHM & [Ca II]/[O I] & [Ni II]/[Fe II] & $f_{\rm [O\,I]}$ & Plat. dur. & Slope & $M_{\rm ZAMS}$ \\
 & (days) & (mag) & (\AA) & & & & (days) & ($\times 0.01$ mag d$^{-1}$) & (M$_\odot$) \\
\hline

2017ivv & 158 & $--$ & $141.56 \pm 42.49$ & $1.35 \pm 0.05$ & $753.63$ & $0.109 \pm 0.001$ & -- & -- & $19.69^{+0.31}_{-5.31}$ \\
2017ivv & 243 & $--$ & $158.72 \pm 3.95$ & $1.72 \pm 0.06$ & $6.47$ & $0.144 \pm 0.002$ & -- & -- & $20.00^{+0.00}_{-5.47}$ \\
2017ivv & 243 & $--$ & $158.72 \pm 3.95$ & $1.72 \pm 0.06$ & $6.47$ & $0.144 \pm 0.002$ & -- & -- & $20.00^{+0.00}_{-5.47}$ \\
2017ivv & 243 & $--$ & $158.72 \pm 3.95$ & $1.72 \pm 0.06$ & $6.47$ & $0.144 \pm 0.002$ & -- & -- & $19.69^{+0.31}_{-5.31}$ \\
2017ivv & 243 & $--$ & $158.72 \pm 3.95$ & $1.72 \pm 0.06$ & $6.47$ & $0.144 \pm 0.002$ & -- & -- & $19.69^{+0.31}_{-5.31}$ \\
2017ivv & 279 & $--$ & $172.71 \pm 5.64$ & $1.81 \pm 0.08$ & $5.15$ & $0.157 \pm 0.001$ & -- & -- & $20.00^{+0.00}_{-5.47}$ \\
2017ivv & 279 & $--$ & $172.71 \pm 5.64$ & $1.81 \pm 0.08$ & $5.15$ & $0.157 \pm 0.001$ & -- & -- & $20.00^{+0.00}_{-5.47}$ \\
2017ivv & 279 & $--$ & $172.71 \pm 5.64$ & $1.81 \pm 0.08$ & $5.15$ & $0.157 \pm 0.001$ & -- & -- & $19.69^{+0.31}_{-5.31}$ \\
2017ivv & 279 & $--$ & $172.71 \pm 5.64$ & $1.81 \pm 0.08$ & $5.15$ & $0.157 \pm 0.001$ & -- & -- & $19.69^{+0.31}_{-5.31}$ \\
2017ivv & 325 & $--$ & $185.67 \pm 5.36$ & $1.41 \pm 0.07$ & $2.07$ & $0.192 \pm 0.001$ & -- & -- & $20.00^{+0.00}_{-5.47}$ \\
2017ivv & 325 & $--$ & $185.67 \pm 5.36$ & $1.41 \pm 0.07$ & $2.07$ & $0.192 \pm 0.001$ & -- & -- & $20.00^{+0.00}_{-5.47}$ \\
2017ivv & 325 & $--$ & $185.67 \pm 5.36$ & $1.41 \pm 0.07$ & $2.07$ & $0.192 \pm 0.001$ & -- & -- & $19.69^{+0.31}_{-5.31}$ \\
2017ivv & 325 & $--$ & $185.67 \pm 5.36$ & $1.41 \pm 0.07$ & $2.07$ & $0.192 \pm 0.001$ & -- & -- & $19.69^{+0.31}_{-5.31}$ \\
2017ivv & 332 & $--$ & $194.44 \pm 10.72$ & $1.40 \pm 0.09$ & $2.88$ & $0.189 \pm 0.009$ & -- & -- & $20.00^{+0.00}_{-5.47}$ \\
2017ivv & 332 & $--$ & $194.44 \pm 10.72$ & $1.40 \pm 0.09$ & $2.88$ & $0.189 \pm 0.009$ & -- & -- & $20.00^{+0.00}_{-5.47}$ \\
2017ivv & 332 & $--$ & $194.44 \pm 10.72$ & $1.40 \pm 0.09$ & $2.88$ & $0.189 \pm 0.009$ & -- & -- & $19.69^{+0.31}_{-5.31}$ \\
2017ivv & 332 & $--$ & $194.44 \pm 10.72$ & $1.40 \pm 0.09$ & $2.88$ & $0.189 \pm 0.009$ & -- & -- & $19.69^{+0.31}_{-5.31}$ \\
2018cuf & 335 & $-16.82$ & $77.90 \pm 0.85$ & $--$ & $0.53$ & $--$ & 72 & 0.396 & $--$ \\
2018gj & 270 & $-17.69$ & $93.68 \pm 1.91$ & $4.49 \pm 0.97$ & $3357.64$ & $0.040 \pm 0.009$ & 65 & 1.214 & $--$ \\
2018hwm & 385 & $-15.27$ & $34.46 \pm 1.92$ & $3.33 \pm 0.39$ & $0.69$ & $0.042 \pm 0.003$ & 129 & 0.351 & $8.97^{+1.18}_{-0.97}$ \\
2018is & 386 & $-15.29$ & $47.01 \pm 3.42$ & $8.69 \pm 1.25$ & $0.30$ & $0.085 \pm 0.002$ & 84 & 0.650 & $11.67^{+2.31}_{-2.55}$ \\
2020jfo & 217 & $-18.02$ & $75.48 \pm 1.14$ & $2.15 \pm 0.17$ & $1.44$ & $0.043 \pm 0.001$ & 49 & 1.127 & $15.26^{+3.28}_{-4.08}$ \\
2020jfo & 256 & $-18.02$ & $74.45 \pm 1.10$ & $1.45 \pm 0.17$ & $1.31$ & $0.081 \pm 0.007$ & 49 & 1.127 & $15.26^{+3.28}_{-4.08}$ \\
2020jfo & 280 & $-18.02$ & $73.19 \pm 1.08$ & $1.70 \pm 0.13$ & $1.58$ & $0.082 \pm 0.003$ & 49 & 1.127 & $15.26^{+3.28}_{-4.08}$ \\
2020jfo & 308 & $-18.02$ & $73.21 \pm 1.25$ & $1.73 \pm 0.12$ & $1.43$ & $0.096 \pm 0.001$ & 49 & 1.127 & $15.26^{+3.28}_{-4.08}$ \\
2020jfo & 351 & $-18.02$ & $73.88 \pm 1.62$ & $1.46 \pm 0.12$ & $1.57$ & $0.127 \pm 0.003$ & 49 & 1.127 & $15.26^{+3.28}_{-4.08}$ \\
2021dbg & 352 & $--$ & $109.14 \pm 2.28$ & $1.81 \pm 0.11$ & $0.80$ & $0.100 \pm 0.002$ & -- & -- & $13.96^{+3.00}_{-3.71}$ \\
2021gmj & 385 & $-15.85$ & $39.05 \pm 0.77$ & $4.26 \pm 0.97$ & $0.41$ & $0.065 \pm 0.013$ & 86 & 0.590 & $--$ \\
2022jox & 240 & $--$ & $--$ & $--$ & $--$ & $--$ & -- & -- & $--$ \\
2022jox & 240 & $--$ & $65.25 \pm 1.51$ & $0.55 \pm 0.04$ & $0.60$ & $--$ & -- & -- & $--$ \\
2023ixf & 260 & $-17.70$ & $116.06 \pm 1.30$ & $1.66 \pm 0.05$ & $1.80$ & $0.114 \pm 0.001$ & 61 & 2.282 & $18.17^{+1.83}_{-5.28}$ \\
ZTF19abqhobb & 163 & $-17.59$ & $122.56 \pm 5.84$ & $1.29 \pm 0.17$ & $0.40$ & $0.035 \pm 0.001$ & -- & -- & $--$ \\
ZTF20aaynrrh & 216 & $-17.48$ & $75.49 \pm 1.14$ & $2.14 \pm 0.21$ & $1.44$ & $0.043 \pm 0.001$ & -- & -- & $15.23^{+3.36}_{-4.01}$ \\
ZTF20aaynrrh & 256 & $-17.48$ & $74.47 \pm 1.10$ & $1.39 \pm 0.14$ & $1.32$ & $0.083 \pm 0.001$ & -- & -- & $15.23^{+3.36}_{-4.01}$ \\
ZTF20aaynrrh & 280 & $-17.48$ & $73.19 \pm 1.08$ & $1.74 \pm 0.14$ & $1.59$ & $0.081 \pm 0.001$ & -- & -- & $15.23^{+3.36}_{-4.01}$ \\
ZTF20aaynrrh & 308 & $-17.48$ & $73.23 \pm 1.25$ & $1.75 \pm 0.12$ & $1.43$ & $0.095 \pm 0.003$ & -- & -- & $15.23^{+3.36}_{-4.01}$ \\
ZTF20aaynrrh & 351 & $-17.48$ & $73.88 \pm 1.62$ & $1.36 \pm 0.11$ & $1.57$ & $0.126 \pm 0.003$ & -- & -- & $15.23^{+3.36}_{-4.01}$ \\
ZTF20abotkfn & 141 & $-17.10$ & $65.65 \pm 1.28$ & $14.49 \pm 1.72$ & $0.46$ & $0.028 \pm 0.001$ & -- & -- & $13.77^{+2.82}_{-2.60}$ \\
ZTF20abotkfn & 211 & $-17.10$ & $59.81 \pm 0.72$ & $1.49 \pm 0.09$ & $0.50$ & $0.056 \pm 0.001$ & -- & -- & $13.77^{+2.82}_{-2.60}$ \\
ZTF21abouuat & 276 & $-17.56$ & $81.06 \pm 0.76$ & $1.47 \pm 0.31$ & $0.37$ & $0.099 \pm 0.020$ & 81 & 0.271 & $--$ \\
ZTF23aaquhaz & 138 & $-17.53$ & $113.78 \pm 14.68$ & $0.84 \pm 0.14$ & $0.66$ & $0.070 \pm 0.001$ & -- & -- & $14.98^{+3.65}_{-4.13}$ \\
ZTF23aaquhaz & 310 & $-17.53$ & $109.72 \pm 1.08$ & $1.88 \pm 0.30$ & $0.83$ & $0.097 \pm 0.015$ & -- & -- & $14.98^{+3.65}_{-4.13}$ \\
ZTF23absdcgi & 337 & $-16.73$ & $103.48 \pm 4.32$ & $1.24 \pm 0.07$ & $0.53$ & $0.167 \pm 0.005$ & -- & -- & $17.97^{+2.03}_{-4.94}$ \\
ZTF23abvgvab & 104 & $-17.12$ & $115.04 \pm 1.12$ & $1.49 \pm 0.17$ & $0.79$ & $0.079 \pm 0.001$ & -- & -- & $--$ \\
\hline
\end{tabular}
\end{table*}

%% file: mzams_threecol_table.tex
\begin{table*}[ht]
\centering
\scriptsize
\caption{Per-spectrum ZAMS mass estimates from the GP inversion. 
Uncertainties include a 10\% systematic added in quadrature.}
\label{tab:mzams_threecol}

\begin{minipage}{0.30\linewidth}
\centering
\begin{tabular}{lcc}
\toprule
SN & Phase & $M_{\rm ZAMS}$ \\
 & (days) & ($M_\odot$) \\
\midrule
SN1998S & 143 & $12.04^{+1.29}_{-1.20}$ \\
SN1998S & 314 & $8.92^{+0.78}_{-0.72}$ \\
SN2004dj & 274 & $14.86^{+2.95}_{-3.87}$ \\
SN2004et & 355 & $13.02^{+2.98}_{-3.20}$ \\
SN2013by & 287 & $17.32^{+2.68}_{-4.97}$ \\
SN2013ej & 435 & $12.92^{+3.68}_{-3.57}$ \\
SN2015bs & 421 & $20.00^{+0.00}_{-6.82}$ \\
SN2017ivv & 332 & $20.00^{+0.00}_{-5.47}$ \\
SN2018zd & 182 & $13.25^{+2.33}_{-2.20}$ \\
SN2018zd & 228 & $12.00^{+2.30}_{-1.81}$ \\
SN2018zd & 291 & $11.77^{+2.43}_{-2.29}$ \\
SN2018zd & 306 & $14.45^{+3.37}_{-3.71}$ \\
SN2018zd & 315 & $13.67^{+3.18}_{-3.36}$ \\
SN2018zd & 338 & $11.47^{+2.47}_{-2.06}$ \\
SN2018zd & 399 & $12.42^{+2.31}_{-2.87}$ \\
ASASSN15oz & 342 & $15.74^{+3.34}_{-4.26}$ \\
SN1990E & 253 & $14.73^{+3.81}_{-3.27}$ \\
SN1990E & 269 & $16.35^{+3.65}_{-4.33}$ \\
SN1990E & 304 & $15.70^{+3.18}_{-4.34}$ \\
SN1990E & 330 & $16.34^{+3.32}_{-5.14}$ \\
SN1990E & 330 & $16.04^{+3.49}_{-4.84}$ \\
SN1990E & 330 & $16.34^{+3.32}_{-5.14}$ \\
SN1990E & 330 & $16.04^{+3.49}_{-4.84}$ \\
SN1990Q & 320 & $13.38^{+3.30}_{-3.15}$ \\
SN1991G & 355 & $11.40^{+2.52}_{-2.22}$ \\
SN1992H & 386 & $13.67^{+3.45}_{-3.69}$ \\
SN1992ad & 225 & $13.40^{+2.83}_{-2.52}$ \\
SN1992ad & 286 & $12.84^{+3.46}_{-3.47}$ \\
SN1992ad & 287 & $14.81^{+3.32}_{-3.81}$ \\
SN1993K & 295 & $17.60^{+2.40}_{-5.34}$ \\
SN1996W & 252 & $13.73^{+3.30}_{-3.38}$ \\
SN1996W & 295 & $13.66^{+3.69}_{-3.40}$ \\
SN1997D & 207 & $15.97^{+4.03}_{-3.90}$ \\
SN1997D & 350 & $15.77^{+3.09}_{-4.95}$ \\
SN1999em & 317 & $12.88^{+2.97}_{-3.03}$ \\
SN2002hh & 396 & $9.71^{+1.62}_{-1.52}$ \\
SN2003B & 275 & $17.36^{+2.64}_{-4.99}$ \\
SN2003gd & 130 & $14.12^{+2.41}_{-2.12}$ \\
SN2003gd & 250 & $14.04^{+3.31}_{-3.48}$ \\
SN2004A & 286 & $13.81^{+3.50}_{-3.42}$ \\
SN2004dj & 136 & $14.42^{+2.98}_{-2.68}$ \\
SN2004dj & 164 & $12.43^{+1.84}_{-1.53}$ \\
SN2004dj & 199 & $14.00^{+3.18}_{-2.90}$ \\
SN2004dj & 226 & $14.63^{+3.39}_{-3.18}$ \\
SN2004dj & 253 & $14.07^{+3.14}_{-3.09}$ \\
SN2004dj & 290 & $14.78^{+3.65}_{-3.83}$ \\
SN2004et & 162 & $12.18^{+1.52}_{-1.37}$ \\
SN2004et & 201 & $13.13^{+2.39}_{-2.05}$ \\
SN2004et & 226 & $12.96^{+2.56}_{-2.70}$ \\
SN2004et & 258 & $13.20^{+3.05}_{-2.39}$ \\
SN2004et & 282 & $14.01^{+2.99}_{-3.61}$ \\
SN2004et & 313 & $14.97^{+3.37}_{-4.31}$ \\
SN2004et & 353 & $12.95^{+2.90}_{-3.07}$ \\
SN2004et & 354 & $12.66^{+3.13}_{-2.88}$ \\
SN2005ay & 285 & $14.15^{+3.41}_{-3.09}$ \\
SN2005cs & 304 & $13.73^{+3.45}_{-3.49}$ \\
SN2005cs & 334 & $11.21^{+2.17}_{-1.98}$ \\
SN2007aa & 384 & $15.40^{+4.18}_{-4.03}$ \\
\bottomrule
\end{tabular}
\end{minipage}
\hspace{0.015\linewidth}
\begin{minipage}{0.30\linewidth}
\centering
\begin{tabular}{lcc}
\toprule
SN & Phase & $M_{\rm ZAMS}$ \\
 & (days) & ($M_\odot$) \\
\midrule
SN2008cn & 340 & $15.00^{+3.37}_{-4.83}$ \\
SN2008ex & 285 & $14.64^{+3.16}_{-3.81}$ \\
SN2009N & 411 & $10.41^{+1.89}_{-2.00}$ \\
SN2009dd & 232 & $13.54^{+2.82}_{-2.72}$ \\
SN2009dd & 408 & $8.78^{+1.27}_{-0.78}$ \\
SN2009ib & 219 & $12.62^{+2.58}_{-2.31}$ \\
SN2009ib & 262 & $17.85^{+2.15}_{-5.31}$ \\
SN2012A & 86 & $12.63^{+1.47}_{-1.11}$ \\
SN2012A & 86 & $12.30^{+0.98}_{-0.92}$ \\
SN2012A & 408 & $12.46^{+2.81}_{-3.00}$ \\
SN2012aw & 369 & $14.32^{+3.28}_{-3.29}$ \\
SN2012ch & 357 & $12.35^{+2.80}_{-3.09}$ \\
SN2012ec & 403 & $16.03^{+3.97}_{-4.59}$ \\
SN2012ho & 168 & $11.84^{+1.46}_{-1.27}$ \\
SN2012ho & 229 & $12.18^{+2.30}_{-1.98}$ \\
SN2012ho & 249 & $14.14^{+3.31}_{-3.38}$ \\
SN2012ho & 252 & $13.85^{+3.40}_{-3.28}$ \\
SN2012ho & 265 & $14.01^{+3.29}_{-3.21}$ \\
SN2012ho & 318 & $14.33^{+3.16}_{-3.88}$ \\
SN2012ho & 318 & $14.07^{+3.50}_{-3.64}$ \\
SN2013by & 160 & $15.64^{+3.89}_{-3.39}$ \\
SN2013by & 292 & $17.86^{+2.14}_{-5.50}$ \\
SN2013fs & 270 & $19.93^{+0.07}_{-6.06}$ \\
SN2014G & 137 & $13.08^{+2.02}_{-1.80}$ \\
SN2014G & 187 & $15.97^{+3.33}_{-3.78}$ \\
SN2014G & 341 & $18.28^{+1.72}_{-5.15}$ \\
SN2014cx & 350 & $19.23^{+0.77}_{-5.68}$ \\
SN2015bs & 420 & $19.28^{+0.72}_{-7.38}$ \\
SN2016aqf & 202 & $11.80^{+1.95}_{-1.61}$ \\
SN2016aqf & 259 & $11.64^{+2.03}_{-1.68}$ \\
SN2016aqf & 285 & $12.83^{+2.91}_{-3.14}$ \\
SN2016aqf & 303 & $13.88^{+2.99}_{-3.30}$ \\
SN2016aqf & 330 & $11.98^{+2.40}_{-2.25}$ \\
SN2016aqf & 351 & $12.79^{+2.87}_{-3.08}$ \\
SN2016bkv & 257 & $16.99^{+3.01}_{-4.43}$ \\
SN2016bkv & 436 & $9.08^{+1.26}_{-1.08}$ \\
SN2017eaw & 250 & $13.83^{+3.20}_{-3.23}$ \\
SN2017ivv & 110 & $16.44^{+3.48}_{-3.16}$ \\
SN2017ivv & 136 & $19.51^{+0.49}_{-4.80}$ \\
SN2017ivv & 158 & $20.00^{+0.00}_{-5.01}$ \\
SN2017ivv & 243 & $20.00^{+0.00}_{-4.80}$ \\
SN2017ivv & 279 & $20.00^{+0.00}_{-6.35}$ \\
SN2017ivv & 325 & $20.00^{+0.00}_{-5.89}$ \\
SN2017ivv & 332 & $19.69^{+0.31}_{-5.31}$ \\
SN2018gj & 270 & $11.13^{+1.71}_{-1.65}$ \\
SN2018hwm & 385 & $8.97^{+1.18}_{-0.97}$ \\
SN2018is & 386 & $11.67^{+2.31}_{-2.55}$ \\
SN2020jfo & 217 & $12.29^{+2.41}_{-2.02}$ \\
SN2020jfo & 256 & $15.39^{+3.62}_{-4.26}$ \\
SN2020jfo & 280 & $14.59^{+3.12}_{-4.01}$ \\
SN2020jfo & 308 & $15.15^{+3.02}_{-3.86}$ \\
SN2020jfo & 351 & $15.26^{+3.28}_{-4.08}$ \\
SN2021dbg & 352 & $13.96^{+3.00}_{-3.71}$ \\
SN2021gmj & 385 & $10.55^{+1.79}_{-1.99}$ \\
SN2023ixf & 260 & $18.17^{+1.83}_{-5.28}$ \\
ZTF19abqhobb & 163 & $12.73^{+1.84}_{-1.93}$ \\
ZTF20aaynrrh & 216 & $12.36^{+2.37}_{-1.93}$ \\
ZTF20aaynrrh & 256 & $15.65^{+3.39}_{-4.17}$ \\
\bottomrule
\end{tabular}
\end{minipage}
\hspace{0.015\linewidth}
\begin{minipage}{0.30\linewidth}
\centering
\begin{tabular}{lcc}
\toprule
SN & Phase & $M_{\rm ZAMS}$ \\
 & (days) & ($M_\odot$) \\
\midrule
ZTF20aaynrrh & 280 & $14.65^{+3.18}_{-3.74}$ \\
ZTF20aaynrrh & 308 & $14.68^{+3.23}_{-3.99}$ \\
ZTF20aaynrrh & 351 & $15.23^{+3.36}_{-4.01}$ \\
ZTF20abotkfn & 141 & $12.41^{+1.46}_{-1.36}$ \\
ZTF20abotkfn & 211 & $13.77^{+2.82}_{-2.60}$ \\
ZTF21abouuat & 276 & $16.23^{+3.67}_{-4.16}$ \\
ZTF22abssiet & 175 & $11.05^{+1.14}_{-0.81}$ \\
ZTF22abssiet & 222 & $13.62^{+2.84}_{-3.14}$ \\
ZTF22abtjefa & 136 & $11.82^{+1.40}_{-1.11}$ \\
ZTF22abtjefa & 158 & $12.56^{+1.72}_{-1.53}$ \\
ZTF22abtjefa & 169 & $12.28^{+1.72}_{-1.58}$ \\
ZTF22abtjefa & 215 & $12.14^{+2.13}_{-1.99}$ \\
ZTF22abtjefa & 282 & $12.71^{+3.00}_{-2.58}$ \\
ZTF22abtjefa & 332 & $13.73^{+3.39}_{-3.25}$ \\
ZTF22abtjefa & 366 & $12.85^{+2.68}_{-3.05}$ \\
ZTF22abtjefa & 393 & $13.41^{+3.17}_{-3.15}$ \\
ZTF22abvaetz & 391 & $8.62^{+0.93}_{-0.62}$ \\
ZTF22abyivoq & 258 & $11.38^{+1.59}_{-1.85}$ \\
ZTF22abyivoq & 287 & $11.98^{+2.33}_{-2.44}$ \\
ZTF22abyivoq & 340 & $11.19^{+2.02}_{-2.11}$ \\
ZTF22abyivoq & 369 & $11.46^{+2.44}_{-2.42}$ \\
ZTF22abyivoq & 439 & $11.07^{+2.14}_{-2.18}$ \\
ZTF23aabksje & 387 & $9.85^{+1.81}_{-1.53}$ \\
ZTF23aabksje & 432 & $8.44^{+1.05}_{-0.44}$ \\
ZTF23aackjhs & 296 & $11.89^{+2.77}_{-2.52}$ \\
ZTF23aanxrjm & 231 & $12.79^{+2.80}_{-2.49}$ \\
ZTF23aanxrjm & 264 & $10.79^{+1.87}_{-1.41}$ \\
ZTF23aanxrjm & 309 & $11.08^{+2.03}_{-1.76}$ \\
ZTF23aanxrjm & 347 & $11.24^{+2.37}_{-2.23}$ \\
ZTF23aaquhaz & 138 & $17.67^{+2.33}_{-4.31}$ \\
ZTF23aaquhaz & 310 & $14.98^{+3.65}_{-4.13}$ \\
ZTF23abgmhgw & 165 & $12.41^{+1.72}_{-1.51}$ \\
ZTF23abgmhgw & 210 & $12.13^{+2.27}_{-1.89}$ \\
ZTF23abnogui & 142 & $12.75^{+1.70}_{-1.56}$ \\
ZTF23abnogui & 148 & $13.43^{+2.06}_{-2.07}$ \\
ZTF23abnogui & 172 & $14.44^{+3.25}_{-2.63}$ \\
ZTF23abnogui & 190 & $13.48^{+2.94}_{-2.72}$ \\
ZTF23abnogui & 262 & $11.60^{+2.31}_{-1.75}$ \\
ZTF23abpbuha & 115 & $16.93^{+3.07}_{-3.90}$ \\
ZTF23absdcgi & 337 & $17.97^{+2.03}_{-4.94}$ \\
ZTF23absscow & 250 & $8.00^{+0.00}_{-0.00}$ \\
ZTF23absscow & 342 & $10.88^{+2.39}_{-1.71}$ \\
ZTF23abvgvab & 104 & $18.75^{+1.25}_{-5.04}$ \\
ZTF24aaasazz & 147 & $13.41^{+2.41}_{-2.06}$ \\
ZTF24aaasazz & 353 & $10.28^{+1.78}_{-1.76}$ \\
ZTF24aabppgn & 180 & $11.62^{+1.27}_{-1.25}$ \\
ZTF24aaejecr & 147 & $12.35^{+1.68}_{-1.41}$ \\
ZTF24aaejecr & 291 & $14.56^{+3.63}_{-3.84}$ \\
ZTF24aaejecr & 348 & $14.05^{+3.15}_{-3.84}$ \\
ZTF24aaezido & 180 & $14.03^{+2.98}_{-2.65}$ \\
ZTF24aaplfjd & 132 & $11.94^{+1.03}_{-1.03}$ \\
ZTF24aaplfjd & 184 & $11.16^{+1.16}_{-1.18}$ \\
ZTF24aaplfjd & 394 & $10.50^{+2.00}_{-1.95}$ \\
ZTF24abtczty & 149 & $16.50^{+3.50}_{-3.79}$ \\
ZTF24abtczty & 334 & $10.91^{+1.94}_{-2.11}$ \\
ZTF24abtczty & 393 & $10.51^{+2.50}_{-1.94}$ \\
\bottomrule
\end{tabular}
\end{minipage}
\end{table*}